\documentclass[pdflatex,sn-basic]{sn-jnl}

\jyear{2021}%

\theoremstyle{thmstyleone}%
\theoremstyle{thmstyletwo}%

\theoremstyle{thmstylethree}%

\usepackage{graphicx}	

\def\NOTE#1{{\textcolor{black}{#1}}}

\begin{document}

\title[Unappreciated cross-helicity effects]{Unappreciated cross-helicity effects in plasma physics: Anti-diffusion effects in dynamo and momentum transport}


\author*[1,2]{\fnm{Nobumitsu} \sur{Yokoi}}\email{nobyokoi@iis.u-tokyo.ac.jp}

\affil*[1]{\orgdiv{Institute of Industrial Science}, \orgname{University of Tokyo}, \orgaddress{\street{Komaba, Meguro}, \city{Tokyo}, \postcode{153-8505}, \country{Japan}}}

\affil[2]{\orgdiv{Nordic Institute for Theoretical Physics}, \orgname{Stockholm University and KTH}, \orgaddress{\street{Hannes Alfv\'{e}n v\"{a}g}, \city{Stockholm}, \postcode{106 91}, \country{Sweden}}}


\abstract{The cross helicity (velocity--magnetic-field correlation) effects in the magnetic-field induction and momentum transport in the magnetohydrodynamic (MHD) turbulence are investigated with the aid of the multiple-scale renormalized perturbation expansion analysis, which is a theoretical framework for the strongly non-linear and inhomogeneous turbulence. The outline of the theory is presented with reference to the role of the cross-interaction response functions between the velocity and magnetic field. In this formulation, the expressions of the turbulent fluxes: the turbulent electromotive force (EMF) in the mean induction equation and the Reynolds and turbulent Maxwell stresses in the momentum equation are obtained. Related to the expression of EMF, the physical origin of the cross-helicity effect in dynamos, as well as other dynamo effects, is discussed. Properties of dynamo and momentum transport are determined by the spatiotemporal distribution of turbulence. In order to understand the actual role of the turbulent cross helicity, its transport equations is considered. Several generation mechanisms of cross helicity are discussed with illustrative examples. On the basis of the cross-helicity production mechanisms, its effect in stellar dynamos is discussed. The role of cross helicity in the momentum transport and global flow generation is also argued. One of the situations where the cross-helicity effect both in magnetic-field induction and global flow generation play an important role is the turbulent magnetic reconnection. Characteristic features of turbulence effects in fast reconnection are reviewed with special emphasis on the role of cross helicity in localizing the effective resistivity. Finally, a remark is addressed on an approach that elucidates the structure generation and sustainment in extremely strong turbulence. An appropriate formulation for the anti-diffusion effect, which acts against the usual diffusion effect, is needed. Turbulence modeling approach based on such an analytical formulation is also argued in comparison with the conventional heuristic modeling. The importance of the self-consistent framework treating the non-linear interaction between the mean field and turbulence is stressed as well.}




\maketitle

\section{Introduction}\label{sec1}

Astrophysical and geophysical and plasma physics phenomena, such as the formation of jets on giant planets, the eleven-year solar activity cycle, the generation of Earth's magnetism, the large-scale vortical and magnetic-field structures in interstellar medium, and jets from massive black holes, are characterized by the interaction processes of vast range of spatial and temporal scales. As such these amazing phenomena are almost invariably extremely turbulent. Magnetohydrodynamics (MHD) provides a good framework for understanding such turbulence with the mutual interaction between the fluid flows and magnetic fields. However, because of the vast range of scales it contains from the largest energy-containing scales to the smallest dissipation scales both for viscosity and resistivity, direct computations of the turbulent astrophysical and geophysical flows are simply impossible, even with using most sophisticated algorithms optimized for massively parallel computers. Also, because of the strong non-linear coupling among the modes or scales, we cannot expect any obvious scale separation in strong turbulence. Then, for understanding the nature of turbulent MHD flows, developing a sophisticated statistical analytical theory and modeling turbulence on the basis of the theoretical results are of central importance. Here in this article, an approach for developing such a theory will be introduced. Such a theory will have a considerable impact on constructing a turbulence model that predicts the properties of amazing phenomena we encounter in astrophysics, geophysics, and plasma physics, beyond the conventional heuristic turbulence modeling.

In the theoretical formulation of large-scale structure formation, breakage of mirror-symmetry plays an essential role. Such breakages are represented by pseudo-scalars such as the kinetic, magnetic (current), cross, and other generalized helicities. The statistical properties of helical turbulence and their consequences were reviewed in another review paper \citep{pou2022}, and will not be treated in this paper. Also the basic characteristics of the cross helicity including the conservation, topological interpretation, pseudo-scalar properties, boundedness, relation to Alfv\'{e}n waves, etc.\ were treated in other article \citep{yok2013a}, so the descriptions related to such characteristics will not be repeated in this article, either.  Here, we focus our attention on two aspects of the turbulent MHD transport with primary importance: the magnetic-field induction (dynamos) and the linear and angular momentum transport (global flow generation).

One of the main subjects in the dynamo studies is to explore the effects of turbulence in the mean induction equation. This type of studies is often called {\it the  mean-field dynamo theory}, which has historically gained several connotations. They include (i) kinematic approach; (ii) Ansatz for the turbulent electromotive (EMF) expression; (iii) dynamo coefficients as constant or prescribed parameters; (iv) exclusive $\alpha$--$\Omega$ dynamo scenario for dynamo process; (v) azimuthal averaging as a substitute for the ensemble averaging; and (vi) incompressible treatment of turbulent motions. Some of these treatments have played a role to consider dynamos in a simplified manner. However, none of them is a basic ingredient of the mean-field dynamo theory. In the following, we briefly refer to these treatments (assumptions and approximations) one by one.
	
\noindent{\it (i) Kinematic approach:}
In the kinematic approach, the velocity field in the induction equation is treated as a prescribed one. The evolution of the magnetic field is subject to 
the induction equation with a fixed velocity. The feedback of the magnetic field to the velocity through the Lorentz force ${\bf{j} \times {\bf{b}}}$ in the momentum equation is neglected [${\bf{b}}$: magnetic field, ${\bf{j}} (= \nabla \times {\bf{b}})$: electric-current density]. This kinematic approach of the dynamo has a long history and has provided a good tool for understanding the basic properties of dynamo actions, especially in some simple configuration \citep{tob2021}. However, the dynamo process in the real world is usually non-linear, and the back-reactions to the velocity field by the generated magnetic fields play a crucial role in the turbulence generation and consequently to the dynamo process. For instance, this importance of the feed-back has been stressed in the context of the magnetorotational instability (MRI). Even a very week magnetic field, coupled with a specific configuration of rotation, can contribute to the turbulence generation \citep{bal1998}. Then, resultant dynamo properties are determined by the turbulent motions at least at an developing stage of this MRI process. In this sense, the kinematic approach has only limited relevance to the dynamo process.

\noindent{\it (ii) Ansatz for the turbulent electromotive force (EMF):} Another assumption which has been very often adopted in the mean-field dynamo studies is the Ansatz that the turbulent EMF $\langle {{\bf{u}}' \times {\bf{b}}'} \rangle$ is linearly related to the mean magnetic field ${\bf{B}}$ and its spatial derivatives (${\bf{u}}'$: velocity fluctuation, ${\bf{b}}'$: magnetic-field fluctuation). This ansatz, based on the assumption of the linear relationship between the fluctuating magnetic field ${\bf{b}}'$ and the mean magnetic field ${\bf{B}}$, greatly reduces the complexity of the mean magnetic-field evolution, with the kinematic treatment. However, in order to obtain an expression of the turbulent EMF $\langle {{\bf{u}}' \times {\bf{b}}'} \rangle$ in fully non-linear regime of turbulence, we have to directly treat the equations of the velocity and magnetic-field fluctuations, ${\bf{u}}'$ and ${\bf{b}}'$. These fluctuation fields depend not only on the mean magnetic field ${\bf{B}}$ and its spatial derivatives, but also on the mean velocity ${\bf{U}}$ and its spatial derivatives. In addition, the system of equations of ${\bf{u}}'$ and ${\bf{b}}'$ is fully non-linear in the presence of the fluctuating Lorentz force. As these considerations indicate, the dynamo arguments based on the Ansatz are too simple, and their applicability is limited to very specific simple situations such as the kinematic cases at very low magnetic Reynolds number ($Rm = U L/\eta$, $U$: characteristic velocity, $L$: characteristic length scale, $\eta$: magnetic diffusivity or resistivity). This point will be argued in detail in Section~3.1.

\noindent{\it (iii) Dynamo coefficients as constant or prescribed parameters:} Related to the Ansatz mentioned above, the proportional coefficients for the mean magnetic field ${\bf{B}}$ and its derivatives, $\alpha$, $\eta_{\rm{T}}$, etc.\ are often treated as adjustable parameters. However, this is nothing but the crudest approximation. The transport coefficients for the turbulent fluxes must reflect the statistical properties of turbulence. As this consequence, these transport coefficients are not constants but vary in space and time reflecting the spatiotemporal distributions of the turbulent statistical quantities such as the turbulent energy, turbulent helicity, turbulent cross helicity, the energy dissipation rate, etc.

\noindent{\it (iv) $\alpha$--$\Omega$ framework:}
The usual mean-field dynamo model is constructed in the framework of the $\alpha$--$\Omega$ dynamo. In this framework, the turbulent EMF is assumed to be constituted by the $\alpha$ effect proportional to the mean magnetic field ${\bf{B}}$ and by the turbulent magnetic diffusivity $\eta_{\rm{T}}$ effect proportional to the curl of ${\bf{B}}$. Another crucial ingredient is the inhomogeneous mean velocity that induces the mean magnetic-field component vertical to the original field ($\Omega$ effect). As will be discussed later in Section~3.1, the inhomogeneity of the mean velocity alters not only the mean magnetic-field component but also the properties of fluctuation fields. The equations of the velocity fluctuation ${\bf{u}}'$ and the magnetic-field fluctuation ${\bf{b}}'$ depend on the mean velocity shear. If we retain these mean velocity shears in the ${\bf{u}}'$ and ${\bf{b}}'$ equations, the cross-correlation between ${\bf{u}}'$ and ${\bf{b}}'$, or the turbulent cross helicity $\langle {{\bf{u}}' \cdot {\bf{b}}'} \rangle$ enters the turbulent EMF expression as the coefficient of the mean velocity shear. We see a marked difference between this cross-helicity effect and the usual $\alpha$ effect, which is expressed by the correlation between the velocity and vorticity fluctuations, $\langle {{\bf{u}}' \cdot \mbox{\boldmath$\omega$}'} \rangle$ [$\mbox{\boldmath$\omega$}' (= \nabla \times {\bf{u}}')$: vorticity fluctuation] and the correlation between the magnetic-field fluctuation and the electric-current $\langle {{\bf{b}}' \cdot {\bf{j}}} \rangle$ [${\bf{j}}' (= \nabla \times {\bf{b}}')$: electric-current density fluctuation] coupled with the mean magnetic-field $\bf{B}$. The cross-helicity dynamo effect, which is based on the dependence of the velocity and magnetic-field fluctuations on the mean velocity shear, may pave the way for expanding the applicability of the mean-field dynamo theories to the inhomogeneous turbulence.

\noindent{\it (v) Azimuthal averaging:}
In the mean-field dynamo studies, the azimuthal averaging is often adopted as a substitute for the ensemble averaging. This sometimes severely limits the applicability of the mean-field theories to situations where non-axisymmetric global structures play an essential role in the dynamo process \citep{sch2003}. One way to adopt an averaging procedure other than the azimuthal averaging is to adopt a subgrid-scale (SGS) filtering which allows us to retain large-scale non-axisymmetric structures.

\noindent{\it (vi) Incompressible treatment of turbulence:}
Another ubiquitous assumption in the mean-field dynamo studies is assuming the incompressible turbulence. One of the justifications for the treatment of incompressible turbulence lies in the point that the magnetic induction equation does not explicitly depend on the density. However, statistical properties and dynamic behavior of turbulence is essentially different between the incompressible and compressible cases. There is no reason to discard compressibility in dynamo on phenomena with a high density variance $\langle {\rho'{}^2} \rangle$ ($\rho'$: density fluctuation) \citep{yok2018a,yok2018b}. Such a high density variance corresponds to a high Mach number case. We ubiquitously encounter astrophysical phenomena with a high Mach number, including the star-formation region in the interstellar medium and the interior of massive stars.

None of the above assumptions/approximations are intrinsic condition of the mean-field dynamo theory. Hereafter, the term ``mean-field dynamo'' theory contains some of the above connotations, and leads to misunderstanding on the scope of the theory. Therefore, we prefer using the term ``turbulent dynamo'' to  ``mean-field dynamo'' in the following.

Another important aspect of the turbulent MHD transport in the astrophysical and geophysical plasma context is the turbulent momentum fluxes. The linear and angular momentum transport due to turbulence is represented by the Reynolds stress $\langle {{\bf{u}}' {\bf{u}}'} \rangle = \{ \langle {u'{}^i u'{}^j} \rangle \}$ and the turbulent Maxwell stress $\langle {{\bf{b}}' {\bf{b}}'} \rangle = \{ {\langle {b'{}^i b'{}^j} \rangle} \}$. They are sole quantities directly express the linear and angular momentum transports in the mean momentum equation. In astrophysical and geophysical as well as fusion plasma flow phenomena, turbulence is considered to play an important role in the angular momentum transport. An example is an accretion disk surrounding a central massive astrophysical body such as black hole. In order for the rotating plasma gas to accrete to the central object, the angular momentum should be transported by turbulence \citep{sha1973}.

The angular momentum transport in the stellar convection is another important topic. It has been recently recognized that the current numerical simulations do not capture some basic characteristics of the solar convection. The convective velocity amplitude at large horizontal scales observed by helioseismic investigations is much smaller than the one predicted by global convection simulations \citep{han2012,han2016,pro2021}. The differential rotation profile obtained by the global numerical simulations is not realistic in the sense that the azimuthal rotational velocity in the equator region does not show the solar-like prograde profile as compared with the one in the higher latitude region, but does show the retrograde profile, if the solar values of luminosity (energy transfer rate) and rotation rate are adopted in the simulation. Also, if the large-scale convection motions are actually small, how such weak flows can transfer the solar luminosity and mean differential rotation rate observed in the helioseismology? These discrepancies between the numerical simulations and observations are called the convection conundrum \citep{sch2020}.

In the stellar convection problems, the conventional expressions for the turbulent fluxes (turbulent momentum, heat, and mass fluxes) on the basis of the gradient-diffusion approximation are known to be inappropriate since they give too destructive or dissipative effects on the fluxes. From the viewpoint of theoretical investigation of the turbulent fluxes, there may be several possibilities to resolve this convection conundrum. One possible way is to introduce the effect of coherent fluctuating motions such as plumes, thermals, and jets in the convective turbulence. Implementation of the coherent-structure effect into the modeling of convective turbulence has been considered important to capture the transport properties of turbulence \citep{ras1998,bra2016,gre2020}. Recently, the relevance of the non-equilibrium effect of turbulence along the plume motions has been pointed out in the context of the turbulence modeling of stellar convection \citep{yok2022}. Since the non-equilibrium effect along the coherent fluctuating motions alters the length scale and timescale of turbulence, it directly affects the transport properties of turbulence.

The second possible way is to implement the rotation or global vorticity effect. \NOTE{Rotation is considered to suppress the convective motion and enhance the thermal transport at relatively smaller scales \citep{vas2021}.} One of the studies to incorporate the rotation effect into the linear and angular momentum transport is through the kinetic-helicity inhomogeneity. It was found that, in the Reynolds stress expression, the absolute vorticity $\mbox{\boldmath$\Omega$}_\ast (\equiv \mbox{\boldmath$\Omega$} + 2 \mbox{\boldmath$\omega$}_{\rm{F}})$ [$\mbox{\boldmath$\Omega$} (= \nabla \times {\bf{U}})$: the mean vorticity, $\mbox{\boldmath$\omega$}_{\rm{F}}$: rotation] is coupled with the inhomogeneous turbulent helicity represented by the gradient of turbulent helicity, $\nabla \langle {{\bf{u}}' \cdot \mbox{\boldmath$\omega$}'} \rangle$ \citep{yok1993}. Unlike the eddy-viscosity effect, which arises from the turbulent energy coupled with the mean velocity strain rate (the symmetric part of the mean velocity shear) and destroys the large-scale inhomogeneous flow structure by strong effective viscosity, this inhomogeneous turbulent helicity effect arises from the presence of non-uniform kinetic helicity coupled with the mean absolute vorticity (the anti-symmetric part of the mean velocity shear). The latter effect is expected to contribute to generating and sustaining non-trivial large-scale flow structures such as prograde differential rotation in the solar convection zone \citep{yok2023}. Since one of the generation mechanisms of turbulent helicity is coupling of a rotation and inhomogeneities of energy, density, etc.\ along the rotation axis, inhomogeneous turbulent helicity (spatial distribution and segregation) is ubiquitously present in a rotating spherical configuration \citep{dua2016,ran2023}. In this sense, the inhomogeneous helicity effect is expected to play an important role in the angular momentum transport in the stellar convection zone.

	The third possible way is to incorporate the magnetic shear effect into the Reynolds and turbulent Maxwell stresses. In this effect, the transport coefficient is expressed by the turbulent cross helicity $\langle {{\bf{u}}' \cdot {\bf{b}}'} \rangle$. The presence of the magnetic fluctuation ${\bf{b}}'$ in statistical correlation with the velocity fluctuation ${\bf{u}}'$ is a crucial ingredient of this cross-helicity effect in the linear and angular momentum transport.

	These theoretical consideration suggests that, in addition to the usual eddy-viscosity effect coupled with the mean velocity strain, the non-equilibrium effect coupled with the coherent component of fluctuations (plumes, thermals, and jets), the inhomogeneous helicity effect coupled with the mean absolute vorticity (rotation and large-scale vortical motion), and the cross-helicity effect coupled with the mean magnetic-field strain may play some role in the linear and angular momentum transport. These effects have not been well explored in the previous studies of the turbulent fluxes in the stellar convection. The first two effects have been argued in the other papers: the non-equilibrium effect along the plume motions in the stellar convection in \citet{yok2022} and the inhomogeneous turbulent helicity effect in \citet{yok1993,yok2016a,yok2023}. In the present article, we will discuss the cross-helicity effect coupled with the global magnetic-field shear in the momentum transport, later in Section~6.

	The organization of this article is as follows. Following Introduction, the theoretical formulation for the inhomogeneous MHD turbulence is presented in Section~2. Our theoretical framework: the two-scale direct-interaction approximation, a multiple-scale renormalized perturbation expansion theory is briefly outlined. Special reference is put to the distinction of the simple self-response Green's functions and the cross-interaction Green's functions. As examples of the application of the theory, the analytical expressions for the turbulent electromotive force in the mean magnetic induction equation and the Reynolds and turbulent Maxwell stresses in the mean momentum equation are presented. In Section~3, the cross-helicity effect in the dynamo is discussed. First, the physical origin of the cross-helicity effect, as well as the counterparts of the kinetic and electric-current helicity effects and the turbulent magnetic diffusivity effect, will be argued. Some numerical validations of the cross-helicity effect and related dynamo are presented. The properties of cross-helicity dynamo depend on the spatiotemporal evolution of the turbulent cross helicity. In order to better understand the cross helicity evolution, the transport equation of the turbulent cross helicity is discussed in Section~4. The representative production mechanisms of the turbulent cross helicity are illustrated with special reference to the field configurations leading to the cross-helicity generation. In addition, the evaluation of the cross-helicity dissipation rate is presented through the derivation of the equation of the cross-helicity dissipation rate. In Section~5, the cross helicity effect is applied to the oscillatory stellar dynamo. Unlike the kinetic and current helicities, the cross helicity is expected to change its sign across the reversal of magnetic field. This property of cross helicity is fully investigated in the framework of the mean-field dynamo equations. 
\NOTE{
	In Section~6, another important aspect of the cross helicity effects, contribution in the momentum equation, is discussed. In the presence of non-trivial mean magnetic-field configuration, the turbulent cross helicity plays an important role in the linear and angular momenta transport. The physical origin of the flow generation is discussed with reference to the role of the fluctuating Lorentz force in the momentum equation. Both of these effects of turbulent cross-helicity: the magnetic induction and flow generation are expected to arise in the magnetic reconnection phenomena. In Section~7, the turbulent effects in magnetic reconnection are argued. The cross helicity, which counterbalances the turbulent magnetic diffusivity effect, contributes to the localization of the effective diffusivity, leading to the fast reconnection. The concluding remarks are given in Section~8, where a special emphasis is given in the counter-diffusion effect in the turbulent EMF and the Reynolds and turbulent Maxwell stresses.
}

\section{Theoretical formulation}\label{sec:2}
In this section, we present the basic procedure of a multiple-scale renormalized perturbation expansion theory, the two-scale direct-interaction approximation (TSDIA) \citep{yos1984,yok2020}. To be specific, we discuss the incompressible magnetohydrodynamic (MHD) turbulence \citep{yos1990,ham2008,yok2013a}, which is favored for presenting the basic properties of MHD turbulent transport It should be understood, however, the following formulation can be readily adapted to the strongly compressible MHD turbulence as well. The calculations are cumbersome, but are straightforward \citep{yok2018a,yok2018b}.

\subsection{Fundamental equations}\label{sec:2.1}
In order to show the basic theoretical formulation for a fluid plasma turbulence, we present here the fundamental equations of MHD for the incompressible or non-variable density fluid.

	In the non-variable density case, a plasma fluid obeys the equations of incompressible magnetohydrodynamics, which are constituted by the momentum equation:
\begin{equation}
	\frac{\partial {\bf{u}}}{\partial t}
	+ ({\bf{u}} \cdot \nabla) {\bf{u}}
	= ({\bf{b}} \cdot \nabla) {\bf{b}}
	- \nabla p_{\rm{M}} 
	- 2 \mbox{\boldmath$\omega$}_{\rm{F}} \times {\bf{u}}
	+ \nu \nabla^2 {\bf{u}},
	\label{eq:inst_vel_eq}
\end{equation}
magnetic induction equation:
\begin{equation}
	\frac{\partial {\bf{b}}}{\partial t}
	+ ({\bf{u}} \cdot \nabla) {\bf{b}}
	= ({\bf{b}} \cdot \nabla) {\bf{u}}
	+ \eta \nabla^2 {\bf{b}},
	\label{eq:inst_mag_ind_eq}
\end{equation}
and the solenoidal conditions of the velocity and magnetic field:
\begin{equation}
	\nabla \cdot {\bf{u}}
	= \nabla \cdot {\bf{b}}
	= 0,
	\label{eq:inst_sol_conds}
\end{equation}
where ${\bf{u}}$ is the velocity, ${\bf{b}}$ the magnetic field, $p_{\rm{M}} (= p + {\bf{b}}^2/2)$ the MHD pressure ($p$: gas pressure), $\mbox{\boldmath$\omega$}_{\rm{F}}$ the angular velocity of a rotation, $\nu$ the kinematic viscosity, and $\eta$ the magnetic diffusivity. Here, the magnetic field is measured in the Alfv\'{e}n speed unit defined by ${\bf{b}} = {\bf{b}}_\ast / (\mu_0 \rho)^{1/2}$ (${\bf{b}}_\ast$: magnetic field measured in the natural unit, $\mu$: magnetic permeability, $\rho$: density of fluid).

\subsection{Mean and fluctuation in multiple-scale analysis}
\subsubsection{Mean and fluctuation}
	In order to see the turbulent effect on the mean fields, we divide a field quantity $f$ into the mean $F$ and the fluctuation around it, $f'$, as
\begin{equation}
	f = F + f',\;\; F = \langle {f} \rangle
	\label{eq:rey_decomp}
\end{equation}
with
\begin{subequations}
\begin{equation}
	f = ({\bf{u}}, {\bf{b}}, p, \mbox{\boldmath$\omega$}, {\bf{j}}),
	\label{eq:f_defs}
\end{equation}
\begin{equation}
	F = ({\bf{U}}, {\bf{B}}, P, \mbox{\boldmath$\Omega$}, {\bf{J}}),
	\label{eq:F_defs}
\end{equation}
\begin{equation}
	f' = ({\bf{u}}', {\bf{b}}', p', \mbox{\boldmath$\omega$}', {\bf{j}}').
	\label{eq:f'_defs}
\end{equation}
\end{subequations}
Here, $\mbox{\boldmath$\omega$} (= \nabla \times {\bf{u}})$ is the vorticity, ${\bf{j}} (= \nabla \times {\bf{b}})$ is the electric-current density, and $\langle {\cdots} \rangle$ denotes the ensemble average.

	Under this decomposition, the equations for the mean velocity ${\bf{U}}$ and the mean magnetic field ${\bf{B}}$ are written as
\begin{equation}
	\frac{\partial {\bf{U}}}{\partial t}
	+ ({\bf{U}} \cdot \nabla) {\bf{U}}
	= ({\bf{B}} \cdot \nabla) {\bf{B}}
	- \nabla \cdot \mbox{\boldmath${\cal{R}}$}
	- \nabla P_{\rm{M}} 
	- 2 \mbox{\boldmath$\omega$}_{\rm{F}} \times {\bf{U}}
	+ \nu \nabla^2 {\bf{U}},
	\label{eq:mean_U_eq}
\end{equation}
\begin{equation}
	\frac{\partial {\bf{B}}}{\partial t}
	+ ({\bf{U}} \cdot \nabla) {\bf{B}}
	= ({\bf{B}} \cdot \nabla) {\bf{U}}
	- \langle {({\bf{u}}' \cdot \nabla) {\bf{b}}'} \rangle
	+ \langle {({\bf{b}}' \cdot \nabla) {\bf{u}}'} \rangle
	+ \eta \nabla^2 {\bf{B}},
	\label{eq:mean_B_eq}
\end{equation}
with the solenoidal conditions
\begin{equation}
	\nabla \cdot {\bf{U}}
	= \nabla \cdot {\bf{B}}
	= 0,
	\label{eq:mean_sol_conds}
\end{equation}
Alternatively, the mean magnetic field equation (\ref{eq:mean_B_eq}) is rewritten in the rotational form as
\begin{equation}
	\frac{\partial {\bf{B}}}{\partial t}
= \nabla \times ({\bf{U}} \times {\bf{B}})
+ \nabla \times {\bf{E}}_{\rm{M}}
+ \eta \nabla^2 {\bf{B}}.
	\label{eq:mean_B_rot_eq}
\end{equation}
In (\ref{eq:mean_U_eq}), $\mbox{\boldmath${\cal{R}}$} = \{ {{\cal{R}}^{ij}} \}$ is the Reynolds and turbulent Maxwell stresses defined by
\begin{equation}
	{\cal{R}}^{ij}
	\equiv \langle {
		u'{}^i u'{}^j
		- b'{}^i b'{}^j
	} \rangle
	\label{eq:rey_turb_max_str_def}
\end{equation}
and $P_{\rm{M}}$ is the mean part of the total MHD pressure $p_{\rm{M}} \equiv p + {\bf{b}}^2/2$, defined by
\begin{equation}
	P_{\rm{M}} 
	= P + {\bf{B}}^2/2 + \langle {{\bf{b}}'{}^2} \rangle/2.	
	\label{eq:P_M_def}
\end{equation}
In (\ref{eq:mean_B_rot_eq}), ${\bf{E}}_{\rm{M}}$ is the turbulent electromotive force (EMF) defined by
\begin{equation}
	{\bf{E}}_{\rm{M}}
	\equiv \langle {
		{\bf{u}}' \times {\bf{b}}'
	} \rangle.
	\label{turb_emf_def}
\end{equation}

	On the other hand, the equations of the velocity fluctuation ${\bf{b}}'$ and the magnetic-field fluctuation ${\bf{b}}'$ are written as
\begin{eqnarray}
	\frac{\partial {\bf{u}}'}{\partial t}
	&+& ({\bf{U}} \cdot \nabla) {\bf{u}}'
	= - ({\bf{u}}' \cdot \nabla) {\bf{U}}
	+ ({\bf{B}} \cdot \nabla) {\bf{b}}'
	+ ({\bf{b}}' \cdot \nabla) {\bf{B}}
	\nonumber\\
	&-& ({\bf{u}}' \cdot \nabla) {\bf{u}}'
	+ ({\bf{b}}' \cdot \nabla) {\bf{b}}'
	+ \nabla \cdot \mbox{\boldmath${\cal{R}}$}
	- \nabla p'_{\rm{M}} 
	- 2 \mbox{\boldmath$\omega$}_{\rm{F}} \times {\bf{u}}'
	+ \nu \nabla^2 {\bf{u}}',
	\label{eq:fluct_u_eq}
\end{eqnarray}
\begin{eqnarray}
	\frac{\partial {\bf{b}}'}{\partial t}
	&+& ({\bf{U}} \cdot \nabla) {\bf{b}}'
	= - ({\bf{u}}' \cdot \nabla) {\bf{B}}
	+ ({\bf{B}} \cdot \nabla) {\bf{u}}'
	+ ({\bf{b}}' \cdot \nabla) {\bf{U}}
	\nonumber\\
	&-& ({\bf{u}}' \cdot \nabla) {\bf{b}}'
	+ ({\bf{b}}' \cdot \nabla) {\bf{u}}'
	+ \langle {({\bf{u}}' \cdot \nabla) {\bf{b}}'} \rangle
	- \langle {({\bf{b}}' \cdot \nabla) {\bf{u}}'} \rangle
	+ \eta \nabla^2 {\bf{b}}'
	\label{eq:fluct_b_eq}
\end{eqnarray}
with the solenoidal conditions
\begin{equation}
	\nabla \cdot {\bf{u}}'
	= \nabla \cdot {\bf{b}}'
	= 0.
	\label{eq:fluct_sol_conds}
\end{equation}
In (\ref{eq:fluct_u_eq}), $p_{\rm{M}}'$ is the fluctuating MHD pressure defined by $p_{\rm{M}}' = p_{\rm{M}} - P_{\rm{M}}$.

\subsubsection{Multiple-scale analysis}
	Considering that mean fields vary slowly at large scales while fluctuations do fast at small scales, we introduce two scales: slow and fast variables as
\begin{equation}
	\mbox{\boldmath$\xi$} (= {\bf{x}}),\;
	{\bf{X}} (= \delta {\bf{x}}),\;
	\tau (= t),\;
	T (= \delta t),
	\label{eq:fast_slow_vars}
\end{equation}
where $\delta$ is the scale parameter. If $\delta$ is small, ${\bf{X}}$ and $T$ change substantially only when ${\bf{x}}$ and $t$ vary considerably. In this sense, ${\bf{X}}$ and $T$, which are suitable for describing the slow and large variations, are called slow variables. On the other hand, $\mbox{\boldmath$\xi$}$ and $\tau$ are called the fast variables. The scale parameter $\delta$ is not necessarily small, but if $\delta$ is small ($\delta \ll 1$), there is a large scale separation between the slow and fast variables. Because of the introduction of two scales defined by (\ref{eq:fast_slow_vars}), the spatial and temporal derivatives are expressed as
\begin{equation}
	\nabla_{\bf{x}} 
	= \nabla_{\mbox{\boldmath$\xi$}} 
	+ \delta \nabla_{\bf{X}};\;\;
	\frac{\partial}{\partial t}
	= \frac{\partial}{\partial \tau}
	+ \delta \frac{\partial}{\partial T}.
	\label{eq:der_expan_def}
\end{equation}
This means that the derivatives with respect to the slow variables ${\bf{X}}$ and $T$ show up with a scale parameter $\delta$. Expansions with respect to $\delta$ are derivative expansions. With these slow and fast variables, a field quantity $f$ is expressed as
\begin{equation}
	f({\bf{x}};t)
	= F({\bf{X}};T)
	+ f'(\mbox{\boldmath$\xi$},{\bf{X}}; \tau,T).
	\label{eq:fld_decomp_in_multiple}
\end{equation}
Note that the fluctuating field $f'$ depends on the slow variables ${\bf{X}}$ and $T$ as well as on the fast variables $\mbox{\boldmath$\xi$}$ and $\tau$. Such a dependence of fluctuating fields on the slow variables is of essential importance for describing inhomogeneous turbulence.

\subsubsection{Mean- and fluctuation-field equations}
	In this two-scale formulation, the equations of the fluctuating velocity ${\bf{u}}'$ is written as
\begin{eqnarray}
	\frac{\partial u'{}^i}{\partial \tau}
	&+& U^j \frac{\partial u'{}^i}{\partial \xi^j}
	+ \frac{\partial}{\partial \xi^j} 
		(u'{}^j u'{}^i - b'{}^j b'{}^i)
	+ \frac{\partial p'_{\rm{M}}}{\partial \xi^i}
	- \nu \frac{\partial^2 u'{}^i}{\partial \xi^j \partial \xi^j}
	- B^j \frac{\partial b'{}^i}{\partial \xi^j}
	\nonumber\\
	&=& \delta \left[ {
		b'{}^j \frac{\partial B^i}{\partial X^j}
		- u'{}^j \left( {
			\frac{\partial U^i}{\partial X^j}
			+ \epsilon^{jik} \Omega_0^k
		} \right)
	+ B^j \frac{\partial b'{}^i}{\partial X^j}
	- \frac{\widetilde{D} u'{}^i}{DT}
	} \right.
	\nonumber\\
	&& \left. { \hspace{25pt}
	- \frac{\partial}{\partial X^j} \left( {
		u'{}^j u'{}^i - b'{}^j b'{}^i
		- \langle {u'{}^j u'{}^i - b'{}^j b'{}^i} \rangle
		} \right)
		- \frac{\partial p'_{\rm{M}}}{\partial X^i}
	} \right],
	\label{eq:ts_fluct_u_eq}
\end{eqnarray}
and the solenoidal condition:
\begin{equation}
	\frac{\partial u'{}^j}{\partial \xi^j}
	+ \delta \frac{\partial u'{}^j}{\partial X^j}
	= 0.
	\label{eq:ts_fluct_u_sol_cond}
\end{equation}
The counterparts of the fluctuating magnetic field ${\bf{b}}'$ is written as
\begin{eqnarray}
	\frac{\partial b'{}^i}{\partial \tau}
	&+& U^j \frac{\partial b'{}^i}{\partial \xi^j}
	+ \frac{\partial}{\partial \xi^j} 
		(u'{}^j b'{}^i - b'{}^j u'{}^i)
	- \eta \frac{\partial^2 b'{}^i}{\partial \xi^j \partial \xi^j}
	- B^j \frac{\partial u'{}^i}{\partial \xi^j}
	\nonumber\\
	&=& \delta \left[ {
		u'{}^j \frac{\partial B^i}{\partial X^j}
		+ b'{}^j \left( {
		\frac{\partial U^i}{\partial X^j}
		+ \epsilon^{jik} \Omega_0^k
		} \right)
	+ B^j \frac{\partial u'{}^i}{\partial X^j}
	- \frac{\widetilde{D} b'{}^i}{DT}
	} \right.
	\nonumber\\
	&& \left. { \hspace{25pt}
	- \frac{\partial}{\partial X^j} \left( {
		u'{}^j b'{}^i - b'{}^j u'{}^i
		- \langle {u'{}^j b'{}^i - b'{}^j u'{}^i} \rangle
    } \right)
	} \right],
	\label{eq:ts_fluct_b_eq}
\end{eqnarray}
and the solenoidal condition:
\begin{equation}
	\frac{\partial b'{}^j}{\partial \xi^j}
	+ \delta \frac{\partial b'{}^j}{\partial X^j}
	= 0.
	\label{eq:ts_fluct_b_sol_cond}
\end{equation}
Here, in order to keep the material derivatives to be objective, we adopt a co-rotational derivative
\begin{equation}
	\frac{\widetilde{D} u'{}^i}{DT}
	= \frac{\partial u'{}^i}{\partial t}
	+ U^j \frac{\partial u'{}^i}{\partial x^j}
	+ \epsilon^{jik} \Omega_0^k u'{}^i
	\label{eq:corot_deriv_def}
\end{equation}
with
\begin{equation}
	\mbox{\boldmath$\Omega$}_0
	= \mbox{\boldmath$\omega$}_{\rm{F}}/\delta,
	\label{eq:Omega_0_def}
\end{equation}
in place of the Lagrange or advective derivative
\begin{equation}
	\frac{Du'{}^i}{DT}
	= \frac{\partial u'{}^i}{\partial t}
	+ U^j \frac{\partial u'{}^i}{\partial x^j},
	\label{eq:lag_deriv_def}
\end{equation}
which is not objective with respect to a rotation. Note that the mean velocity gradient in the fluctuation equations:
\begin{equation}
	\frac{\partial U^i}{\partial X^j}
	+ \epsilon^{jik} \Omega_0^k
	= \frac{1}{2} \left( {
		\frac{\partial U^i}{\partial X^j}
		+ \frac{\partial U^j}{\partial X^i}
	} \right)
	+ \frac{1}{2} \left( {
		\frac{\partial U^i}{\partial X^j}
		- \frac{\partial U^j}{\partial X^i}
		+ 2 \epsilon^{jik} \Omega_0^k
	} \right)
	\label{eq:object_mean_vel_grad}
\end{equation}
is objective since the both the strain-rate tensor and the absolute-vorticity tensor are objective \citep{thi2001,ham2006}.

	A field quantity $f'(\mbox{\boldmath$\xi$},{\bf{X}};\tau,T)$ is Fourier transformed with respect to the fast spatial variable $\mbox{\boldmath$\xi$}$ as
\begin{equation}
	f'(\mbox{\boldmath$\xi$},{\bf{X}};\tau,T)
	= \int d{\bf{k}} f({\bf{k}},{\bf{X}};\tau,T) 
		\exp[- i {\bf{k}} \cdot 
			(\mbox{\boldmath$\xi$} - {\bf{U}} \tau)],
	\label{eq:fourier_transform}
\end{equation}
where the Fourier transform of the fast variable is taken in the frame co-moving with the local mean velocity ${\bf{U}}$. Hereafter, for the sake of simplicity of notation, the arguments of the slow variable for the fluctuation field $f(\mbox{\boldmath$\xi$},{\bf{X}};\tau,T)$ is suppressed and just denoted as $f(\mbox{\boldmath$\xi$};\tau)$.

	We apply the Fourier transformation (\ref{eq:fourier_transform}) to (\ref{eq:ts_fluct_u_eq}) and (\ref{eq:ts_fluct_b_eq}) and to the solenoidal conditions (\ref{eq:ts_fluct_u_sol_cond}) and (\ref{eq:ts_fluct_b_sol_cond}). Then we obtain the system of two-scale differential equations as
\begin{eqnarray}
	\frac{\partial u^i({\bf{k}};\tau)}{\partial \tau}
	&+& \nu k^2 u^i({\bf{k}};\tau)
	\nonumber\\
	&-& i M^{ij\ell}({\bf{k}}) \iint d{\bf{p}} d{\bf{q}}\ 
		\delta({\bf{k}} - {\bf{p}} - {\bf{q}}) \times
	\nonumber\\
	&&\hspace{60pt} \left[ {
		u^j({\bf{p}};\tau) u^\ell({\bf{q}};\tau)
		- b^j({\bf{p}};\tau) b^\ell({\bf{q}};\tau)
	} \right]
	\nonumber\\
	&+& i k^j B^j b^i({\bf{k}};\tau)
	\nonumber\\
	&=& \delta \left[ {
	- D^{ij}({\bf{k}}) 
	\frac{\widetilde{D} u^j({\bf{k}};\tau)}{D T_{\rm{I}}}
	- D^{ij}({\bf{k}}) u^m({\bf{k}};\tau) \left( {
		\frac{\partial U^j}{\partial x^m}
		+ \epsilon^{mj\ell} \Omega_0^\ell
		} \right)
	} \right.
	\nonumber\\
	&&  \left. { \hspace{22pt}
	+ B^j \frac{\partial b^i({\bf{k}};\tau)}{\partial X_{\rm{I}}^j}
	+ b^j({\bf{k}};\tau) \frac{\partial B^i}{\partial x^j}
	} \right],
	\label{eq:ts_fluct_u_wave_eq}
\end{eqnarray}
\begin{equation}
	- i k^ j u^j({\bf{k}};\tau)
	+ \delta \frac{\partial u^j({\bf{k}};\tau)}{\partial X^j}
	= 0,
	\label{eq:ts_fluct_u_wave_sol_cond}
\end{equation}
\begin{eqnarray}
	\frac{\partial b^i({\bf{k}};\tau)}{\partial \tau}
	&+& \eta k^2 b^i({\bf{k}};\tau)
	\nonumber\\
	&+& i N^{ij\ell}({\bf{k}}) \iint d{\bf{p}} d{\bf{q}}\ 
	\delta({\bf{k}} - {\bf{p}} - {\bf{q}}) \times
	\nonumber\\
	&&\hspace{60pt} \left[ {
		b^j({\bf{p}};\tau) u^\ell({\bf{q}};\tau)
		- u^j({\bf{p}};\tau) b^\ell({\bf{q}};\tau)
	} \right]
	\nonumber\\
	&+& i k^j B^j u^i({\bf{k}};\tau)
	\nonumber\\
	&=& \delta \left[ {
	- D^{ij}({\bf{k}}) 
		\frac{\widetilde{D} b^j({\bf{k}};\tau)}{D T_{\rm{I}}}
	- D^{ij}({\bf{k}};\tau) b^m({\bf{k}};\tau) \left( {
		\frac{\partial U^j}{\partial x^m}
		+ \epsilon^{mj\ell} \Omega_0^\ell
		} \right)
	} \right.
	\nonumber\\
	&&  \left. { \hspace{22pt}
		+ B^j \frac{\partial u^i({\bf{k}};\tau)}{\partial X_{\rm{I}}^j}
		+ u^j({\bf{k}};\tau) \frac{\partial B^i}{\partial X^j}
	} \right],
	\label{eq:ts_fluct_b_wave_eq}
\end{eqnarray}
\begin{equation}
	- i k^ j b^j({\bf{k}};\tau)
	+ \delta \frac{\partial b^j({\bf{k}};\tau)}{\partial X^j}
	= 0,
	\label{eq:ts_fluct_b_wave_sol_cond}
\end{equation}
where
\begin{equation}
	\left( {\nabla_{{\bf{X}}{\rm{I}}}, \frac{D}{DT_{\rm{I}}}}  \right)
	= \exp \left( {-i {\bf{k}} \cdot {\bf{U}}\tau} \right) 
		\left( {\nabla_{\bf{X}}, \frac{D}{DT}} \right)
		\exp \left( {i {\bf{k}} \cdot {\bf{U}} \tau} \right)
	\label{eq:int_act_repr}
\end{equation}
is the differential operators in the interaction representation. Here in (\ref{eq:ts_fluct_u_wave_eq}) and (\ref{eq:ts_fluct_b_wave_eq}), 
\begin{equation}
	M^{ijk}({\bf{k}})
	= k^j D^{ik}({\bf{k}}) + k^k D^{ij}({\bf{k}}),
	\label{eq:vertex_proj_op_def}
\end{equation}
with the solenoidal projection operator
\begin{equation}
	D^{ij}({\bf{k}}) = \delta^{ij} - \frac{k^i k^j}{k^2},
	\label{eq:project_op_def}
\end{equation}
and
\begin{equation}
	N^{ijk}({\bf{k}}) = k^j \delta^{ik} - k^k \delta^{ij}.
	\label{eq:b_project_op}
\end{equation}
They represent the nonlinear interaction among the different modes.

\subsection{Field equations} 

\subsubsection{Scale-parameter expansion}
We expand a field $f(\mbox{\boldmath$\xi$};\tau)$ with respect to the scale parameter $\delta$, and further expand each field by the external field (the mean magnetic field in this particular case).
\begin{eqnarray}
	f^i({\bf{k}};\tau)
	&=& \sum_{n=0}^{\infty} 
		\delta^n  f_{n}^i({\bf{k}};\tau)
	- \sum_{n=0}^{\infty}
		\delta^{n+1} i \frac{k^i}{k^2} 
		\frac{\partial}{\partial X_{\rm{I}}^{j}} f_{n}^j({\bf{k}};\tau)
	\nonumber\\
	&=& \sum_{n=0}^{\infty} \sum_{m=0}^{\infty}
		\delta^n  f_{nm}^i({\bf{k}};\tau)
	- \sum_{n=0}^{\infty} \sum_{m=0}^{\infty}
		\delta^{n+1} i \frac{k^i}{k^2} 
		\frac{\partial}{\partial X_{\rm{I}}^{j}} f_{nm}^j({\bf{k}};\tau).
	\label{eq:scale_para_expansion}
\end{eqnarray}
In this two-scale formulation, inhomogeneities and anisotropies enter with the scale parameter $\delta$ and the external parameters ${\bf{B}}$ in higher-order fields. The lowest-order fields $f_{00}$ correspond to the homogeneous and isotropic turbulence.

	Using the expansion (\ref{eq:scale_para_expansion}), we write the equations of each order in matrix form. With the abbreviated form of the spectral integral
\begin{equation}
	\int_\Delta
	= \iint d{\bf{p}} d{\bf{q}}\ \delta({\bf{k}} - {\bf{p}} - {\bf{q}}),
	\label{eq:abbrev_spect_int}
\end{equation}
the $f_{00}({\bf{k}};\tau)$ equations are given as
\begin{eqnarray}
	&&{\everymath{\displaystyle}\left( {
	\begin{array}{cc}
		\frac{\partial}{\partial \tau} + \nu k^2 & 0\\
		0 & \frac{\partial}{\partial \tau} + \eta k^2
	\end{array}
	} \right)}
	\left( {
	\begin{array}{cc}
		u_{00}^{i}({\bf{k}};\tau)\\
		b_{00}^{i}({\bf{k}};\tau)\rule{0.ex}{5.ex}
	\end{array}
	} \right)
	\nonumber\\
	&& \hspace{-00pt} {\everymath{\displaystyle}
	+ i \left( {
	\begin{array}{cc}
		- M^{ij\ell}({\bf{k}}) \int_\Delta u_{00}^j({\bf{p}};\tau) 
		& M^{ij\ell}({\bf{k}}) \int_\Delta b_{00}^j({\bf{p}};\tau)\\
			N^{ij\ell}({\bf{k}}) \int_\Delta b_{00}^j({\bf{p}};\tau) 
  		& - N^{ij\ell}({\bf{k}}) \int_\Delta u_{00}^j({\bf{p}};\tau)
	\end{array}
	} \right)
	\left( {
	\begin{array}{c}
		u_{00}^{\ell}({\bf{q}};\tau)\\
		b_{00}^{\ell}({\bf{q}};\tau)\rule{0.ex}{5.ex}
	\end{array}
	} \right)}
	\nonumber\\
	&&= \left( {
	\begin{array}{c}
		0\\
		0\rule{0.ex}{5.ex}
	\end{array}
	} \right),
	\label{eq:f00_eq}
\end{eqnarray}
the $f_{01}({\bf{k}};\tau)$ equations are given as
\begin{eqnarray}
	&&{\everymath{\displaystyle}\left( {
	\begin{array}{cc}
		\frac{\partial}{\partial \tau} + \nu k^2 & 0\\
		0 & \frac{\partial}{\partial \tau} + \eta k^2
	\end{array}
	} \right)}
	\left( {
	\begin{array}{cc}
		u_{01}^{i}({\bf{k}};\tau)\\
		b_{01}^{i}({\bf{k}};\tau)\rule{0.ex}{5.ex}
	\end{array}
	} \right)
	\nonumber\\
	&& {\everymath{\displaystyle}
	+ i \left( {
	\begin{array}{cc}
		- 2M^{ij\ell}({\bf{k}}) \int_\Delta u_{00}^j({\bf{p}};\tau) 
		& 2M^{ij\ell}({\bf{k}}) \int_\Delta b_{00}^j({\bf{p}};\tau)\\
  			N^{ij\ell}({\bf{k}}) \int_\Delta b_{00}^j({\bf{p}};\tau) 
  		& - N^{ij\ell}({\bf{k}}) \int_\Delta u_{00}^j({\bf{p}};\tau)
	\end{array}
	} \right)
	\left( {
	\begin{array}{c}
		u_{01}^{\ell}({\bf{q}};\tau)\\
		b_{01}^{\ell}({\bf{q}};\tau)\rule{0.ex}{5.ex}
	\end{array}
	} \right)}
	\nonumber\\
	&& = - i k^j B^j 
	\left( {
	\begin{array}{cc}
		0 & 1\\
		1 & 0 \rule{0.ex}{5.ex}
	\end{array}
	} \right)
	\left( {
	\begin{array}{c}
		u_{00}^i({\bf{k}};\tau)\\
		b_{00}^i({\bf{k}};\tau)\rule{0.ex}{5.ex}
	\end{array}
	} \right)
	\equiv \left( {
	\begin{array}{c}
		F_{01u}^i\\
		F_{01b}^i\rule{0.ex}{5.ex}
	\end{array}
	} \right),
	\label{eq:f01_eq}
\end{eqnarray}
and the $f_{10}({\bf{k}};\tau)$ equations are
\begin{eqnarray}
	&&{\everymath{\displaystyle}\left( {
	\begin{array}{cc}
		\frac{\partial}{\partial \tau} + \nu k^2 & 0\\
		0 & \frac{\partial}{\partial \tau} + \eta k^2
	\end{array}
	} \right)}
	\left( {
	\begin{array}{cc}
		u_{10}^{i}({\bf{k}};\tau)\\
		b_{10}^{i}({\bf{k}};\tau)\rule{0.ex}{5.ex}
	\end{array}
	} \right)
	\nonumber\\
	&& {\everymath{\displaystyle}
	+ i \left( {
	\begin{array}{cc}
		- 2M^{ij\ell}({\bf{k}}) \int_\Delta u_{00}^j({\bf{p}};\tau) 
		& 2M^{ij\ell}({\bf{k}}) \int_\Delta b_{00}^j({\bf{p}};\tau)\\
			N^{ij\ell}({\bf{k}}) \int_\Delta b_{00}^j({\bf{p}};\tau) 
  		& - N^{ij\ell}({\bf{k}}) \int_\Delta u_{00}^j({\bf{p}};\tau)
	\end{array}
	} \right)
	\left( {
	\begin{array}{c}
		u_{10}^{\ell}({\bf{q}};\tau)\\
		b_{10}^{\ell}({\bf{q}};\tau)\rule{0.ex}{5.ex}
	\end{array}
	} \right)}
	\nonumber\\
	&& = {\everymath{\displaystyle} B^j \frac{\partial}{\partial X_{\rm{I}}^j}
	\left( {
	\begin{array}{cc}
		0
		& 1\\
		1
		& 0 \rule{0.ex}{5.ex}
	\end{array}
	} \right)
	\left( {
	\begin{array}{c}
		u_{00}^i({\bf{k}})\\
		b_{00}^i({\bf{k}})\rule{0.ex}{5.ex}
	\end{array}
	} \right)}
	- D^{ij}({\bf{k}}) \frac{\widetilde{D}}{DT_{\rm{I}}}
		{\everymath{\displaystyle}\left( {
	\begin{array}{cc}
		1
		& 0\\
		0
		& 1 \rule{0.ex}{5.ex}
	\end{array}
	} \right)
	\left( {
	\begin{array}{c}
		u_{00}^j({\bf{k}})\\
		b_{00}^j({\bf{k}})\rule{0.ex}{5.ex}
	\end{array}
	} \right)}
	\nonumber\\
	&& + {\everymath{\displaystyle}\left( {
	\begin{array}{cc}
		- D^{ij}({\bf{k}}) \left( {
		\frac{\partial U^j}{\partial X^\ell}
		+ \epsilon^{\ell jn} \Omega_0^n
	} \right)
	& D^{ij}({\bf{k}}) \frac{\partial B^j}{\partial X^\ell}\\
	- D^{ij}({\bf{k}}) \frac{\partial B^j}{\partial X^\ell}
	& D^{ij}({\bf{k}}) \left( {
    \frac{\partial U^j}{\partial X^\ell} 
    + \epsilon^{\ell jn} \Omega_0^n
	} \right)
	\end{array}
	} \right)
	\left( {
	\begin{array}{c}
		u_{00}^\ell({\bf{k}};\tau)\\
		b_{00}^\ell({\bf{k}};\tau)\rule{0.ex}{5.ex}
	\end{array}
	} \right)}
	\nonumber\\
	&& \equiv \left( {
	\begin{array}{c}
		F_{10u}^i\\
		F_{10b}^i \rule{0.ex}{5.ex}
	\end{array}
	} \right),
	\label{eq:f10_eq}
\end{eqnarray}
where, $F_{01u}$, $F_{01b}$, $F_{10u}$, and $F_{10b}$ denote each component of the second right-hand sides (r.h.s.) of (\ref{eq:f01_eq}) and (\ref{eq:f10_eq}). They can be regarded as the forcing for the evolution equations of $f_{01}({\bf{k}};\tau)$ and $f_{10}({\bf{k}};\tau)$, respectively.

\subsubsection{Introduction of Green's functions}
	For the purpose of solving these differential equations, we introduce the Green's functions. We consider the responses of the velocity and magnetic field to infinitesimal perturbations of the velocity and magnetic field $\delta {\bf{u}}$ and $\delta{\bf{b}}$. Such infinitesimal perturbations arise from the external stirring force $\delta {\bf{F}}$, for instance for the velocity perturbation $\delta {\bf{u}}$ as
\begin{equation}
	\delta u^i 
	= \int d{\bf{k}}' \int_{-\infty}^t\!\!\!\! dt'\ 
	{\cal{G}}^{ij}({\bf{k}},{\bf{k}}';t,t')\ \delta F^j ({\bf{k}}';t')
	\label{eq:stirring_force_vel}
\end{equation}
with a Green's function in the wave number space. As this form shows, the Green's function itself is a random variable, which changes from one realization to realization of ${\bf{u}}$.

	It follows from (\ref{eq:f00_eq}) that the equations of the infinitesimal perturbations can be written as
\begin{eqnarray}
	&&{\everymath{\displaystyle}\left( {
	\begin{array}{cc}
		\frac{\partial}{\partial \tau} + \nu k^2 & 0\\
		0 & \frac{\partial}{\partial \tau} + \eta k^2
	\end{array}
	} \right)}
	\left( {
	\begin{array}{cc}
		\delta u_{00}^{i}({\bf{k}};\tau)\\
		\delta b_{00}^{i}({\bf{k}};\tau)\rule{0.ex}{5.ex}
	\end{array}
	} \right)
	\nonumber\\
	&& {\everymath{\displaystyle}
	+ i \left( {
	\begin{array}{cc}
		- 2M^{ij\ell}({\bf{k}}) \int_\Delta u_{00}^j({\bf{p}};\tau) 
		& 2M^{ij\ell}({\bf{k}}) \int_\Delta b_{00}^j({\bf{p}};\tau)\\
			N^{ij\ell}({\bf{k}}) \int_\Delta b_{00}^j({\bf{p}};\tau) 
		& - N^{ij\ell}({\bf{k}}) \int_\Delta u_{00}^j({\bf{p}};\tau)
	\end{array}
	} \right)
	\left( {
	\begin{array}{c}
		\delta u_{00}^{\ell}({\bf{q}};\tau)\\
		\delta b_{00}^{\ell}({\bf{q}};\tau)\rule{0.ex}{5.ex}
	\end{array}
	} \right)}
	\nonumber\\
	&&= \left( {
	\begin{array}{c}
		0\\
		0\rule{0.ex}{5.ex}
	\end{array}
	} \right).
	\label{eq:infinitesimal_u00_pert}
\end{eqnarray}
Here, we should note that the nonlinear convolution terms in (\ref{eq:f00_eq}) have led to the linear contribution in the form
\begin{equation}
	\widehat{{\bf{u}}{\bf{u}}}\;\;
	\rightarrow\;\;
	2\widehat{{\bf{u}} \delta{\bf{u}}},\;\;\;\; 
	\widehat{{\bf{b}}{\bf{b}}}\;\;
	\rightarrow\;\;
	2\widehat{{\bf{b}} \delta{\bf{b}}}.
	\label{eq:linealize_for_G}
\end{equation}
In order to treat mutual interaction among the velocity and magnetic field, we consider four Green's functions; the Green function $G_{uu}$ representing the response of the velocity field ${\bf{u}}$ to the velocity perturbation ${\bf{u}}$, $G_{ub}$ the response of ${\bf{u}}$ to the magnetic perturbation ${\bf{b}}$, $G_{bu}$ the response of ${\bf{b}}$ to the velocity perturbation ${\bf{u}}$, and $G_{bb}$ the response of magnetic field ${\bf{b}}$ to the magnetic perturbation ${\bf{b}}$. 

	From the left-hand side (l.h.s.) of (\ref{eq:infinitesimal_u00_pert}) we construct the system of equations representing the responses to the infinitesimal forcing. It follows that these four Green's functions should be defined by their evolution equations as
\begin{eqnarray}
	&&{\everymath{\displaystyle}\left( {
	\begin{array}{cc}
		\frac{\partial}{\partial \tau} + \nu k^2 & 0\\
		0 & \frac{\partial}{\partial \tau} + \eta k^2
	\end{array}
	} \right)}
	\left( {
	\begin{array}{cc}
		G_{uu}^{ij} & G_{ub}^{ij}\\
		G_{bu}^{ij} & G_{bb}^{ij}\rule{0.ex}{5.ex}
	\end{array}
	} \right)
	\nonumber\\
	&& {\everymath{\displaystyle}
	+ i \left( {
	\begin{array}{cc}
		- 2M^{ikm} \int_\Delta u_{00}^k & 2M^{ikm} \int_\Delta b_{00}^k\\
		N^{ikm} \int_\Delta b_{00}^k & - N^{ikm} \int_\Delta u_{00}^k
	\end{array}
	} \right)
	\left( {
	\begin{array}{cc}
		G_{uu}^{mj} & G_{ub}^{mj}\\
		G_{bu}^{mj} & G_{bb}^{mj}\rule{0.ex}{5.ex}
	\end{array}
	} \right)}
	\nonumber\\
	&&= \delta^{ij} \delta(\tau - \tau') \left( {
	\begin{array}{cc}
		1 & 0\\
		0 & 1\rule{0.ex}{5.ex}
	\end{array}
	} \right).
	\label{eq:greens_fn_def}
\end{eqnarray}
Reflecting the structure of the MHD equations and the field expansion (\ref{eq:scale_para_expansion}), the left-hand sides (l.h.s.) of (\ref{eq:f01_eq}) and (\ref{eq:f10_eq}) or the differential operators to the $f_{01}({\bf{k}};\tau)$ and $f_{10}({\bf{k}};\tau)$ fields are in the same form as the equations of fluctuations to the infinitesimal perturbations (\ref{eq:infinitesimal_u00_pert}). Considering that the r.h.s.\ of (\ref{eq:f01_eq}) and (\ref{eq:f10_eq}) are the force terms, we formally solve $f_{01}$ and $f_{10}$ fields with the aid of the Green's functions. The $f_{01}$ fields are expressed as
\begin{equation}
	\left( {
	\begin{array}{c}
		u_{01}^i\\
		b_{01}^i \rule{0.ex}{5.ex}
	\end{array}
	} \right)
	= \int_{-\infty}^\tau \!\!\! d\tau_1 
	\left( {
	\begin{array}{cc}
		G_{uu}^{ij} & G_{ub}^{ij}\\
		G_{bu}^{ij} & G_{bb}^{ij} \rule{0.ex}{5.ex}
    \end{array}
	} \right)
	\left( {
	\begin{array}{c}
		F_{01u}^j\\
		F_{01b}^j \rule{0.ex}{5.ex}
	\end{array}
	} \right),
	\label{eq:f01_formal_sol}
\end{equation}
or explicitly written as
\begin{eqnarray}
	u_{01}^i({\bf{k}};\tau)
	&=& \int d\tau_1 G_{uu}^{ij} F_{01u}^j
	+ \int d\tau_1 G_{ub}^{ij} F_{01b}^j
	\nonumber\\
	&=& \int_{-\infty}^\tau \!\!\! d\tau_1
	G_{ub}^{ij}({\bf{k}};\tau,\tau_1) \left[ {
	- i k^m B^m b_{00}^j({\bf{k}};\tau)
	} \right],
	\label{u01_formal_sol}
\end{eqnarray}
\begin{eqnarray}
	b_{01}^i({\bf{k}};\tau)
	&=& \int d\tau_1\ G_{bu}^{ij} F_{01u}^j
	+ \int d\tau_1\ G_{bb}^{ij} F_{01b}^j
	\nonumber\\
	&=& \int_{-\infty}^\tau \!\!\! d\tau_1
	G_{bu}^{ij}({\bf{k}};\tau,\tau_1) \left[ {
	- i k^m B^m u_{00}^j({\bf{k}};\tau)
	} \right].
	\label{eq:b01_formal_sol}
\end{eqnarray}
Note that ${\bf{u}}_{01}$ and ${\bf{b}}_{01}$ are expressed in terms of ${\bf{b}}_{00}$ and ${\bf{u}}_{00}$ coupled with the mean magnetic field ${\bf{B}}$, respectively. Consequently, ${\bf{u}}_{01}$ and ${\bf{b}}_{01}$ multiplied by ${\bf{b}}_{00}$ and ${\bf{u}}_{00}$ in an external product manner will not contribute to the EMF.

	On the other hand, the $f_{10}$ fields are expressed as
\begin{equation}
	\left( {
	\begin{array}{c}
		u_{10}^i\\
		b_{10}^i \rule{0.ex}{5.ex}
	\end{array}
	} \right)
	= \int_{-\infty}^\tau \!\!\! d\tau_1 
	\left( {
	\begin{array}{cc}
		G_{uu}^{ij} & G_{ub}^{ij}\\
		G_{bu}^{ij} & G_{bb}^{ij} \rule{0.ex}{5.ex}
	\end{array}
	} \right)
	\left( {
	\begin{array}{c}
		F_{10u}^j\\
		F_{10b}^j \rule{0.ex}{5.ex}
	\end{array}
	} \right),
	\label{eq:f10_formal_sol}
\end{equation}
or explicitly written as
\begin{eqnarray}
	&&u_{10}^i({\bf{k}};\tau)
	= \int d\tau_1\ G_{uu}^{ij} F_{10u}^j
	+ \int d\tau_1\ G_{ub}^{ij} F_{10b}^j
	\nonumber\\
	&&= \int_{-\infty}^\tau\!\!\! d\tau_1\ 
	G_{uu}^{ij}({\bf{k}};\tau,\tau_1)
	\left[ {
	-D^{jk}({\bf{k}}) 
		\frac{\widetilde{D} u_{00}^k({\bf{k}};\tau_1)}{DT_{\rm{I}}}
	} \right.
	\nonumber\\
	&&\hspace{90pt} - D^{jk}({\bf{k}}) \left( {
    	\frac{\partial U^k}{\partial X^m} + \epsilon^{mkn} \Omega_0^n
	} \right) u_{00}^m({\bf{k}};\tau_1)
	\nonumber\\
	&&\hspace{90pt}+ \left. {
	B^k \frac{\partial b_{00}^i({\bf{k}};\tau_1)}{\partial X_{\rm{I}}^k}
	+ D^{jk}({\bf{k}}) \frac{\partial B^k}{\partial X^m} 
		b_{00}^m({\bf{k}};\tau_1)
	} \right]
	\nonumber\\
	&&+ \int_{-\infty}^\tau\!\!\! d\tau_1\ 
		G_{ub}^{ij}({\bf{k}};\tau,\tau_1) \left[ {
		B^k 
		\frac{\partial u_{00}^i({\bf{k}};\tau_1)}{\partial X_{\rm{I}}^k} 
		- D^{jk}({\bf{k}}) 
		\frac{\partial B^k}{\partial X^m} u_{00}^m({\bf{k}};\tau_1)
	} \right.
	\nonumber\\
	&& \hspace{10pt}\left. {
	- D^{jk}({\bf{k}}) 
	\frac{\widetilde{D} b_{00}^k({\bf{k}};\tau_1)}{DT_{\rm{I}}}
	+ D^{jk}({\bf{k}}) \left( {
	\frac{\partial U^k}{\partial X^m}
	+ \epsilon^{mkn} \Omega_0
	} \right) b_{00}^m({\bf{k}};\tau_1)
	} \right],
	\label{eq:u10_formal_sol}
\end{eqnarray}
\begin{eqnarray}
	&&b_{10}^i({\bf{k}};\tau)
	= \int d\tau_1 G_{bu}^{ij} F_{10u}^j
	+ \int d\tau_1 G_{bb}^{ij} F_{10b}^j
	\nonumber\\
	&&= \int_{-\infty}^\tau\!\!\! d\tau_1\ 
	G_{bu}^{ij}({\bf{k}};\tau,\tau_1)
	\left[ {
		-D^{jk}({\bf{k}}) 
		\frac{\widetilde{D} u_{00}^k({\bf{k}};\tau_1)}{DT_{\rm{I}}}
	}\right.
	\nonumber\\
	&&	\hspace{90pt}- D^{jk}({\bf{k}}) \left( {
		\frac{\partial U^k}{\partial X^m} + \epsilon^{mkn} \Omega_0^n
	} \right) u_{00}^m({\bf{k}};\tau_1)
	\nonumber\\
	&&\hspace{90pt}+ \left. {
		B^k \frac{\partial b_{00}^i({\bf{k}};\tau_1)}{\partial X_{\rm{I}}^k}
		+ D^{jk}({\bf{k}}) \frac{\partial B^k}{\partial X^m} 
			b_{00}^m({\bf{k}};\tau_1)
	} \right]
	\nonumber\\
	&&+ \int_{-\infty}^\tau\!\!\! d\tau_1\ 
		G_{bb}^{ij}({\bf{k}};\tau,\tau_1) \left[ {
		B^k 
		\frac{\partial u_{00}^i({\bf{k}};\tau_1)}{\partial X_{\rm{I}}^k} 
	- D^{jk}({\bf{k}}) 
		\frac{\partial B^k}{\partial X^m} u_{00}^m({\bf{k}};\tau_1)
	} \right.
	\nonumber\\
	&& \hspace{10pt}\left. {
	- D^{jk}({\bf{k}}) 
	\frac{\widetilde{D} b_{00}^k({\bf{k}};\tau_1)}{DT_{\rm{I}}}
	+ D^{jk}({\bf{k}}) \left( {
	\frac{\partial U^k}{\partial X^m}
	+ \epsilon^{mkn} \Omega_0
	} \right) b_{00}^m({\bf{k}};\tau_1)
	} \right].
	\label{eq:b10_formal_sol}
\end{eqnarray}

\subsubsection{Statistical assumption on the basic fields}
	We assume that the basic or lowest-order fields are homogeneous and isotropic. 
\begin{equation}
	\frac{\left\langle {
		\vartheta_{00}^i({\bf{k}};\tau) \chi_{00}^j({\bf{k}}';\tau')
	} \right\rangle}{\delta({\bf{k}} + {\bf{k}}')}
	= D^{ij}({\bf{k}}) Q_{\vartheta\chi}({\bf{k}};\tau,\tau')
	+ \frac{i}{2} \frac{k^\ell}{k^2} \epsilon^{ij\ell}
		H_{\vartheta\chi}({\bf{k}};\tau,\tau'),
	\label{eq:stat_assump_basic_fld}
\end{equation}
where $\mbox{\boldmath$\vartheta$}_{00}$ and $\mbox{\boldmath$\chi$}_{00}$ represent one of ${\bf{u}}_{00}$ and ${\bf{b}}_{00}$, and the indices $\vartheta$ and $\chi$ do one of $u$ and $b$. The Green's functions are written as
\begin{equation}
	\langle {G_{\vartheta\chi}^{ij}({\bf{k}};\tau,\tau')} \rangle
	= D^{ij}({\bf{k}}) G_{\vartheta\chi}({\bf{k}};\tau,\tau').
	\label{eq:stat_assump_green_fn}
\end{equation}

	The spectral functions, $Q_{uu}$, $Q_{bb}$, $Q_{ub}$, $H_{uu}$, $H_{bb}$, $H_{ub}$, and $H_{bu}$, are related to the turbulent statistical quantities (the turbulent kinetic energy, magnetic energy, cross helicity, kinetic helicity, electric-current helicity, torsional correlations between velocity and magnetic field) of the basic or lowest-order fields as
\begin{equation}
	\int d{\bf{k}}\ Q_{uu}(k;\tau,\tau) 
	= \langle {{\bf{u}}'_{00}{}^2} \rangle/2,
	\label{eq:Quu_def}
\end{equation}
\begin{equation}
	\int d{\bf{k}}\ Q_{bb}(k;\tau,\tau) 
	= \langle {{\bf{b}}'_{00}{}^2} \rangle/2,
	\label{eq:Q_bb_def}
\end{equation}
\begin{equation}
	\int d{\bf{k}}\ Q_{ub}(k;\tau,\tau) 
	= \langle {{\bf{u}}'_{00} \cdot {\bf{b}}'_{00}} \rangle,
	\label{eq:Qub_def}
\end{equation}
\begin{equation}
	\int d{\bf{k}}\ H_{uu}(k;\tau,\tau) 
	= \langle {{\bf{u}}'_{00} \cdot \mbox{\boldmath$\omega$}'_{00}} \rangle,
	\label{eq:Huu_def}
\end{equation}
\begin{equation}
	\int d{\bf{k}}\ H_{bb}(k;\tau,\tau) 
	= \langle {{\bf{b}}'_{00} \cdot {\bf{j}}'_{00}} \rangle,
	\label{eq:Hbb_def}
\end{equation}
\begin{equation}
	\int d{\bf{k}}\ H_{ub}(k;\tau,\tau) 
	= \langle {{\bf{u}}'_{00} \cdot {\bf{j}}'_{00}} \rangle,
	\label{eq:Hub_def}
\end{equation}
\begin{equation}
	\int d{\bf{k}}\ H_{bu}(k;\tau,\tau) 
	= \langle {{\bf{b}}'_{00} \cdot \mbox{\boldmath$\omega$}'_{00}} \rangle.
	\label{eq:Hbu_def}
\end{equation}

\subsection{Calculation of the electromotive force}
The turbulent electromotive force (EMF) is expressed in terms of the wave-number representation of the velocity and magnetic-field as
\begin{equation}
	E_{\rm{M}}^i
	\equiv \epsilon^{ijk} \langle {u'{}^j b'{}^k} \rangle
	= \epsilon^{ijk} \int d{\bf{k}}\ 
	\langle {u^j({\bf{k}};\tau) b^k({\bf{k}}';\tau)} \rangle 
	/ \delta({\bf{k}} + {\bf{k}}').
	\label{eq:emf_calc}
\end{equation}
Using the results of (\ref{eq:f01_formal_sol})-(\ref{eq:b10_formal_sol}), we calculate the velocity--magnetic-field correlation up to the $f_{01}g_{00}$ and $f_{10} g_{00}$ orders as
\begin{equation}
	\langle {u^j b^k} \rangle
	= \langle {u_{00}^j b_{00}^k} \rangle
	+ \langle {u_{01}^j b_{00}^k} \rangle 
	+ \langle {u_{00}^j b_{01}^k} \rangle
	+ \delta \langle {u_{10}^j b_{00}^k} \rangle
	+ \delta \langle {u_{00}^j b_{10}^k} \rangle
	+ \cdots.
	\label{eq:emf_calc_by_pert}
\end{equation}
In the direct-interaction approximation (DIA) formalism, the lowest-order spectral functions $Q_{uu}$, $Q_{bb}$, $Q_{ub}$, $H_{uu}$, $H_{bb}$, $H_{ub}$, and $H_{bu}$, and the lowest-order Green's functions $G_{uu}$, $G_{bb}$, $G_{ub}$, and $G_{bu}$ are replaced with their exact counterparts, $\tilde{Q}_{uu}$, $\tilde{Q}_{bb}$, $\cdots$, and $\tilde{G}_{uu}$, $\tilde{G}_{bb}$, $\cdots$, respectively as
\begin{eqnarray}
	&&Q_{uu} \to \widetilde{Q}_{uu},\;\; 
	Q_{bb} \to \widetilde{Q}_{bb},\;\; 
	Q_{ub} \to \widetilde{Q}_{ub},\;\;
	\nonumber\\ 
	&&H_{uu} \to \widetilde{H}_{uu},\;\; 
	H_{bb} \to \widetilde{H}_{ub},\;\; 
	H_{ub} \to \widetilde{H}_{ub},\;\; 
	H_{bu} \to \widetilde{Q}_{bu}
	\nonumber\\
	&&G_{uu} \to \widetilde{G}_{uu},\;\;
	G_{bb} \to \widetilde{G}_{bb},\;\;
	G_{ub} \to \widetilde{G}_{ub},\;\;
	G_{bu} \to \widetilde{G}_{bu}.
	\label{eq:propagator_renormal}
\end{eqnarray}
Under this renormalization procedure on the propagators (spectral and response functions), important turbulent correlation functions are calculated. For the sake of simplicity, hereafter, the tilde denoting the exact propagator will be suppressed. Namely, the exact propagators are denoted without tilde.

	Here we present the final results of the turbulent EMF as
\begin{equation}
	\langle {{\bf{u}}' \times {\bf{b}}'} \rangle
	= \alpha {\bf{B}}
	- (\beta + \zeta) \nabla \times {\bf{B}}
	- (\nabla \zeta) \times {\bf{B}}
	+ \gamma \nabla \times {\bf{U}},
	\label{eq:emf_expression}
\end{equation}
where transport coefficients $\alpha$, $\beta$, $\zeta$, and $\gamma$ are given as
\begin{equation}
	\alpha 
	= \frac{1}{3} \left[ {
	- I\{ {G_{bb}, H_{uu}} \}
	+ I\{ {G_{uu},H_{bb}} \}
	- I\{ {G_{bu},H_{ub}} \}
	+ I\{ {G_{ub},H_{bu}} \}
	} \right],
	\label{eq:alpha_exp}
\end{equation}
\begin{equation}
	\beta 
	= \frac{1}{3} \left[ {
	I\{ {G_{bb}, Q_{uu}} \}
	+ I\{ {G_{uu},Q_{bb}} \}
	- I\{ {G_{bu},Q_{ub}} \}
	- I\{ {G_{ub},Q_{bu}} \}
	} \right],
	\label{eq:beta_exp}
\end{equation}
\begin{equation}
	\zeta 
	= \frac{1}{3} \left[ {
	I\{ {G_{bb}, Q_{uu}} \}
	- I\{ {G_{uu},Q_{bb}} \}
	+ I\{ {G_{bu},Q_{ub}} \}
	- I\{ {G_{ub},Q_{bu}} \}
	} \right],
	\label{eq:zeta_exp}
\end{equation}
\begin{equation}
	\gamma 
	= \frac{1}{3} \left[ {
	I\{ {G_{bb}, Q_{ub}} \}
	+ I\{ {G_{uu},Q_{bu}} \}
	- I\{ {G_{bu},Q_{uu}} \}
	- I\{ {G_{ub},Q_{bb}} \}
	} \right]
	\label{eq:gamma_exp}
\end{equation}
with the abbreviate form of the spectral and time integral
\begin{equation}
	I\{{A,B}\}
	= \int d{\bf{k}} \int_{-\infty}^{\tau}\!\!\! d\tau_1
		A(k;\tau,\tau_1) B(k;\tau,\tau_1).
	\label{eq:IAB_def}
\end{equation}

\subsection{Reynolds and turbulent Maxwell stresses}
	The Reynolds and turbulent Maxwell stresses in the mean momentum equation is defined by
\begin{equation}
	{\cal{R}}^{ij}
	= \langle {u'{}^i u'{}^j} \rangle
	- \langle {b'{}^i b'{}^j} \rangle.
	\label{eq:rey_turb_maxwell_def}
\end{equation}
In a similar manner as for the turbulent electromotive force ${\bf{E}}_{\rm{M}}$, $\mbox{\boldmath${\cal{R}}$}$ can be calculated from ${\bf{u}}'$ and ${\bf{b}}'$ with
\begin{equation}
	\langle {u^j u^k} \rangle
	= \langle {u_{00}^j u_{00}^k} \rangle
	+ \langle {u_{01}^j u_{00}^k} \rangle 
	+ \langle {u_{00}^j u_{01}^k} \rangle
	+ \delta \langle {u_{10}^j u_{00}^k} \rangle
	+ \delta \langle {u_{00}^j u_{10}^k} \rangle
	+ \cdots,
	\label{eq:rey_str_calc_by_pert}
\end{equation}
\begin{equation}
	\langle {b^j b^k} \rangle
	= \langle {b_{00}^j b_{00}^k} \rangle
	+ \langle {b_{01}^j b_{00}^k} \rangle 
	+ \langle {b_{00}^j b_{01}^k} \rangle
	+ \delta \langle {b_{10}^j b_{00}^k} \rangle
	+ \delta \langle {b_{00}^j b_{10}^k} \rangle
	+ \cdots.
	\label{eq:turb_max_str_calc_by_pert}
\end{equation}
The expression of $\mbox{\boldmath${\cal{R}}$}$ is given as
\begin{equation}
	\langle {u'{}^i u'{}^j - b'{}^i b'{}^j} \rangle_{\rm{D}}
	= - \nu_{\rm{K}} {\cal{S}}^{ij}
	+ \nu_{\rm{M}} {\cal{M}}^{ij}
	+ [\Gamma^i \Omega_\ast^j + \Gamma^j \Omega_\ast^i]_{\rm{D}},
	\label{eq:rey_turb_max_str_exp}
\end{equation}
where $\mbox{\boldmath$\Omega$}_\ast (= \nabla \times {\bf{U}} + 2 \mbox{\boldmath$\omega$}_{\rm{F}})$ is the absolute vorticity (the relative vorticity and the angular velocity of rotation), the strain rate of the mean velocity $\mbox{\boldmath${\cal{S}}$} = \{ {{\cal{S}}^{ij}} \}$ and that of the mean magnetic field $\mbox{\boldmath${\cal{M}}$} = \{ {{\cal{M}}^{ij}} \}$ are defined by
\begin{equation}
	{\cal{S}}^{ij}
	= \frac{\partial U^j}{\partial x^i}
	+ \frac{\partial U^i}{\partial x^j}
	- \frac{2}{3} \delta^{ij} \nabla \cdot {\bf{U}},
	\label{eq:mean_vel_strn_def}
\end{equation}
\begin{equation}
	{\cal{M}}^{ij}
	= \frac{\partial B^j}{\partial x^i}
	+ \frac{\partial B^i}{\partial x^j}
	- \frac{2}{3} \delta^{ij} \nabla \cdot {\bf{B}}
	= \frac{\partial B^j}{\partial x^i}
	+ \frac{\partial B^i}{\partial x^j},
	\label{eq:mean_mag_strn_def}
\end{equation}
respectively, and the suffix ${\rm{D}}$ denotes the deviatoric part of a tensor as
\begin{equation}
	{\cal{A}}^{ij}_{\rm{D}}
	= {\cal{A}}^{ij} 
	- \frac{1}{3} \delta^{ij} {\cal{A}}^{\ell\ell}.
	\label{eq:dev_tensor_def}
\end{equation}
The transport coefficients $\nu_{\rm{K}}$, $\nu_{\rm{M}}$, and $\mbox{\boldmath$\Gamma$}$ in (\ref{eq:rey_turb_max_str_exp}) are expressed as
\begin{equation}
	\nu_{\rm{K}} = \frac{7}{5} \beta
	= \frac{7}{15} \left[ {
	I\{ {G_{bb}, Q_{uu}} \}
	+ I\{ {G_{uu},Q_{bb}} \}
	- I\{ {G_{bu},Q_{ub}} \}
	- I\{ {G_{ub},Q_{bu}} \}
	} \right],
	\label{eq:nuK_exp}
\end{equation}
\begin{equation}
	\nu_{\rm{M}}
	= \frac{7}{5} \gamma 
	= \frac{7}{15} \left[ {
	I\{ {G_{bb}, Q_{ub}} \}
	+ I\{ {G_{uu},Q_{bu}} \}
	- I\{ {G_{bu},Q_{uu}} \}
	- I\{ {G_{ub},Q_{bb}} \}
	} \right],
	\label{eq:nuM_exp}
\end{equation}
\begin{equation}
	\mbox{\boldmath$\Gamma$}
	= \frac{1}{15} \left( {
	I_{-1} \{ {G_{uu}+G_{bb}, \nabla H_{uu}} \}
	- I_{-1} \{ {G_{ub}, \nabla H_{bu}} \}
	} \right).
	\label{eq:Gamma_exp}
\end{equation}
Here, $\nu_{\rm{K}}$ is the eddy viscosity coupled with the mean velocity strain $\mbox{\boldmath${\cal{S}}$} = \{ {{\cal{S}}^{ij}} \}$, and is determined mainly by the turbulent MHD energy as the first two terms of the right-most equation (\ref{eq:nuK_exp}) show. On the other hand, $\nu_{\rm{M}}$ couples with the mean magnetic-field strain $\mbox{\boldmath${\cal{M}}$} = \{ {{\cal{M}}^{ij}} \}$, and is mainly determined by the turbulent cross helicity $\langle {{\bf{u}}' \cdot {\bf{b}}'} \rangle$ as the first two terms of the right-most equation (\ref{eq:nuM_exp}) show. The transport coefficient $\mbox{\boldmath$\Gamma$}$ couples with the absolute vorticity $\mbox{\boldmath$\Omega$}_\ast$, and is determined by the gradient of the turbulent helicity $\langle {{\bf{u}}' \cdot \mbox{\boldmath$\omega$}} \rangle$.

	The expression of $\mbox{\boldmath${\cal{R}}$}$ (\ref{eq:rey_turb_max_str_exp}) indicates that, in addition to the eddy or turbulent viscosity $\nu_{\rm{K}}$ coupled with the mean velocity strain (symmetric part of the mean velocity shear), we also have other effects in the Reynolds and turbulent Maxwell stresses. One is the cross-helicity-related effect $\nu_{\rm{M}}$ coupled with the mean magnetic-field strain, and the other is the inhomogeneous kinetic helicity effect $\mbox{\boldmath$\Gamma$}$ coupled with the mean absolute vorticity (anti-symmetric part of the mean velocity shear).

	The role of the Reynolds and turbulent Maxwell stresses and the consequence of the expression (\ref{eq:rey_turb_max_str_exp}) shall be further discussed later in Sections~6 and 7 with special reference to the large-scale flow generation and the linear and angular momentum transport.

\subsection{Symmetric and anti-symmetric response function effects}

\subsubsection{Standard self-interaction response function effects}
In our formulations, we have four Green's functions, $G_{uu}$, $G_{bb}$, $G_{ub}$, and $G_{bu}$, whose definitions are given in the evolution equations (\ref{eq:greens_fn_def}). If we assume that the cross-interaction Green's functions $G_{ub}$ and $G_{bu}$ vanish as
\begin{equation}
	G_{ub} = G_{bu} = 0,
	\label{eq:vanishing_Gub_Gbu}
\end{equation}
the usual EMF for the incompressible turbulence is recovered. As the simplest possible model for the EMF, apart from the model constants, we have
\begin{equation}
	\alpha 
	= - \tau_b \langle {{\bf{u}}' \cdot \mbox{\boldmath$\omega$}} \rangle
	+ \tau_u \langle {{\bf{b}}' \cdot {\bf{j}}'} \rangle
	\equiv \alpha_{\rm{S}},
	\label{eq:alpha_S_def}
\end{equation}
\begin{equation}
	\beta 
	= \tau_b \langle {{\bf{u}}'{}^2} \rangle/2
	+ \tau_u \langle {{\bf{b}}'{}^2} \rangle/2
	\equiv \beta_{\rm{S}},
	\label{eq:beta_S_def}
\end{equation}
\begin{equation}
	\zeta 
	= \tau_b \langle {{\bf{u}}'{}^2} \rangle/2
	- \tau_u \langle {{\bf{b}}'{}^2} \rangle/2
	\equiv \zeta_{\rm{S}},
	\label{eq:zeta_S_def}
\end{equation}
\begin{equation}
	\gamma 
	= (\tau_b + \tau_u) \langle {{\bf{u}}' \cdot {\bf{b}}'} \rangle
	\equiv \gamma_{\rm{S}},
	\label{eq:gamma_S_def}
\end{equation}
where $\tau_u$ and $\tau_b$ are timescales associated with the Green's function $G_{uu}$ and $G_{bb}$, and they are evaluated as
\begin{equation}
	\tau_u f({\bf{k}},\tau)
	\sim \int_{-\infty}^\tau\!\!\! d\tau'\ 
	G_{uu}({\bf{k}};\tau,\tau') f({\bf{k}};\tau'),
	\label{eq:tauu_Guu_rel}
\end{equation}
\begin{equation}
	\tau_b f({\bf{k}},\tau)
	\sim \int_{-\infty}^\tau\!\!\! d\tau'\ 
	G_{bb}({\bf{k}};\tau,\tau') f({\bf{k}};\tau'),
	\label{eq:taub_Gbb_rel}
\end{equation}
respectively. With special emphasis on the simplest dynamo model, we denote these transport coefficients with suffix ${\rm{S}}$ as $\alpha_{\rm{S}}$, $\beta_{\rm{S}}$, $\zeta_{\rm{S}}$, and $\gamma_{\rm{S}}$, as the right-most sides of (\ref{eq:alpha_S_def})-(\ref{eq:gamma_S_def}) show.

	The alpha effect, $\alpha {\bf{B}}$, the first term in (\ref{eq:emf_expression}), depends on the kinetic helicity density $\langle {{\bf{u}}' \cdot \mbox{\boldmath$\omega$}'} \rangle$ and the electric-current helicity density $\langle {{\bf{b}}' \cdot {\bf{j}}'} \rangle$. The physical origin of the kinetic helicity effect and the current helicity effect will be discussed in the following section (Section~3). Equation~(\ref{eq:alpha_S_def}) shows that, if the kinetic helicity $\langle {{\bf{u}}' \cdot \mbox{\boldmath$\omega$}'} \rangle$ and current helicity $\langle {{\bf{b}}' \cdot {\bf{j}}'} \rangle$ have the same sign, their effects are suppressed with each other. This is the reason why the current-helicity effect is often argued as a correction to the alpha effect, leading to the suppression or saturation of the alpha effect. However, this is no the case if the kinetic and current helicities are generated by each production mechanism due to large-scale inhomogeneities. This point will be discussed in Section~3. We should note that the timescale associated with the kinetic helicity contribution is the magnetic one $\tau_b$, while the counterpart associated with the electric-current helicity is the kinetic one $\tau_u$.

	The turbulent magnetic diffusivity effect coupled with the mean electric-current density ${\bf{J}} (= \nabla \times {\bf{B}})$, the second term in (\ref{eq:emf_expression}), arises from the turbulent kinetic and magnetic energies. Since the magnetic energy $\langle{{\bf{b}}'{}^2}\rangle$ contributions in $\beta$ and $\zeta$ completely cancel with each other, the magnetic diffusivity $\beta + \zeta$ depends solely on the turbulent kinetic energy $\langle {{\bf{u}}'{}^2} \rangle$. It has been theoretically pointed out that the turbulent magnetic energy $\langle {{\bf{b}}'{}^2} \rangle$ contributes to the turbulent magnetic diffusivity in anisotropic turbulence \citep{rog2001} and in the presence of compressibility \citep{yok2018a}.

	The magnetic pumping effect $- (\nabla \zeta) \times {\bf{B}}$, the third term in (\ref{eq:emf_expression}), depends on the gradient of the MHD residual energy $\langle {{\bf{u}}'{}^2 - {\bf{b}}'{}^2} \rangle/2$. For the same mean magnetic field ${\bf{B}}$, the direction of the pumping effect alters depending on the direction of $\nabla \zeta$.

	The cross helicity effect $\gamma \mbox{\boldmath$\Omega$}_\ast$, the fourth term in (\ref{eq:emf_expression}), arises from the turbulent cross helicity $\langle {{\bf{u}}' \cdot {\bf{b}}'} \rangle$. This effect is expected to play an important role in dynamo in the presence of the global vortical motion or rotation represented by the mean absolute vorticity $\mbox{\boldmath$\Omega$}_\ast (= \nabla \times {\bf{U}} + 2 \mbox{\boldmath$\omega$}_{\rm{F}})$ ($\mbox{\boldmath$\omega$}_{\rm{F}}$: angular velocity of rotation).

\subsubsection{Effects of cross-interaction response functions}
	If we retain the contributions from the cross-interaction Green's functions $G_{ub}$ and $G_{bu}$, the third and fourth terms in each of (\ref{eq:alpha_exp})-(\ref{eq:alpha_exp}), we have additional contributions to the dynamo transport coefficients $\alpha$, $\beta$, $\zeta$, and $\gamma$. In order to get a clear picture on the cross-interaction Green's functions, we first consider the counterpart of the cross-interaction in the Els\"{a}sser-variable formulation \citep{yos1990,yok2013a}.

	In the Els\"{a}sser-variable formulation with $\mbox{\boldmath$\phi$}= {\bf{u}} + {\bf{b}}$ and $\mbox{\boldmath$\psi$}= {\bf{u}} - {\bf{b}}$, we introduce four Green's functions, $G_{\phi\phi}$, $G_{\psi\psi}$, $G_{\phi\phi}$, $G_{\phi\psi}$, and $G_{\psi\phi}$. In this formulation, the transport coefficients of the turbulent EMF are expressed as
\begin{equation}
	\alpha
	= - I\{ {G_{\rm{S}}, H_{uu}} \} 
	+ I\{ {G_{\rm{S}}, H_{bb}} \}
	- I\{ {G_{\rm{A}}, H_{ub}} \}
	+ I\{ {G_{\rm{A}}, H_{bu}} \},
	\label{eq:elsasser_alpha_exp}
\end{equation}
\begin{equation}
	\beta
	= I\{ {G_{\rm{S}}, Q_{uu}} \} 
	+ I\{ {G_{\rm{S}}, Q_{bb}} \}
	- I\{ {G_{\rm{A}}, Q_{ub}} \}
	- I\{ {G_{\rm{A}}, Q_{bu}} \},
	\label{eq:elsasser+beta_exp}
\end{equation}
\begin{equation}
	\zeta
	= I\{ {G_{\rm{S}}, Q_{uu}} \} 
	- I\{ {G_{\rm{S}}, Q_{bb}} \}
	+ I\{ {G_{\rm{A}}, Q_{ub}} \}
	- I\{ {G_{\rm{A}}, Q_{bu}} \},
	\label{eq:elsasser_zeta_exp}
\end{equation}
\begin{equation}
	\gamma
	= I\{ {G_{\rm{S}}, Q_{ub}} \} 
	+ I\{ {G_{\rm{S}}, Q_{bu}} \}
	- I\{ {G_{\rm{A}}, Q_{uu}} \}
	- I\{ {G_{\rm{A}}, Q_{bb}} \},
	\label{eq:elsasser_gamma_exp}
\end{equation}
where $G_{\rm{S}}$ and $G_{\rm{A}}$ denote the mirror-symmetric and anti-mirror-symmetric parts of $G_{\phi\phi}$ and $G_{\psi\psi}$, defined by
\begin{equation}
	G_{\rm{S}} = \frac{G_{\phi\phi} + G_{\psi\psi}}{2},
	\label{eq:GS_def}
\end{equation}
\begin{equation}
	G_{\rm{A}} = \frac{G_{\phi\phi} - G_{\psi\psi}}{2},
	\label{eq:GA_def}
\end{equation}
respectively.

	Comparing (\ref{eq:elsasser_alpha_exp})-(\ref{eq:elsasser_gamma_exp}) with (\ref{eq:alpha_exp})-(\ref{eq:gamma_exp}), we see that the self-interaction part $G_{uu}$ and $G_{bb}$ corresponds to the symmetric part, $G_{\rm{S}}$, while the cross-interaction part, $G_{ub}$ and $G_{bu}$, does the anti-symmetric part, $G_{\rm{A}}$. We see from (\ref{eq:GA_def}) that the anti-symmetric part $G_{\rm{A}}$ is connected to the difference between the timescales associated with the Alfv\'{e}n waves propagating in the counter-parallel and pro-parallel directions along the magnetic field. The imbalance between the counter- and pro-propagating Alfv\'{e}n waves results in a non-vanishing turbulent cross helicity. These points suggest that non-vanishing $G_{ub}$ and $G_{bu}$ are linked to the turbulent cross helicity.

\NOTE{
	Following the above consideration, the dynamo coefficients $d = (\alpha, \beta, \zeta, \gamma)$ are constituted of $d_{\rm{S}}$ and $d_{\rm{X}}$ as
\begin{equation}
	d
	= d_{\rm{S}}
	+ d_{\rm{X}},
	\label{eq:dS_dX}
\end{equation}
where $d_{\rm{S}}$ is the standard $d$ coefficients defined by (\ref{eq:alpha_S_def})-(\ref{eq:gamma_S_def}).
The cross-interaction parts of the dynamo coefficients, $\alpha_{\rm{X}}$, $\beta_{\rm{X}}$, $\zeta_{\rm{X}}$, and $\gamma_{\rm{X}}$, are expressed from the third and fourth terms of (\ref{eq:alpha_exp})-(\ref{eq:gamma_exp}) as
\begin{equation}
	\alpha_{\rm{X}}
	= \frac{1}{3} \left[ {
	- I\{ {G_{bu},H_{ub}} \}
	+ I\{ {G_{ub},H_{bu}} \}
	} \right],
	\label{eq:alphaX_exp}
\end{equation}
\begin{equation}
	\beta_{\rm{X}}
	= \frac{1}{3} \left[ {
	- I\{ {G_{bu},Q_{ub}} \}
	- I\{ {G_{ub},Q_{bu}} \}
	} \right],
	\label{eq:betaX_exp}
\end{equation}
\begin{equation}
	\zeta_{\rm{X}}
	= \frac{1}{3} \left[ {
	+ I\{ {G_{bu},Q_{ub}} \}
	- I\{ {G_{ub},Q_{bu}} \}
	} \right],
	\label{eq:zetaX_exp}
\end{equation}
\begin{equation}
	\gamma_{\rm{X}}
	= \frac{1}{3} \left[ {
	- I\{ {G_{bu},Q_{uu}} \}
	+ I\{ {G_{ub},Q_{bb}} \}
	} \right].
	\label{eq:gammaX_exp}
\end{equation}
}

\NOTE{
These cross-interaction dynamo coefficients may be modeled as
\begin{equation}
	\alpha_{\rm{X}}
	= - \tau_{{\rm{X}}bu} \Upsilon 
		\langle {{\bf{u}}' \cdot {\bf{j}}'} \rangle
	+ \tau_{{\rm{X}}ub} \Upsilon 
		\langle {\mbox{\boldmath$\omega$}' \cdot {\bf{b}}'} \rangle,
	\label{eq:alphaX_model}
\end{equation}
\begin{equation}
	\beta_{\rm{X}}
	= - \tau_{{\rm{X}}bu} \Upsilon 
		\langle {{\bf{u}}' \cdot {\bf{b}}'} \rangle
	- \tau_{{\rm{X}}ub} \Upsilon 
		\langle {{\bf{u}}' \cdot {\bf{b}}'} \rangle,
	\label{eq:betaX_model}
\end{equation}
\begin{equation}
	\zeta_{\rm{X}}
	= + \tau_{{\rm{X}}bu} \Upsilon 
		\langle {{\bf{u}}' \cdot {\bf{b}}'} \rangle
	- \tau_{{\rm{X}}ub} \Upsilon 
		\langle {{\bf{u}}' \cdot {\bf{b}}'} \rangle,
	\label{eq:zetaX_model}
\end{equation}
\begin{equation}
	\gamma_{\rm{X}}
	= - \tau_{{\rm{X}}bu} \Upsilon 
		\langle {{\bf{u}}'{}^2} \rangle
	+ \tau_{{\rm{X}}ub} \Upsilon 
		\langle {{\bf{b}}'{}^2} \rangle,
	\label{eq:gammaX_model}
\end{equation}
where $\tau_{{\rm{X}}bu}$ and $\tau_{{\rm{X}}ub}$ are timescales associated with the Green's functions $G_{bu}$ and $G_{ub}$, respectively.
}

For instance, the transport coefficient $\alpha_{\rm{X}}$ as well as $\alpha_{\rm{S}}$ is a pseudo-scalar. Since both $\langle {{\bf{u}}' \cdot {\bf{j}}'} \rangle$ and $\langle {\mbox{\boldmath$\omega$}' \cdot {\bf{b}}'} \rangle$ are pure-scalars, we need a pseudo-scalar factor $\Upsilon$ in (\ref{eq:alphaX_model}). This is a direct consequence of the fact that the Green's functions $G_{bu}$ and $G_{ub}$ are pseudo-scalar functions as their definitions (\ref{eq:greens_fn_def}) show. As the simplest possible candidate, we adopt a non-dimensional pseudoscalar quantity defined by
\begin{equation}
	\Upsilon
	= \frac{\langle {{\bf{u}}' \cdot {\bf{b}}'} \rangle}
    	{\langle {{\bf{u}}'{}^2 + {\bf{b}}'{}^2} \rangle/2},
	\label{eq:Upsilon_def1}
\end{equation}
or alternatively we may adopt
\begin{equation}
	\Upsilon
	= \frac{\langle {{\bf{u}}' \cdot {\bf{b}}'} \rangle}
    	{ \sqrt{
			\langle {{\bf{u}}'{}^2} \rangle \langle {{\bf{b}}'{}^2} \rangle
		}}.
	\label{eq:Upsilon_def2}
\end{equation}

The $\alpha_{\rm{X}}$ expression (\ref{eq:alphaX_exp}) and its model (\ref{eq:alphaX_model}) suggest that there are some conditions for this cross-interaction response effect to work. First we have to remark that the torsional cross correlations $\langle {{\bf{u}}' \cdot {\bf{j}}'} \rangle$ and $\langle {\mbox{\boldmath$\omega$}' \cdot {\bf{b}}'} \rangle$ are related to each other as
\begin{equation}
	- \langle {{\bf{u}}' \cdot {\bf{j}}'} \rangle
	+ \langle {\mbox{\boldmath$\omega$}' \cdot {\bf{b}}'} \rangle
	= \nabla \cdot \langle {{\bf{u}}' \times {\bf{b}}'} \rangle.
	\label{eq:uj_omegab_rel}
\end{equation}
In homogeneous turbulence, the r.h.s.\ of (\ref{eq:uj_omegab_rel}) vanishes, and we have
\begin{equation}
	\langle {{\bf{u}}' \cdot {\bf{j}}'} \rangle
	= \langle {\mbox{\boldmath$\omega$}' \cdot {\bf{b}}'} \rangle.
	\label{eq:tors_cross_cor_rel_homo}
\end{equation}
In this case, we have no contribution to $\alpha_{\rm{X}}$ for $\tau_{bu} = \tau_{ub}$. This suggests the conditions for the effect of the cross-interaction response functions to work:

\noindent(i) Even if we have no timescale difference ($\tau_{bu} = \tau_{ub}$), non-zero flux of the turbulent EMF across the boundary $\nabla \cdot \langle {{\bf{u}}' \times {\bf{b}}'} \rangle$ may lead to a finite $\alpha_{\rm{X}}$;

\noindent(ii) Even in homogeneous turbulence, where $\langle {{\bf{u}}' \cdot {\bf{j}}'} \rangle = \langle {\mbox{\boldmath$\omega$}' \cdot {\bf{b}}'} \rangle$, if there is a timescale difference between $\tau_{bu}$ and $\tau_{ub}$, we have a non-zero $\alpha_{\rm{X}}$; 

\noindent(iii) The pseudo-scalar factor $\Upsilon$ coupled with the timescale $\tau_{bu}$ and $\tau_{ub}$, should represent $G_{bu}$ and $G_{ub}$. The torsional cross correlations may be regarded as the combination of the cross helicity and the helicities as
\begin{equation}
	\langle {{\bf{u}}' \cdot {\bf{j}}'} \rangle
	\propto \langle {{\bf{u}}' \cdot {\bf{b}}'} 
			\rangle \langle {{\bf{b}}' \cdot {\bf{j}}'} \rangle,
	\label{eq:uj_ub-bj}
\end{equation}
\begin{equation}
	\langle {\mbox{\boldmath$\omega$}' \cdot {\bf{b}}'} \rangle
	\propto \langle {\mbox{\boldmath$\omega$}' \cdot {\bf{u}}'} 
			\rangle \langle {{\bf{u}}' \cdot {\bf{b}}'} \rangle,
	\label{eq:omegab_omegau-ub}
\end{equation}
This suggests that the coexistence of the cross helicity and current helicity and the coexistence of the cross helicity and kinetic helicity may be favorable conditions for the cross-interaction response effect $\alpha_{\rm{X}}$. In this sense, the turbulent cross helicity may play a key role in this cross-interaction effect. 
\NOTE{
As an important possibility, the cross-interaction response effect in the $\alpha$ effect, $\alpha_{\rm{X}}$, is investigated in the non-equilibrium or non-stationary turbulence in \citet{miz2023}.
}

We should note that this cross-interaction effect arises from the formulation with the response functions. In addition to the velocity and magnetic-field fluctuations and their correlation functions, the equations of the responses are also treated. As a result of this formulation, the response of the velocity fluctuation to the magnetic disturbance and the response of the magnetic-field fluctuation to the velocity disturbance enter the expressions of the turbulent fluxes. The structure of this formulation should be further explored.

\NOTE{
These cross-interaction effects may show a strong relevance especially under some conditions, such as the turbulence with non-equilibrium, non-stationary, and breakage of some symmetry. However, such conditions are rather specific, we only treat the standard or self-interaction response-function effects in the following sections. This does not deny the potential importance of the cross-interaction effects.
}

\section{Cross-helicity effect in dynamos}
\NOTE{
In the previous section, with the aid of analytical formulation for the inhomogeneous turbulence: the multiple-scale renormalized perturbation expansion theory, the expressions of the turbulent electromotive force (EMF) $\langle {{\bf{u}}' \times {\bf{b}}'} \rangle$ in the mean induction equation and the Reynolds and turbulent Maxwell stresses $\langle { {\bf{u}}' {\bf{u}}'  - {\bf{b}}' {\bf{b}}'} \rangle$ in the mean momentum equation, are derived. The turbulent cross-helicity effect gets into the $\langle {{\bf{u}}' \times {\bf{b}}'} \rangle$ expression coupled with the inhomogeneous mean velocity, and into the $\langle {{\bf{u}}' {\bf{u}}' - {\bf{b}}' {\bf{b}}'} \rangle$ expression coupled with the inhomogeneous mean magnetic field. In this section, the cross-helicity effect in the mean magnetic-field induction is argued. This argument is followed by the argument of the cross-helicity evolution in Section~4, and the cross-helicity effect in stellar dynamos in Section~5. The cross-helicity effect in the momentum equation will be argued in Section~6.
}

\subsection{Cross helicity and global flow effect in dynamos}
\begin{figure}[htb]
  \centering
  \includegraphics[width= 0.45 \columnwidth]{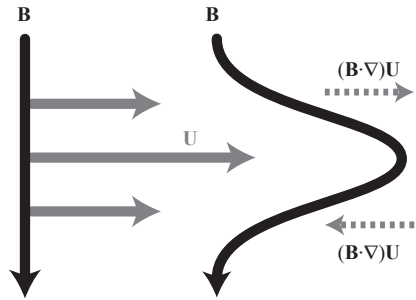}
  \caption{Differential rotation effect ($\Omega$ effect).}
    \label{fig:omega_effect}
\end{figure}

The mean magnetic induction equation is written as
\begin{equation}
	\frac{\partial {\bf{B}}}{\partial t}
	= \nabla \times ({\bf{U}} \times {\bf{B}})
	+ \nabla \times \langle {{\bf{u}}' \times {\bf{b}}'} \rangle
	+ \eta \nabla^2 {\bf{B}},
	\label{eq:mean_B_ind_eq_sec3}
\end{equation}
or equivalently
\begin{equation}
	\frac{\partial {\bf{B}}}{\partial t}
	+ ({\bf{U}} \cdot \nabla) {\bf{B}}
	= ({\bf{B}} \cdot \nabla) {\bf{U}}
	- (\nabla \cdot {\bf{U}}) {\bf{B}}
	+ \nabla \times \langle {{\bf{u}}' \times {\bf{b}}'} \rangle
	+ \eta \nabla^2 {\bf{B}}.
	\label{eq:mean_B_ind_eq_2_sec3}
\end{equation}
The first term on the right-hand side (r.h.s.) of (\ref{eq:mean_B_ind_eq_2_sec3}) represents the differential-rotation effect. If the mean velocity ${\bf{U}}$ is inhomogeneous along the mean magnetic field ${\bf{B}}$, $({\bf{B}} \cdot \nabla) {\bf{U}}$, it contributes to the generation of the mean magnetic-field component in the direction of the mean velocity (Fig.~\ref{fig:omega_effect}). This differential-rotation effect plays an important role in dynamo process, and is called the $\Omega$ (Omega) effect. It is considered to produce the azimuthal or toroidal component of the mean magnetic field from the latitudinal or poloidal one (and vice versa) through the non-uniform mean velocity effect.

	The second term of the r.h.s. of (\ref{eq:mean_B_ind_eq_sec3}) and the third term on the r.h.s. of (\ref{eq:mean_B_ind_eq_2_sec3}) contain the contribution of the turbulent electromotive force (EMF) defined by
\begin{equation}
	{\bf{E}}_{\rm{M}}
	\equiv \langle {{\bf{u}}' \times {\bf{b}}'} \rangle.
	\label{eq:emf_def_sec3}
\end{equation}
This ${\bf{E}}_{\rm{M}}$ is the sole (direct) turbulence effect in the mean magnetic induction equation [(\ref{eq:mean_B_ind_eq_sec3}) and (\ref{eq:mean_B_ind_eq_2_sec3})], and is the quantity of central importance in the turbulent dynamo study.

	Unlike the treatment in the $\Omega$ effect, the non-uniform or inhomogeneous mean velocity effect has been neglected in considering turbulence. As we saw in (\ref{eq:fluct_u_eq}) and (\ref{eq:fluct_b_eq}), the equations of the fluctuating velocity and magnetic field are given as
\begin{equation}
	\frac{\partial {\bf{u}}'}{\partial t}
	+ ({\bf{U}} \cdot \nabla) {\bf{u}}'
	= ({\bf{B}} \cdot \nabla) {\bf{b}}'
	+ ({\bf{b}}' \cdot \nabla) {\bf{B}}
	- ({\bf{u}}' \cdot \nabla) {\bf{U}}
	+ \cdots,
	\label{eq:fluct_u_eq_sec3}
\end{equation}
\begin{equation}
	\frac{\partial {\bf{b}}'}{\partial t}
	+ ({\bf{U}} \cdot \nabla) {\bf{b}}'
	= ({\bf{B}} \cdot \nabla) {\bf{u}}'
	- ({\bf{u}}' \cdot \nabla) {\bf{B}}
	+ ({\bf{b}}' \cdot \nabla) {\bf{U}}
	+ \cdots,
	\label{eq:fluct_b_eq_sec3}
\end{equation}
The evaluation of EMF is obtained from the equations of the fluctuating velocity and magnetic field [(\ref{eq:fluct_u_eq_sec3}) and (\ref{eq:fluct_b_eq_sec3})]. We multiply (\ref{eq:fluct_u_eq_sec3}) and (\ref{eq:fluct_b_eq_sec3}) by ${\bf{b}}'$ and ${\bf{u}}'$ in the vector product manner, respectively, and add them. After taking the ensemble average, we obtain the evolution equation of the ELM as
\begin{eqnarray}
	\frac{DE_{\rm{M}}^i}{Dt}
	&=& \left\langle {
		\frac{1}{\overline{\rho}} b'{}^k \epsilon^{ijk} 
		\frac{\partial b'{}^j}{\partial x^\ell}
		- u'{}^k \epsilon^{ijk} 
		\frac{\partial u'{}^j}{\partial x^\ell}
	} \right\rangle B^\ell
	- \left\langle {
		u'{}^\ell u'{}^k 
		+ \frac{b'{}^\ell b'{}^k}{\mu_0 \overline{\rho}}
	} \right\rangle \mu_0 \epsilon^{ijk} \frac{\partial B^j}{\partial x^\ell}
	\nonumber\\
	&&+ \left\langle {
		u'{}^\ell b'{}^k + b'{}^\ell u'{}^k
	} \right\rangle \epsilon^{ijk} \frac{\partial U^j}{\partial x^\ell}
	+ {\rm{H.T.}},
	\label{eq:emf_eq}
\end{eqnarray}
Equation~(\ref{eq:emf_eq}) with the inhomogeneous mean velocity or the $\nabla {\bf{U}}$-term dropped corresponds to the Ansatz that the EMF is expressed in terms of the mean magnetic field and its derivatives such as
\begin{equation}
	\langle {{\bf{u}}' \times {\bf{b}}'} \rangle^i
	= \alpha^{i\ell} B^\ell
	+ \beta^{ij\ell} \frac{\partial B^j}{\partial x^\ell}
	+ \cdots,
	\label{eq:Ansatz}
\end{equation}
where $\alpha^{i\ell}$ and $\beta^{ij\ell}$ are transport coefficients. In other words, the adoption of the usual Ansatz corresponds to the assumption of the no mean velocity inhomogeneity effect in the EMF. As (\ref{eq:fluct_u_eq_sec3}) and (\ref{eq:fluct_b_eq_sec3}) show, the velocity and magnetic-field fluctuations depend on the large-scale inhomogeneity of the velocity. If we retain the mean velocity inhomogeneity effects, the third or $\nabla {\bf{U}}$ terms should show up in the ELM expression. 

	In Section~2, we obtained the expression of the turbulent EMF (\ref{eq:emf_expression}) in an elaborated formulation. However, in order to get an intuitive view of the physical origins of the dynamo effects, here we use much more simplified arguments with assuming the simplest statistics on turbulence. If we assume that turbulent field is homogeneous and isotropic, the two-point two-time turbulent velocity and magnetic-field correlations are expressed in the generic form as
\begin{eqnarray}
	&&\langle {
		\phi'{}^i({\bf{x}};t) \psi'{}^j({\bf{x}}';t')
	} \rangle
	= \langle {
		\phi'{}^i({\bf{0}};t) \psi'{}^j({\bf{r}};t')
	} \rangle
	\nonumber\\
	&&= g(r;t,t') \delta^{ij}
	+ \frac{f(r;t,t') - g(r;t,t')}{r^2} r^i r^j
	+ h(r;t,t') \epsilon^{ij\ell} \frac{r^\ell}{r},
	\label{eq:two_time_two_point_cor_exp}
\end{eqnarray}
where $r = \|{\bf{r}}\|$ with ${\bf{r}} = {\bf{x}}' - {\bf{x}}$ is the distance between two points, and $\mbox{\boldmath$\phi$}'$ and $\mbox{\boldmath$\psi$}'$ denote either one of the velocity and magnetic field, ${\bf{u}}'$ and ${\bf{b}}'$, respectively. Here, $g$, $f$, and $h$ are the longitudinal, transverse, and cross correlation functions, respectively.

	If we substitute (\ref{eq:two_time_two_point_cor_exp}) with $\mbox{\boldmath$\phi$}'$ and $\mbox{\boldmath$\psi$}'$ being ${\bf{u}}'$ and ${\bf{b}}'$ into (\ref{eq:emf_eq}), we have
\begin{eqnarray}
	&&\frac{DE_{\rm{M}}^i}{Dt}
	= \frac{1}{3} \left\langle {
    	\frac{1}{\overline{\rho}} b'{}^k \epsilon^{k\ell j} 
		\frac{\partial b'{}^j}{\partial x^\ell}
    	- u'{}^k \epsilon^{k\ell j} 
		\frac{\partial u'{}^j}{\partial x^\ell}
	} \right\rangle B^i
	\nonumber\\
	&&- \frac{1}{3} \left\langle {
		u'{}^k u'{}^k 
		+ \frac{b'{}^k b'{}^k}{\mu_0 \overline{\rho}}
	} \right\rangle \mu_0 \epsilon^{i\ell j} 
		\frac{\partial B^j}{\partial x^\ell}
	+ \frac{2}{3} \left\langle {
		u'{}^k b'{}^k
	} \right\rangle \epsilon^{i\ell j} \frac{\partial U^j}{\partial x^\ell}
	+ {\rm{H.T.}},
	\label{eq:two_time_two_point_cor_exp_iso}
\end{eqnarray}
This suggests that the EMF is expressed as
\begin{equation}
	{\bf{E}}_{\rm{M}}
	= \alpha {\bf{B}}
	- \beta \mu_0 {\bf{J}}
	+ \gamma \mbox{\boldmath$\Omega$}.
	\label{eq:emf_exp_sec3}
\end{equation}
The transport coefficients $\alpha$, $\beta$, and $\gamma$ are expressed as
\begin{equation}
	\alpha = \tau_\alpha H,
	\label{eq:alpha_model_sec3}
\end{equation}
\begin{equation}
	\beta = \tau_\beta K,
	\label{eq:beta_model_sec3}
\end{equation}
\begin{equation}
	\gamma = \tau_\gamma W.
	\label{eq:gamma_model_sec3}
\end{equation}
Here, $H$, $K$, and $W$ are the turbulent residual helicity, the turbulent MHD energy, and the turbulent cross helicity defined by
\begin{equation}
	H = \left\langle {
		- {\bf{u}}' \cdot \mbox{\boldmath$\omega$}'
		+ {\bf{b}}' \cdot {\bf{j}}'/\overline{\rho}
	} \right\rangle,
	\label{eq:H_def_sec3}
\end{equation}
\begin{equation}
	K = \left\langle {
		{\bf{u}}'{}^2
		+ {\bf{b}}'{}^2/(\mu_0 \overline{\rho})
	} \right\rangle/2,
	\label{eq:K_def_sec3}
\end{equation}
\begin{equation}
	W = \left\langle {
		{\bf{u}}' \cdot {\bf{b}}'
	} \right\rangle,
	\label{eq:W_def_sec3}
\end{equation}
and $\tau_\alpha$, $\tau_\beta$, and $\tau_\gamma$ are the time scales associated with the turbulent residual helicity, turbulent MHD energy, and the turbulent cross helicity, respectively. Equation~(\ref{eq:emf_exp_sec3}) shows that in the presence of the non-uniform mean velocity, the cross-helicity or $W$-related term, in addition to the usual helicity or $H$-related term and the energy or $K$-related term, should be included in the expression of the EMF.

	Note that more thorough and detailed expressions of the EMF can be obtained by more elaborated dynamo theories than the above simple argument with resorting to (\ref{eq:emf_eq}) with the homogeneous and isotropic assumption (\ref{eq:two_time_two_point_cor_exp}). However, the point here is that retaining the non-uniform mean velocity effects in the fluctuating velocity and magnetic field, the turbulent cross-helicity coupled with the mean vortical motion emerges in the EMF expression.

\subsection{Physical origins of the turbulent effects in dynamos}
If turbulence possesses some statistical properties, the effective fluxes associated with the turbulence contribute to the dynamo-related transport through the turbulent EMF in coupling with a mean-field configuration. In the following, we scrutinize what physical processes are the essence of these turbulence effects by examining each term of the evolution equation of the velocity and magnetic-field fluctuations. Of course, we should be cautious in the use of such an argument, since it relies on the consideration of one particular term. It may occur that other terms completely or substantially cancels the effect. However, it is true such an argument is useful to grasp a feel of the physical origin of the effect.

\subsubsection{$\alpha$ effect: Kinetic and current helicity effect}
\begin{figure}[htb]
  \centering
  \includegraphics[width= 0.5 \columnwidth]{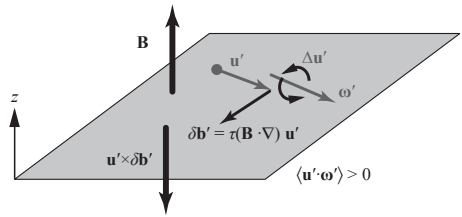}
  \caption{$\alpha$ effect due to the turbulent kinetic helicity.}
    \label{fig:alpha_kin_effect}
\end{figure}

Let us consider a fluid element fluctuating in the mean magnetic field ${\bf{B}}$ (Fig.~\ref{fig:alpha_kin_effect}). Here, we assume a positive kinetic helicity in turbulence, $\langle {{\bf{u}}' \cdot \mbox{\boldmath$\omega$}'} \rangle > 0$. Namely, the fluctuating velocity and magnetic field are statistically aligned with each other. In the presence of the mean magnetic field ${\bf{B}}$, the velocity fluctuation associated with the fluctuating vorticity varies along ${\bf{B}}$. Due to this ${\bf{u}}'$ variation along the mean magnetic field ${\bf{B}}$, from the first term on the r.h.s.\ of (\ref{eq:fluct_b_eq_sec3}), the magnetic-field fluctuation $\delta {\bf{b}}'$ is induced as
\begin{equation}
	\delta {\bf{b}}'
	= \tau_b ({\bf{B}} \cdot \nabla) {\bf{u}}',
	\label{eq:delb_ind_var_along_B}
\end{equation}
where $\tau_b$ is the time scale of ${\bf{b}}'$ evolution. The electromotive force (EMF) due to this effect is expressed as
\begin{equation}
	\langle {{\bf{u}}' \times \delta {\bf{b}}'} \rangle
	= \tau_b \langle {
		{\bf{u}}' \times ({\bf{B}} \cdot \nabla) {\bf{u}}'
	} \rangle,
	\label{eq:u_times_delb}
\end{equation}
whose direction is antiparallel to the mean magnetic field ${\bf{B}}$ for the positive turbulent kinetic helicity $\langle {{\bf{u}}' \cdot \mbox{\boldmath$\omega$}'} \rangle > 0$. The direction of the EMF is parallel to ${\bf{B}}$ for the negative turbulent kinetic helicity $\langle {{\bf{u}}' \cdot \mbox{\boldmath$\omega$}'} \rangle < 0$.

\begin{figure}[htb]
  \centering
  \includegraphics[width= 0.5 \columnwidth]{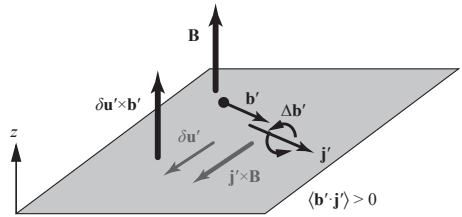}
  \caption{$\alpha$ effect due to the turbulent electric-current helicity.}
    \label{fig:alpha_mag_effect}
\end{figure}

	On the other hand, if we assume a positive current helicity in turbulence, $\langle {{\bf{b}}' \cdot {\bf{j}}'} \rangle > 0$, the fluctuating magnetic field ${\bf{b}}'$ is statistically aligned with the electric-current density ${\bf{j}}'$ (Fig.~\ref{fig:alpha_mag_effect}). In an entirely similar manner as in the turbulent kinetic helicity effect mentioned above, in association with the fluctuating electric-current density ${\bf{j}}'$, the fluctuating magnetic field varies along the mean magnetic field ${\bf{B}}$. From the first term or $({\bf{B}} \cdot \nabla) {\bf{b}}'$, the fluctuating velocity $\delta {\bf{u}}'$ is induced as
\begin{equation}
	\delta {\bf{u}}'
	= \tau_u ({\bf{B}} \cdot \nabla) {\bf{b}}',
	\label{eq:delu_ind_var_along_B}
\end{equation}
where $\tau_u$ is the time scale of ${\bf{u}}'$ evolution due to this mean magnetic field effect. In this case, we can alternatively consider the effect of the fluctuating Lorentz force ${\bf{j}}' \times {\bf{B}}$ associated with the mean magnetic field ${\bf{B}}$. The fluctuating velocity $\delta {\bf{u}}'$ is induced as
\begin{equation}
	\delta {\bf{u}}'
	= \tau_u {\bf{j}}' \times {\bf{B}}.
	\label{eq:delu_due_to_j_times_B}
\end{equation}
This is equivalent to (\ref{eq:delu_ind_var_along_B}). This induced velocity fluctuation $\delta {\bf{b}}'$ combined with the fluctuating magnetic field ${\bf{b}}'$ constitutes the electromotive force as
\begin{equation}
	\langle {\delta {\bf{u}}' \times {\bf{b}}'} \rangle^i
	= \tau_u \langle {({\bf{j}}' \times {\bf{B}}) \times {\bf{b}}'} \rangle^i
	= \tau_u \langle {{\bf{b}}' \cdot {\bf{j}}'} \rangle B^i.
	\label{eq:delu_times_b}
\end{equation}
Here, we dropped the contribution from the fluctuations along the mean magnetic field ${\bf{B}}$ as $- \langle {({\bf{b}}' \cdot {\bf{B}}) {\bf{j}}'} \rangle = - \langle {b'_{\parallel B} {\bf{j}}'} \rangle \|{\bf{B}}\|$, where $b'_{\parallel B} = {\bf{b}}' \cdot {\bf{B}}/\|{\bf{B}}\|$ is the ${\bf{b}}'$ component parallel to ${\bf{B}}$. The direction of the EMF is parallel to the mean magnetic field ${\bf{B}}$ for positive turbulent current helicity $\langle {{\bf{b}}' \cdot {\bf{j}}'} \rangle > 0$, and is antiparallel to ${\bf{B}}$ for negative turbulent current helicity $\langle {{\bf{b}}' \cdot {\bf{j}}'} \rangle < 0$. This current helicity effect on the $\alpha$ dynamo was first pointed out by \citet{pou1976} on the basis of a closure calculation of MHD turbulence with the aid of the eddy-damped quasi-normal Markovianized (EDQNM) approximation. In the derivation, the timescale for the kinetic-helicity effect, $\tau_b$ (\ref{eq:u_times_delb}), and the timescale for the current-helicity effect, $\tau_u$ (\ref{eq:delu_times_b}), were not distinguished.

These physical pictures show that if the turbulent kinetic and current helicities have a sign same with each other, these two helicity effects counterbalance each other. In this sense, the current-helicity in the $\alpha$ dynamo is often argued that this magnetic correction to the $\alpha$ dynamo represents the saturation of the $\alpha$ dynamo due to the magnetic-field effect. However, the sign of the current helicity can be opposite to that of the kinetic helicity. In such a case, the current-helicity effect may enhance the magnetic field generated by the kinetic-helicity effect. In the case of inhomogeneous turbulence, which is ubiquitous in the real world, the spatiotemporal evolution of the current helicity $\langle {{\bf{b}}' \cdot {\bf{j}}'} \rangle$ depends on the production rates of the current helicity directly related to the mean-field inhomogeneities. Actually, it is reported that, in the numerical simulation of the solar convective zone, at some deep region, the magnitude of the turbulent current helicity is dominantly larger than the counterpart of the turbulent kinetic helicity. This suggests that mean magnetic field can be generated by the current-helicity effect at the depth even in the absence of the turbulent kinetic helicity there. Such arguments should be done on the basis of the evolution equation of the kinetic and current helicities.

\subsubsection{Turbulent magnetic diffusivity: Turbulent energy effect}
Let us consider a fluid element located in a mean electric-current density ${\bf{J}}$ (Figure~3.4). In this case, the fluid element fluctuates in a non-uniform mean magnetic field associated with the mean electric-current density ${\bf{J}}$. Because of the second term or $- ({\bf{u}}' \cdot \nabla) {\bf{B}}$ on the r.h.s.\ of (\ref{eq:fluct_b_eq_sec3}), the magnetic-field fluctuation $\delta {\bf{b}}'$ is induced in the direction of the non-uniform ${\bf{B}}$ but in the sense that it counterbalances the increase of ${\bf{B}}$ as
\begin{equation}
	\delta {\bf{b}}' 
	= - \tau_b ({\bf{u}}' \cdot \nabla) {\bf{B}}.
	\label{eq:delb_var_B_along_u}
\end{equation}
The EMF arising from the non-uniform mean magnetic field is given by
\begin{equation}
	\langle {{\bf{u}}' \times \delta {\bf{b}}'} \rangle^i
	= - \tau_b \langle {
		{\bf{u}}' \times [({\bf{u}}' \cdot \nabla) {\bf{B}}]
	} \rangle^i
	= - \tau_b \langle {u'{}^j u'{}^\ell} \rangle \epsilon^{ijk} 
		\frac{\partial B^k}{\partial x^\ell}.
	\label{eq:u_times_delb_sec3}
\end{equation}
If we assume that the velocity fluctuation is isotropic as $\langle {u'{}^j u'{}^\ell} \rangle = \langle {{\bf{u}}'{}^2} \rangle \delta^{j\ell}/3$, (\ref{eq:u_times_delb_sec3}) is reduced to
\begin{equation}
	\langle {{\bf{u}}' \times \delta {\bf{b}}'} \rangle^i
	= - \frac{1}{3} \tau_b \langle {{\bf{u}}'{}^2} \rangle \epsilon^{i\ell k}
		\frac{\partial B^k}{\partial x^\ell}
	= - \frac{1}{3} \tau_b \langle {{\bf{u}}'{}^2} \rangle
		(\nabla \times {\bf{B}})^i.
	\label{eq:u_times_delb_iso}
\end{equation}
In the presence of velocity fluctuation, the turbulent electromotive force (EMF) due to the non-uniform mean magnetic field $\nabla {\bf{B}}$ is in the direction antiparallel to the mean electric-current density ${\bf{J}} (= \nabla \times {\bf{B}})$. The transport coefficient is proportional to the intensity of fluctuation or the turbulent kinetic energy, $\langle {{\bf{u}}'{}^2} \rangle$.

\begin{figure}[htb]
  \centering
  \includegraphics[width= 0.5 \columnwidth]{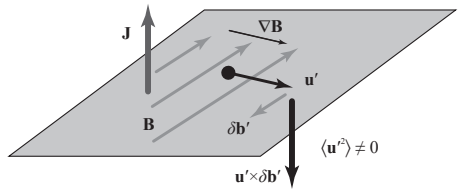}
  \caption{Turbulent magnetic diffusivity due to the velocity fluctuation.}
    \label{fig:beta_kin_effect}
\end{figure}

\begin{figure}[htb]
  \centering
  \includegraphics[width= 0.5 \columnwidth]{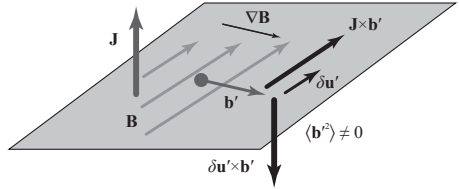}
  \caption{Turbulent magnetic diffusivity due to the magnetic fluctuation.}
    \label{fig:beta_mag_effect}
\end{figure}

	In a similar manner, we argue the effect of the magnetic fluctuation on the turbulent magnetic diffusivity. First we consider turbulence with magnetic-field fluctuation ${\bf{b}}'$ in a non-uniform mean magnetic field ${\bf{B}}$ (Fig.~\ref{fig:beta_mag_effect}). In the presence of the mean electric-current density ${\bf{J}} (= \nabla \times {\bf{B}})$, the fluid element is subject to the fluctuating Lorentz force associated with the mean electric-current density, ${\bf{J}} \times {\bf{b}}'$. Then the fluctuating velocity $\delta {\bf{u}}'$ due to ${\bf{J}}$ is induced as
\begin{equation}
	\delta {\bf{u}}'
	= \tau_u {\bf{J}} \times {\bf{b}}'.
	\label{eq:del_u_due_to_J_times_b}
\end{equation}
The EMF due to the magnetic-field fluctuation in the presence of ${\bf{J}}$ is expressed as
\begin{equation}
	\langle {\delta {\bf{u}}' \times {\bf{b}}'} \rangle
	= \langle {\tau_u ({\bf{J}} \times {\bf{b}}') \times {\bf{b}}'} \rangle
	= - \tau_u \langle {{\bf{b}}'{}^2} \rangle {\bf{J}}.
	\label{eq:delu_times_b_b2J}
\end{equation}
Here, we dropped the contribution from the fluctuating magnetic field component parallel to the mean electric-current density, $\tau_u ({\bf{b}}' \cdot {\bf{J}}) {\bf{b}}' = \tau_u \langle {b'_{\parallel J} {\bf{b}}'} \rangle \|{\bf{J}}\|$, where $b'_{\parallel J} = {\bf{b}}' \cdot {\bf{J}}/\|{\bf{J}}\|$ is the fluctuating magnetic-field component along the mean electric-current density ${\bf{J}}$. Equation~(\ref{eq:delu_times_b_b2J}) implies that the magnetic fluctuation ${\bf{b}}'$ as well as the velocity fluctuation ${\bf{u}}'$ contributes to the turbulent magnetic diffusivity coupled with the mean electric-current density ${\bf{J}}$.

\begin{figure}[htb]
  \centering
  \includegraphics[width= 0.5 \columnwidth]{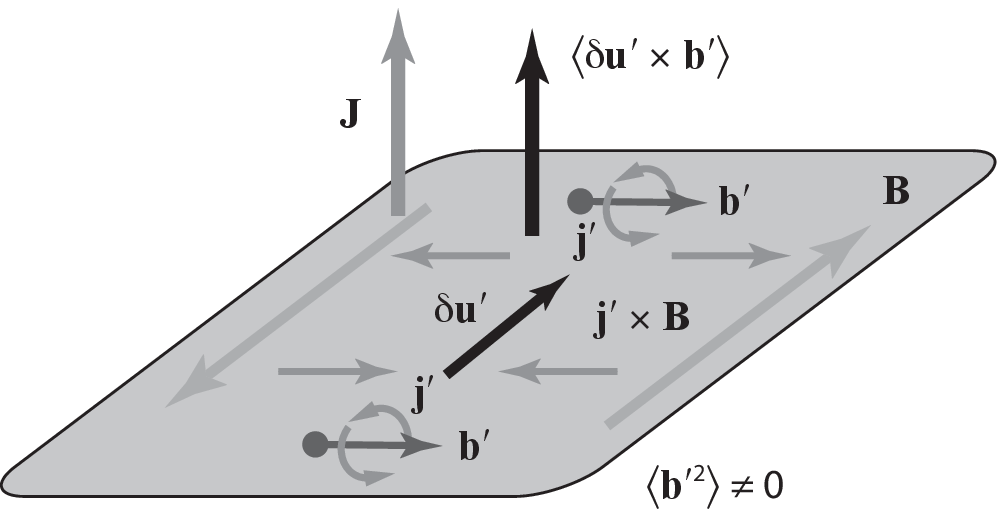}
  \caption{Cancellation of the turbulent magnetic energy effect due to the magnetic pressure effect.}
    \label{fig:cancel_beta_effect}
\end{figure}

	As for this magnetic fluctuation effect, however, we should note the following point. As we see above, in the presence of the mean electric-current density ${\bf{J}}$, the part of fluctuating Lorentz force ${\bf{J}} \times {\bf{b}}'$ induces the fluctuating velocity $\delta {\bf{u}}'$, which plays a key role for the turbulent magnetic diffusivity due to the fluctuating magnetic field. However, this effect can be at least partly canceled by the other part of the fluctuating Lorentz force ${\bf{j}}' \times {\bf{B}}$. If the magnetic fluctuation has a spatial distribution like Fig.~\ref{fig:cancel_beta_effect}, the fluctuating electric-current density associated with ${\bf{b}}'$, $\Delta {\bf{j}}$, also shows a spatial distribution. The associated electric-current density $\Delta {\bf{j}}'$ is in the direction enhancing the original mean electric-current density ${\bf{J}}$ on the side parallel to ${\bf{J}} \times {\bf{b}}'$ with respect to ${\bf{b}}'$ (parallel side), and $\Delta {\bf{j}}'$ in the direction reducing ${\bf{J}}$ on the side antiparallel to ${\bf{J}} \times {\bf{b}}'$ (antiparallel side). Due to this $\Delta {\bf{j}}'$ effect, the magnetic pressure becomes higher in the parallel side than the antiparallel side. In other words, the velocity fluctuation $\delta {\bf{u}}$ induced by the fluctuating Lorentz force $\Delta {\bf{j}}' \times {\bf{B}}$ works for increasing the plasma density on the parallel side, while the flow due to the fluctuating Lorentz force, $\delta {\bf{u}}' = \tau_u \Delta {\bf{j}}' \times {\bf{B}}$, works for decreasing the plasma density on the antiparallel side. This magnetic pressure effect in the combination of the magnetic fluctuation ${\bf{b}}'$ and the non-uniform mean magnetic field ${\bf{B}}$, counterbalances the effect of the Lorentz force ${\bf{J}} \times {\bf{b}}'$. In this sense, the magnetic-field fluctuations ${\bf{b}}'$ effect on the turbulent magnetic diffusivity is not obvious as the counter part of the velocity fluctuations ${\bf{u}}'$. Actually, in the detailed analytical calculation shows that in the solenoidal turbulence, the turbulent magnetic diffusivity does not depend  on the magnetic fluctuation energy $\langle {{\bf{b}}'{}^2} \rangle$. The magnetic-field fluctuation energy becomes relevant to the turbulent magnetic diffusivity in the case of compressible turbulence \citep{yok2018a} and anisotropic situations \citep{rog2001}.

\subsubsection{Turbulent cross-helicity effect}
\begin{figure}[htb]
  \centering
  \includegraphics[width= 0.5 \columnwidth]{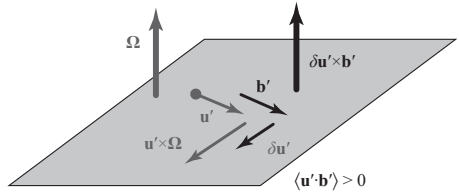}
  \caption{Turbulent cross-helicity effect due to the velocity fluctuation.}
    \label{fig:cross-helicity_effect_vel_fluct}
\end{figure}

	We consider a fluid element that fluctuates in a mean vorticity field $\mbox{\boldmath$\Omega$} (= \nabla \times {\bf{U}})$ (Fig.~\ref{fig:cross-helicity_effect_vel_fluct}). We assume the turbulence has a positive cross helicity $\langle {{\bf{u}}' \cdot {\bf{b}}'} \rangle > 0$, namely, the fluctuating velocity and magnetic field are statistically aligned with each other. Because of the local angular momentum conservation, the fluid is subject to the Coriolis-like force. Then, the velocity fluctuation is induced by the mean vortical motion $\mbox{\boldmath$\Omega$}$ as
\begin{equation}
	\delta {\bf{u}}'
	= \tau_u {\bf{u}}' \times \mbox{\boldmath$\Omega$}.
	\label{delu_u_times_Omega}
\end{equation}	
The electromotive force constituted by this induced $\delta {\bf{u}}'$ and the fluctuating magnetic-field component parallel to the magnetic field ${\bf{b}}'$ is expressed by
\begin{equation}
	\langle {\delta {\bf{u}}' \times {\bf{b}}'} \rangle
	= \langle {
		\tau_u ({\bf{u}}' \times \mbox{\boldmath$\Omega$}) \times {\bf{b}}'
	} \rangle
	= \tau_u \langle {{\bf{u}}' \cdot {\bf{b}}'} \rangle 
		\mbox{\boldmath$\Omega$}.
	\label{eq:delu_times_b_ub_Omega}
\end{equation}	
Here, we dropped the contribution from the magnetic fluctuations along the mean vorticity $- \tau_u \langle {{\bf{u}}' ({\bf{b}}' \cdot \mbox{\boldmath$\Omega$})} \rangle = - \tau_u \langle {b_{\parallel\Omega}} \rangle \|\mbox{\boldmath$\Omega$}\|$, where $b'_{\parallel\Omega}$ is the ${\bf{b}}'$ component parallel to $\mbox{\boldmath$\Omega$}$ defined by ${\bf{b}}' \cdot \mbox{\boldmath$\Omega$}/\|\mbox{\boldmath$\Omega$}\|$ since the statistical average of $b'_{\parallel\Omega} {\bf{u}}'$ is expected to vanish. Equation~(\ref{eq:delu_times_b_ub_Omega}) suggests that EMF due to the cross helicity is aligned with the mean vorticity $\mbox{\boldmath$\Omega$}$. The direction of EMF is determined by the sign of the turbulent cross helicity. The EMF is parallel to the mean vorticity $\mbox{\boldmath$\Omega$}$ for positive turbulent cross helicity $\langle {{\bf{u}}' \cdot {\bf{b}}'} \rangle > 0$, and antiparallel to $\mbox{\boldmath$\Omega$}$ for negative turbulent cross helicity $\langle {{\bf{u}}' \cdot {\bf{b}}'} \rangle < 0$.	

\begin{figure}[htb]
  \centering
  \includegraphics[width= 0.5 \columnwidth]{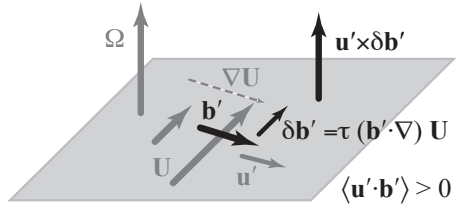}
  \caption{Turbulent cross-helicity effect due to the magnetic fluctuation.}
    \label{fig:cross-helicity_effect_mag_fluct}
\end{figure}

Next, we consider motion of a fluid element with magnetic fluctuations ${\bf{b}}'$ in a non-uniform mean velocity $\nabla {\bf{U}}$ (Fig.~\ref{fig:cross-helicity_effect_mag_fluct}). Also in this case, we assume the cross helicity in turbulence is positive ($\langle {{\bf{u}}' \cdot {\bf{b}}'} \rangle$); the turbulent velocity ${\bf{u}}'$ and turbulent magnetic field ${\bf{b}}'$ are statistically aligned and parallel with each other. We see from the third term or $\tau_b ({\bf{b}}' \cdot \nabla) {\bf{U}}$ on the r.h.s.\ of Eq.~(\ref{eq:fluct_u_eq_sec3}) that the magnetic fluctuation $\delta {\bf{b}}'$ is induced as
\begin{equation}
	\delta {\bf{b}}' = \tau_b ({\bf{b}}' \cdot \nabla) {\bf{U}}.
	\label{eq:delb_b_nabla_U}
\end{equation}
This induced magnetic fluctuation $\delta {\bf{b}}'$ is in the direction parallel to ${\bf{U}}$ if the non-uniform ${\bf{U}}$ increases as the fluctuating magnetic field ${\bf{b}}'$ moves. Multiplying the fluctuating velocity ${\bf{u}}'$ in the vector product manner by Eq.~(\ref{eq:delb_b_nabla_U}) and taking the ensemble averaging, we obtain
\begin{equation}
	\langle {{\bf{u}}' \times \delta {\bf{b}}'} \rangle^i 
	= \langle {
		{\bf{u}}' \times \tau_b ({\bf{b}}' \cdot \nabla) {\bf{U}}
	} \rangle^i
	= \left\langle {
		\epsilon^{ijk} u'{}^j \tau_b b'^\ell 
		\frac{\partial U^k}{\partial x^\ell}
	} \right\rangle
	= \tau_b \langle {u'{}^j b'{}^\ell} \rangle
		\epsilon^{ijk} \frac{\partial U^k}{\partial x^\ell}.
	\label{eq:u_times_delb_ub_gradU}
\end{equation}
If we assume the velocity and magnetic-field correlation is isotropic as $\langle {u'{}^j b'{}^\ell} \rangle = \langle {{\bf{u}}' \cdot {\bf{b}}'} \rangle \delta^{j\ell}/3$, (\ref{eq:u_times_delb_ub_gradU}) is reduced to
\begin{equation}
	\langle {{\bf{u}}' \times \delta {\bf{b}}'} \rangle^i 
	= \tau_b \langle {{\bf{u}}' \cdot {\bf{b}}'} \rangle
	\epsilon^{ijk} \frac{\partial U^k}{\partial x^j}
	= \tau_b \langle {{\bf{u}}' \cdot {\bf{b}}'} \rangle
	(\nabla \times {\bf{U}})^i
	= \tau_b \langle {{\bf{u}}' \cdot {\bf{b}}'} \rangle
	\Omega^i.
	\label{eq:u_times_delb_ub_Omega}
\end{equation}

Since the mean vorticity $\mbox{\boldmath$\Omega$} = \nabla \times {\bf{U}}$ is locally equivalent to the angular velocity of a rotation $\mbox{\boldmath$\omega$}_{\rm{F}} = \mbox{\boldmath$\Omega$}/2$, exactly the same argument can be applied to the fluid element in a rotating system. Hence, we can replace the mean vorticity $\mbox{\boldmath$\Omega$}$ by the mean absolute vorticity $\mbox{\boldmath$\Omega$}_\ast \equiv \mbox{\boldmath$\Omega$} + 2 \mbox{\boldmath$\omega$}_{\rm{F}}$. 

Equations~(\ref{eq:delu_times_b_b2J}) and (\ref{eq:u_times_delb_ub_Omega}) show that in the presence of the mean absolute vorticity $\mbox{\boldmath$\Omega$}_\ast$, the turbulent electromotive force (EMF) is expressed as
\begin{equation}
	\langle {{\bf{u}}' \times {\bf{b}}'} \rangle_\gamma
	= \gamma \mbox{\boldmath$\Omega$}_\ast,
	\label{eq:u_times_b_gamma_Omega}
\end{equation}
where $\gamma$ is the transport coefficient determined by the turbulent cross helicity and turbulence timescale related to the cross helicity, $\tau_\gamma$, expressed by
\begin{equation}
	\gamma 
	= \tau_\gamma \langle {{\bf{u}}' \cdot {\bf{b}}'} \rangle,
	\label{eq:gamma_tau_ub}
\end{equation}
and $\mbox{\boldmath$\Omega$}_\ast$ is the mean absolute vorticity defined by
\begin{equation}
	\mbox{\boldmath$\Omega$}_\ast 
	\equiv \mbox{\boldmath$\Omega$} + 2 \mbox{\boldmath$\omega$}_{\rm{F}}
	\label{eq:absolute_vort_def}
\end{equation}
with the mean relative vorticity $\mbox{\boldmath$\Omega$} = \nabla \times {\bf{U}}$ and the angular velocity of the system rotation $\mbox{\boldmath$\omega$}_{\rm{F}}$.

\subsection{Numerical validation of the cross-helicity effect}

\subsubsection{Kolmogorov flow with imposed uniform magnetic field}
\begin{figure}[htb]
  \centering
  \includegraphics[width= 0.8 \columnwidth]{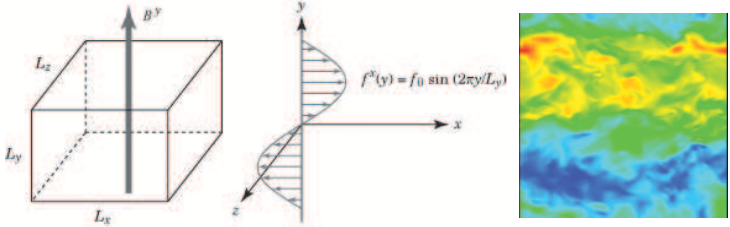}
  \caption{Kolmogorov flow with imposed uniform magnetic field. The triple periodic configuration with the imposed uniform magnetic field in the direction of inhomogeneity ($y$ direction) (left). Sinusoidal forcing for inhomogeneous velocity shear (middle). The contour of the streamwise velocity (right).}
    \label{fig:kol_flow_w_mag_fld_setup}
\end{figure}

In order to validate the cross-helicity effect in dynamo problems, we performed a direct numerical simulation (DNS) of forced magnetohydrodynamic (MHD) turbulence in a three-dimensional periodic box with a uniform magnetic field imposed (Fig.~\ref{fig:kol_flow_w_mag_fld_setup}) \citep{yok2011a}. The size of the box is $(L_x, L_y, L_z)$. The imposed forcing is in the $x$ direction:
\begin{equation}
	{\bf{f}}_{\rm{ext}} = (f^x, f^y, f^z) = (f^x, 0, 0)
	\label{eq:kol_forcing}
\end{equation}
and is inhomogeneous in the $y$ direction:
\begin{equation}
	f^x(y) = f_0 \sin \left( {\frac{2\pi y}{L_y}} \right).
	\label{eq:kol_forcing_sin}
\end{equation}
In this setup, turbulence is generated and sustained by the velocity shear due to the sinusoidal forcing (\ref{eq:kol_forcing}) with (\ref{eq:kol_forcing_sin}). Because of the forcing, the statistics of the turbulence is inhomogeneous in the $y$ direction, but is homogeneous in the $x$ and $z$ directions. The statistics of quantities are represented by averaging in the homogeneous ($x$-$z$) directions. This configuration is known as the Kolmogorov turbulent flow. In order to see the turbulence properties related to dynamo and its transport, we further impose a uniform mean magnetic field in the inhomogeneous direction, namely in the $y$ direction as
\begin{equation}
	{\bf{B}}_0 = (B_0^x, B_0^y, B_0^z) = (0, B_0, 0).
	\label{eq:kol_imposed_B}
\end{equation}
The contour of the streamwise velocity is shown in Figure~\ref{fig:kol_flow_w_mag_fld_setup}. Reflecting the sinusoidal form of the forcing (\ref{eq:kol_forcing}), the value of the streamwise velocity component is positive and negative in the upper half domain ($y > 0$) and lower half domain ($y<0$), respectively. 

	With this setup, we calculate the turbulent correlations and the mean-field quantities that are relevant to the mean-field dynamo. The expression of the turbulent electromotive force (EMF) is theoretically obtained in Section~2, and modeled as
\begin{equation}
	\langle {{\bf{u}}' \times {\bf{b}}'} \rangle
	= \alpha {\bf{B}}
	- \beta {\bf{J}}
	+ \gamma \mbox{\boldmath$\Omega$},
	\label{eq:emf_model_sec3}
\end{equation}
where the transport coefficients $\alpha$, $\beta$, and $\gamma$ are expressed in terms of the turbulent residual helicity, the turbulent energy, and the turbulent cross helicity.

As (\ref{eq:emf_model_sec3}) shows, the relevant mean fields are the mean magnetic field ${\bf{B}}$, the mean electric-current density ${\bf{J}} (= \nabla \times {\bf{B}})$, and the mean relative vorticity $\mbox{\boldmath$\Omega$} (= \nabla \times {\bf{U}})$. On the other hand, the relevant turbulent correlations are the turbulent MHD energy $K$, its dissipation rate $\varepsilon_K (\equiv \varepsilon)$, the turbulent residual helicity $H$, and the turbulent cross helicity $W$. They are defined by
\begin{equation}
	K = \langle {{\bf{u}}'{}^2 + {\bf{b}}'{}^2} \rangle/2,
	\label{eq:K_def_sec3}
\end{equation}
\begin{equation}
	\varepsilon_K
	= \nu \left\langle {
		\left( {\frac{\partial u'{}^j}{\partial x^i}} \right)^2
	} \right\rangle 
	+ \eta \left\langle {
		\left( \frac{\partial b'{}^j}{\partial x^i} \right)^2
	} \right\rangle
	\equiv \varepsilon,
	\label{eq:eps_def_sec3}
\end{equation}
\begin{equation}
	H
	= \left\langle {
	- {\bf{u}}' \cdot \mbox{\boldmath$\omega$}' 
	+ {\bf{b}}' \cdot {\bf{j}}'
	} \right\rangle,
	\label{eq:H_def_sec3}
\end{equation}
\begin{equation}
	W = \left\langle {{\bf{u}}' \cdot {\bf{b}}'} \right\rangle.
	\label{eq:W_def_sec3}
\end{equation}
In terms of these turbulent statistical quantities, the main dynamo transport coefficients are expressed as
\begin{equation}
	\alpha = \tau H,
	\label{eq:alpha_model_tau_H_sec3}
\end{equation}
\begin{equation}
	\beta = \tau K,
	\label{eq:beta_model_tau_K_sec3}
\end{equation}
\begin{equation}
	\gamma = \tau W,
	\label{eq:gamma_model_tau_W_sec3}
\end{equation}
where $\tau$ is the timescale of turbulence evaluated by
\begin{equation}
	\tau = K / \varepsilon.
	\label{eq:tau_model_K_over_eps_sec3}
\end{equation}

\begin{figure}[htb]
  \centering
  \includegraphics[width= 0.4 \columnwidth]{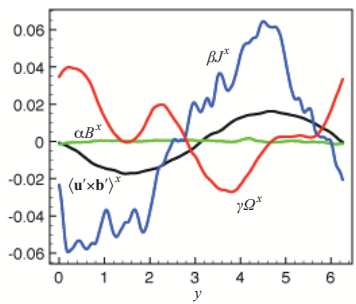}
  \caption{Comparison of the turbulent electromotive force (EMF) with the model terms, $\alpha {\bf{B}}$, $\beta {\bf{J}}$, and $\gamma \mbox{\boldmath$\Omega$}$ in Kolmogorov flow with imposed uniform magnetic field. The spatial distribution with respect to the inhomogeneous direction ($y$ direction) of $\langle {{\bf{u}}' \times {\bf{b}}'} \rangle^x$ (black) is compared with the $\alpha B^x$ (green), $\beta J^x$ (blue), and $\gamma \Omega^x$ (red).}
    \label{fig:kol_flow_w_mag_fld_emf}
\end{figure}

Using the DNS results, we plot the spatial ($y$) distribution of the $x$ component of the EMF $\langle {{\bf{u}}' \times {\bf{b}}'} \rangle$ as well as each term of the r.h.s.\ of Eq.~(\ref{eq:emf_model_sec3}), $\alpha {\bf{B}}$, $\beta {\bf{J}}$, and $\gamma \mbox{\boldmath$\Omega$}$ with (\ref{eq:alpha_model_tau_H_sec3})-(\ref{eq:tau_model_K_over_eps_sec3}) in Fig.~\ref{fig:kol_flow_w_mag_fld_emf}. The DNS of the EMF $\langle {{\bf{u}}' \times {\bf{b}}'} \rangle^x$ shows a sinusoidal distribution in the $y$ direction; negative values for $y<0$ and positive values for $y>0$. Among the model terms, the contribution of the $(\alpha {\bf{B}})^x$ is almost negligible in the whole domain as compared with the other terms, $- \beta {\bf{J}}$ and $\gamma \mbox{\boldmath$\Omega$}$. On the other hand, the turbulent diffusivity $\beta {\bf{J}}$ term plays a dominant role in constituting of the EMF. The spatial ($y$) distribution of $- \beta {\bf{J}}$ is coarsely sinusoidal; negative for $y<0$ and positive for $y>0$. The main balancer for the turbulent magnetic diffusivity $- \beta {\bf{J}}$ is the cross-helicity effect $\gamma \mbox{\boldmath$\Omega$}$. Its spatial distribution is basically positive for $y<0$ and negative for $y>0$. It is very notable that the spatial distribution of the sum of the turbulent magnetic diffusivity and the cross-helicity effect, $- \beta {\bf{J}} + \gamma \mbox{\boldmath$\Omega$}$, roughly agrees with the counterpart of the EMF $\langle {{\bf{u}}' \times {\bf{b}}'} \rangle$. This DNS result shows that, in this configuration, the turbulent EMF is constituted by the turbulent magnetic diffusivity $\beta {\bf{J}}$ and the cross-helicity effect $\gamma \mbox{\boldmath$\Omega$}$, and that the turbulent helicity or $\alpha$ effect does not play any substantial role in the turbulent transport. The profiles of the EMF, whose sign is opposite to that of $\gamma \mbox{\boldmath$\Omega$}$, is similar to the counterpart of the $\beta {\bf{J}}$. This indicates that, in this simulation, the turbulent EMF as a whole works for enhancing the effective diffusivity, not for dynamo. However, it is worth noting that the increased transport or the field destruction by the turbulent magnetic diffusivity $\beta {\bf{J}}$ is certainly suppressed by the dynamo or field-generation/sustainment effect by the cross helicity.

\begin{figure}[htb]
  \centering
  \includegraphics[width= 0.4 \columnwidth]{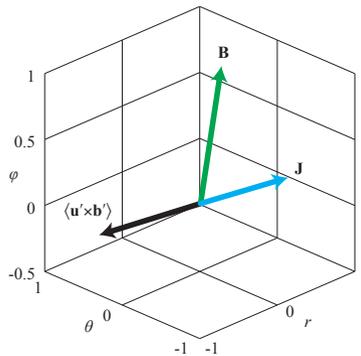}
  \caption{Comparison of the measured turbulent electromotive force (EMF) with the dynamo terms, $\alpha {\bf{B}}$, $\beta {\bf{J}}$, and $\gamma \mbox{\boldmath$\Omega$}$ in Kolmogorov flow with imposed uniform magnetic field. The spatial distribution with respect to the inhomogeneous direction ($y$ direction) of $\langle {{\bf{u}}' \times {\bf{b}}'} \rangle^x$ (black) is compared with the $\alpha B^x$ (green) and $\beta J^x$ (blue). Redrawn from \citet{rah2012}.}
    \label{fig:dynamo_experiment}
\end{figure}

This numerical validation is suggestive for the interpretation of the result of a dynamo experiment. In the liquid sodium dynamo experiment, the turbulent EMF $\langle {{\bf{u}}' \times {\bf{b}}'} \rangle$ was for the first time directly measured by simultaneously measuring three component of the velocity and magnetic field \citep{rah2012}. By comparing with the mean electric-current density term, it was shown that the EMF tends to oppose the local mean electric current, and that the mean magnetic field ${\bf{B}}$ tends to perpendicular to the direction of the EMF. This experimental result also suggests that the helicity or $\alpha$ effect, expressed by $\alpha {\bf{B}}$, does not contribute to the turbulent EMF. This experimental result is consistent with the numerical result of the turbulent transport (Fig.~\ref{fig:dynamo_experiment}) showing the dominance of the turbulent magnetic diffusivity and the irrelevance of the helicity or $\alpha$ effect in dynamo action. We should note that in this liquid sodium experiment a large-scale rotational or poloidal motion is also observed. Since there exists a mean vortical motion and the three-dimensional data of velocity and magnetic field are measured, it is interesting to re-examine the effect of the turbulent cross helicity $\langle {{\bf{u}}' \cdot {\bf{b}}'} \rangle$ which couples with the mean vortical motion $\mbox{\boldmath$\Omega$} (= \nabla \times {\bf{U}})$ as $\gamma \mbox{\boldmath$\Omega$}$.

\subsubsection{Archontis flow}
Another numerical validation was performed in the Archontis flow configuration, which is a generalization of the Arnold--Beltrami--Childress flow but with the cosine terms omitted \citep{sur2009}. As the result of this setup, the Archontis flow configuration is a non-helical one, and suitable to explore the dynamo due to the cross-helicity effect. A net cross helicity with either sign is generated by an instability, and the sign of the cross helicity depends on the initial condition. The direct numerical simulations of this flow show that the cross helicity contributes to inducing a large-scale magnetic field with exponential growth. It turns out that in order to evaluate the cross-helicity effect in dynamo, the mean-field effect should be also considered. This naturally leads us to explore the problem how and how much turbulent cross helicity is generated by the effects of the mean fields. This is the subject of Section~4.

\subsubsection{Cross-helicity and differential rotation effect in spherical shell}
In Section~3.1, we argued the cross helicity and global flow effect in dynamos. We saw the treatment of global-flow inhomogeneity is fairly different between the $\Omega$ (differential rotation) effect and turbulence. If we consider the velocity shear effect also in turbulence, the cross-helicity effect inevitably shows up. Since both the $\Omega$ effect and the cross-helicity effect depend on the inhomogeneous large-scale flow, the relative importance of of these effects in the magnetic field generation process is an interesting subject of the cross-helicity effect in dynamos.

\begin{figure}[htb]
  \centering
  \includegraphics[width= 0.7 \columnwidth]{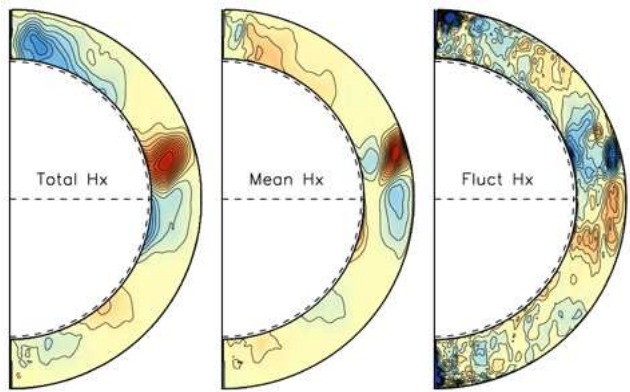}
  \caption{Spatial distribution of the cross helicity in the meridional surface. The total cross helicity (left), the mean cross helicity (middle), and the fluctuating cross helicity (right). Provided by Mark Miesch.}
    \label{fig:ch_distr_mark}
\end{figure}

	The spatial distributions of the total, mean, and fluctuating cross helicities in the direct numerical simulation of a spherical shell mimicking the Sun are plotted in contour in Fig.~\ref{fig:ch_distr_mark}. Here, the total cross helicity is calculated using the instantaneous velocity and magnetic fields, the mean cross helicity is calculated under the azimuthal averaging, and the fluctuating cross helicity is calculated by subtraction of the mean cross helicity from the total one. The spatial distributions of cross helicity is basically antisymmetric with respect to the midplane. This reflects pseudo-scalar property of cross helicity. We also see the signs of the mean and fluctuating cross helicity in each hemisphere are opposite each other. This is not unreasonable since the cross helicity is not a positive-definite quantity and its sign can be altered in scale.

As will be referred to in Section~4, we have the transport equations of the turbulent and mean-field cross helicities. Comparison of the numerical data of the cross helicity with the production, dissipation, and transport rates of the turbulent and mean-field cross helicities in the global simulation of the spherical shell geometry would give a way to understand the physical mechanisms that determine the spatiotemporal distribution of the turbulent cross helicity.

\begin{figure}[htb]
  \centering
  \includegraphics[width= 0.7 \columnwidth]{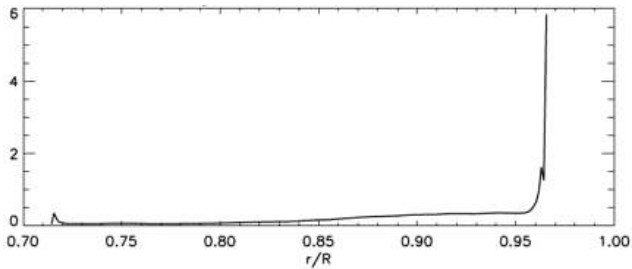}
  \caption{Relative magnitude of the cross-helicity effect $\| {\nabla \times (\gamma \nabla \times {\bf{U}})} \|$ to the differential rotation effect $\| {\nabla \times ({\bf{U}} \times {\bf{B}})} \|$. Here, $R$ is the radius of the Sun, and $r$ is the radial coordinate. Provided by Mark Miesch.}
    \label{fig:rel_ch_effect_mark}
\end{figure}

	Next we evaluate the relative importance of cross-helicity effect to the differential rotation effect in the mean induction equation by calculating
\begin{equation}
	\frac{(\mbox{cross-helicity effect})}
		{(\mbox{differential-rotation effect})}
	= \frac{\| {\nabla \times (\gamma \nabla \times {\bf{U}})} \|}
		{\| {\nabla \times ({\bf{U}} \times {\bf{B}})} \|}.
	\label{eq:rel_importance_ch}
\end{equation}
This ratio is plotted against the radius $r$ in Fig.~\ref{fig:rel_ch_effect_mark}.
This plot implies that the relative importance of the cross-helicity effect to the differential-rotation effect is negligibly small in the deeper region ($0.7 < r/R \lesssim 0.85$), but it becomes $\sim 0.5$ in the upper middle layer ($0.85 \lesssim r/R \lesssim 0.96$). This ratio raises to unity and much more higher in the near surface layer ($0.96 \lesssim r/R$). This increase is because the large-scale vorticity becomes much stronger in the near surface layer. This result shows that the cross-helicity effect is comparable to the differential rotation effect, and is not negligible in the upper middle layer of the Sun. This result further suggests that, in the near surface layer, the cross-helicity effect plays a dominant role in the mean magnetic-field induction.

The ratio (\ref{eq:rel_importance_ch}) may be estimated as
\begin{eqnarray}
	&&\frac{\| {\nabla \times (\gamma \nabla \times {\bf{U}})} \|}
    {\| {\nabla \times ({\bf{U}} \times {\bf{B}})} \|}
\nonumber\\
&&\sim \frac{\langle {{\bf{u}}' \cdot {\bf{b}}'} \rangle}{D ( \partial U / \partial r) B^r}
\frac{\tau_{\rm{turb}}}{\tau_{\rm{mean}}}
\sim \frac{\langle {{\bf{u}}' \cdot {\bf{b}}'} \rangle}{\delta U B^r} Ro^{-1}
= \frac{\langle {{\bf{u}}' \cdot {\bf{b}}'} \rangle}{\delta U B^r}
\frac{K/\varepsilon}{D/\delta U}, 
\end{eqnarray}
where $D$ is the depth of the convection zone, $B^r$ the poloidal magnetic field, $\delta U$ the magnitude of velocity differential rotation, $Ro$ the Rossby number. This suggests that the relative importance of the cross-helicity to the differential rotation can be evaluated by the value of the turbulent cross helicity $\langle {{\bf{u}}' \cdot {\bf{b}}'} \rangle$ and the eddy turn-over time $\tau_{\rm{turb}} (= K / \varepsilon)$, as well as the observable quantities such as the differential rotation velocity $\delta U$, the poloidal magnetic field $B^r$, and the depth of the convection zone $D$. The evaluation of this ratio in several simulation conditions may be interesting subject to explore.

\section{Evolution of the turbulent cross helicity}
In the previous section, we saw that in the presence of cross helicity in turbulence, a mean absolute vorticity (rotation and mean relative vorticity) associated with the non-uniform mean velocity can produce the electromotive force. However, how much cross helicity is present in turbulence is another problem. In this section, we will examine how the turbulent cross helicity is generated by considering the evolution equation of it. 

\subsection{Evolution equation of cross helicity}
As is well known, the total amount of the cross helicity $\int_V {\bf{u}} \cdot {\bf{b}} dV$, as well as the total amount of the MHD energy $\int_V ({\bf{u}}'{}^2 + {\bf{b}}'{}^2)/2 dV$ and the counterpart of the magnetic helicity $\int_V {\bf{a}} \cdot {\bf{b}} dV$, is an inviscid invariant of the incompressible MHD equation, where magnetic field is measured with the Alfv\'{e}n speed unit as ${\bf{b}} = {\bf{b}}_\ast / (\mu_0 \rho)^{1/2}$ ($\mu_0$: magnetic permeability, $\rho$: mass density), ${\bf{a}}$ is the magnetic potential, and $\int_V$ denotes the integral throughout the volume considered, $V$.

We define the local density of the turbulent MHD energy and cross helicity as
\begin{equation}
	K \equiv \langle {{\bf{u}}'{}^2 + {\bf{b}}'{}^2} \rangle/2,
	\label{K_def_sec4}
\end{equation}
\begin{equation}
	W \equiv \langle {{\bf{u}}' \cdot {\bf{b}}'} \rangle,
	\label{eq:W_def_sec4}
\end{equation}
respectively. Due to the conservation property of the total amount of the MHD energy and cross helicity, the evolution equations of the turbulent MHD energy density $K$ and the turbulent cross helicity density $W$ are written in a very simple form as
\begin{equation}
	\frac{DF}{Dt}
	= \left( {{\bf{U}} \cdot \nabla} \right) F
	= P_F
	- \varepsilon_F
	+ T_F,
	\label{eq:K_W_evol_eq}
\end{equation}
where $F = (K \mbox{ or } W)$. In (\ref{eq:K_W_evol_eq}), $P_F$, $\varepsilon_F$, $T_F$ are the production, dissipation, and transport rates of $F$, respectively. The production rate $P_F$ arises from the coupling of the turbulent correlations and the mean-field inhomogeneities. The dissipation rate of $F$, $\varepsilon_F$, comes from the molecular viscosity and magnetic diffusivity or some alternatives such as wave interaction. The transport rate $T_F$ represents fluxes through the boundary. In the case of conservation-related quantities, such fluxes are written in a divergence form.

	The production, dissipation, and transport rates of the turbulent MHD energy $K$ and turbulent cross helicity $W$ are expressed as
\begin{equation}
	P_K
	= - {\cal{R}}^{ij}
		\frac{\partial U^j}{\partial x^i}
	- {\bf{E}}_{\rm{M}} \cdot {\bf{J}},
	\label{eq:P_K_def_sec4}
\end{equation}
\begin{equation}
	\varepsilon_K
= \nu \left\langle {
  \left( {\frac{\partial u'{}^j}{\partial x^i}} \right)^2 
} \right\rangle
+ \eta \left\langle { 
  \left( {\frac{\partial b'{}^j}{\partial x^i}} \right)^2
} \right\rangle
\equiv \varepsilon,
	\label{eq:eps_K_def_sec4}
\end{equation}
\begin{equation}
	T_K
	= {\bf{B}} \cdot \nabla W
	+ \nabla \cdot {\bf{T}}'_K
	\equiv T_K^{(B)}
	+ \nabla \cdot {\bf{T}}'_K,
	\label{eq:T_K_def_sec4}
\end{equation}
\begin{equation}
	P_W
	= - {\cal{R}}^{ij}
	\frac{\partial B^j}{\partial x^i}
	- {\bf{E}}_{\rm{M}} \cdot \mbox{\boldmath$\Omega$},
	\label{eq:P_W_def_sec4}
\end{equation}
\begin{equation}
	\varepsilon_W
	= (\nu + \eta) \left\langle {
		\frac{\partial u'{}^j}{\partial x^i} 
		\frac{\partial b'{}^j}{\partial x^i}
	} \right\rangle,
	\label{eq:eps_W_def_sec4}
\end{equation}
\begin{equation}
	T_W
	= {\bf{B}} \cdot \nabla K
	+ \nabla \cdot {\bf{T}}'_W
	\equiv T_W^{(B)}
	+ \nabla \cdot {\bf{T}}'_W,
	\label{eq:T_W_def_sec4}
\end{equation}
where $\mbox{\boldmath${\cal{R}}$} = \{ {{\cal{R}}^{ij}} \}$ is the Reynolds and turbulent Maxwell stress defined by
\begin{equation}
	{\cal{R}}^{ij}
	= \langle {u'{}^i u'{}^j - b'{}^i b'{}^j} \rangle,
	\label{eq:rey_turb_max_strs_def}
\end{equation}
and ${\bf{E}}_{\rm{M}}$ is the turbulent electromotive force (EMF) defined by (\ref{eq:emf_def_sec3}).

	The production rates $P_K$ and $P_W$ are respectively related to the transfer of the turbulent MHD energy and cross helicity between the large- and small-scales. This point is clearly seen if we write the evolution equations of the mean-field MHD energy ${\cal{K}}$ and the mean-field cross helicity ${\cal{W}}$ defined by
\begin{equation}
	{\cal{K}}
	= ({\bf{U}}^2 + {\bf{B}}^2)/2,
	\label{mean_fld_K_def_sec4}
\end{equation}
\begin{equation}
	{\cal{W}}
	= {\bf{U}} \cdot {\bf{B}},
	\label{mean_fld_W_def_sec4}
\end{equation}
respectively. The equations of ${\cal{K}}$ and ${\cal{W}}$ are written as
\begin{equation}
	\frac{D{\cal{F}}}{Dt}
	\equiv \left( {
		\frac{\partial}{\partial t} 
		+ {\bf{U}} \cdot \nabla
	} \right) {\cal{F}}
	= P_{\cal{F}} 
	- \varepsilon_{\cal{F}} 
	+ T_{\cal{F}},
	\label{eq:mean_K_W_eq_sec4}
\end{equation}
\begin{equation}
	P_{\cal{K}}
	= + {\cal{R}}^{ij} \frac{\partial U^j}{\partial x^i}
	+ {\bf{E}}_{\rm{M}} \cdot {\bf{J}}
	= - P_K,
	\label{eq:mean_fld_P_K_def_sec4}
\end{equation}
\begin{equation}
	\varepsilon_{\cal{K}}
	= \nu \left\langle {
		\left( {\frac{\partial U^j}{\partial x^i}} \right)^2 
	} \right\rangle
	+ \eta \left\langle { 
		\left( {\frac{\partial B^j}{\partial x^i}} \right)^2
	} \right\rangle,
	\label{eq:mean_fld_eps_K_def_sec4}
\end{equation}
\begin{equation}
	T_{\cal{K}}
	= {\bf{B}} \cdot \nabla {\cal{W}}
	+ \nabla \cdot {\bf{T}}'_{\cal{K}}
	\equiv T_{\cal{W}}^{(B)}
	+ \nabla \cdot {\bf{T}}'_{\cal{K}},
	\label{eq:mean_fld_T_K_def_sec4}
\end{equation}
\begin{equation}
	P_{\cal{W}}
	= + {\cal{R}}^{ij} \frac{\partial B^j}{\partial x^i}
	+ {\bf{E}}_{\rm{M}} \cdot \mbox{\boldmath$\Omega$}
	= - P_W,
	\label{eq:mean_fld_P_W_def_sec4}
\end{equation}
\begin{equation}
	\varepsilon_{\cal{W}}
	= (\nu + \eta) \left\langle {
		\frac{\partial U^j}{\partial x^i} 
		\frac{\partial B^j}{\partial x^i}
	} \right\rangle,
	\label{eq:mean_fld_eps_W_def_sec4}
\end{equation}
\begin{equation}
	T_{\cal{W}}
	= {\bf{B}} \cdot \nabla {\cal{K}}
	+ \nabla \cdot {\bf{T}}'_{\cal{W}}
	\equiv T_{\cal{W}}^{(B)}
	+ \nabla \cdot {\bf{T}}'_{\cal{W}}.
	\label{eq:mean_fld_T_W_def_sec4}
\end{equation}
Equations~(\ref{eq:mean_fld_P_K_def_sec4}) and (\ref{eq:mean_fld_P_W_def_sec4}) show that the production rates of the mean-field MHD energy ${\cal{K}}$ and the mean-field cross helicity ${\cal{W}}$ are exactly the same as the counterpart of the turbulent MHD energy $K$ (\ref{eq:P_K_def_sec4}) and the turbulent cross helicity $W$ (\ref{eq:P_W_def_sec4}), but with the opposite signs. As this consequence, these production terms do not contribute to the total amount of that quantity; the sum of the turbulent and mean-field quantities, ${\cal{K}} + K$ and ${\cal{W}} + W$. They contribute just to the transfer of the quantity between the mean and turbulent components. If the sign of the production of a turbulent quantity is positive (or negative), the turbulent quantity increases (or decreases). At the same time, in this case, the sign of the mean-field counterpart is negative (or positive) and the mean-field quantity decreases (or increases). This means that the generation of the turbulent quantity arising from the positive production rate is supplied by the drain or sink of the mean-field counterpart. In this sense, production rates of $K$ and $W$, $P_K$ and $P_W$, represent the cascade properties of the turbulent MHD energy and cross helicity.

The dissipation rates of the mean-field MHD energy and cross helicity, $\varepsilon_{\cal{K}}$ (\ref{eq:mean_fld_eps_K_def_sec4}) and $\varepsilon_{\cal{W}}$ (\ref{eq:mean_fld_eps_W_def_sec4}), are the counterparts of $\varepsilon_K$ (\ref{eq:eps_K_def_sec4}) and $\varepsilon_W$ (\ref{eq:eps_W_def_sec4}). The molecular viscosity $\nu$ and the magnetic diffusivity $\eta$ coupled with the mean-field inhomogeneities $\nabla {\bf{U}}$ and $\nabla {\bf{B}}$ lead to the dissipation. As compared with the inhomogeneities in small scales, $\nabla {\bf{u}}'$ and $\nabla {\bf{b}}'$, these large-scale inhomogeneities are considered to be small. However, in the region where the large-scale field inhomogeneities are fairly large, such as near-wall region and shock vicinity, these dissipation rates associated with the mean-field inhomogeneities can be considerably large and not negligible.

Like the turbulent counterparts $T_K$ and $T_W$, the transport rates of the mean-field quantities, $T_{\cal{K}}$ and $T_{\cal{W}}$ represent the flux through the boundary. The first terms in $T_K$ (\ref{eq:T_K_def_sec4}) and $T_W$ (\ref{eq:T_W_def_sec4}) are originally written in a divergence form as
\begin{equation}
	T_K^{({\rm{B}})}
	= \nabla  \cdot (W {\bf{B}})
	= ({\bf{B}} \cdot \nabla) W,
	\label{eq:T_K_B}
\end{equation}
\begin{equation}
	T_W^{({\rm{B}})}
	= \nabla  \cdot (K {\bf{B}})
	= ({\bf{B}} \cdot \nabla) K.
	\label{eq:T_W_B}
\end{equation}
However, in order to explicitly show that they are related to the turbulence inhomogeneities along the mean magnetic field, they are written as the first terms of (\ref{eq:T_K_def_sec4}) and (\ref{eq:T_W_def_sec4}).

\begin{figure}[htb]
  \centering
  \includegraphics[width= 0.6 \columnwidth]{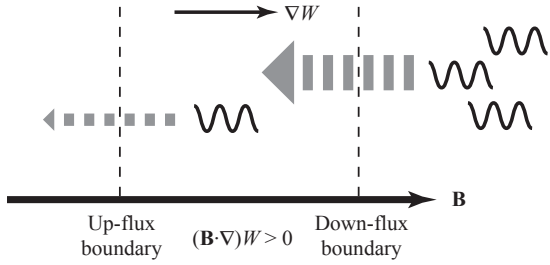}
  \caption{Energy generation due to inhomogeneity of turbulent cross helicity and asymmetry of Alfv\'{e}n wave propagation.}
    \label{fig:en_gen_due_to_cross_helicity}
\end{figure}

The first term of (\ref{eq:T_K_def_sec4}) or (\ref{eq:T_K_B}), $({\bf{B}} \cdot \nabla) W$, is linked to the Alfv\'{e}n wave propagating along a mean magnetic field ${\bf{B}}$. In terms of the Alfv\'{e}n wave, a positive cross helicity $W>0$ represents a dominance of the counter propagating Alfv\'{e}n waves (Alfv\'{e}n wave propagating in the direction antiparallel to the mean magnetic field). Let us consider a domain bounded by two surfaces (Fig.~\ref{fig:en_gen_due_to_cross_helicity}): one is the surface of the upstream direction with respect to the mean magnetic field (up-flux boundary) and the other is the surface of the downstream direction (down-flux boundary). If the magnitude of the turbulent cross helicity increases along the mean magnetic field (down-flux) direction as $({\bf{B}} \cdot \nabla) W > 0$, the counter-propagating Alfv\'{e}n waves are more dominant in the down-flux boundary of ${\bf{B}}$ than the up-flux boundary. This means that the influx Alfv\'{e}n waves at the down-flux boundary are more dominant than the outflux Alfv\'{e}n waves at the up-flux boundary. As this consequence, the total energy influx due to the Alfv\'{e}n waves is positive in the bounded domain, leading to the increase of the turbulent MHD energy. A similar argument applies for the case of negative turbulent cross helicity (dominance of the Alfv\'{e}n waves propagating in the direction parallel to ${\bf{B}}$). These arguments show that the turbulent energy increases in the region where $({\bf{B}} \cdot \nabla) W > 0$ and decreases in the region where $({\bf{B}} \cdot \nabla) W < 0$.

\subsection{Generation mechanisms of the turbulent cross helicity}
The evolution equation of the turbulent cross helicity $W = \langle {{\bf{u}}' \cdot {\bf{b}}'} \rangle$ shows that $W$ can be locally generated in some situations where the turbulent flux and the large-scale structure of the mean field are coupled each other. It follows from (\ref{eq:P_K_def_sec4}) and (\ref{eq:T_K_def_sec4}) that there are two categories in the mechanisms of the turbulent cross-helicity generation. One is originated from the production rate $P_W$ (\ref{eq:P_K_def_sec4}) and the other arises from the transport rate $T_W$ (\ref{eq:T_K_def_sec4}). The production rate $P_W$ is divided into the two parts due to the Reynolds and turbulent Maxwell stress ${\cal{R}}$ and the one due to the turbulent electromotive force ${\bf{E}}_{\rm{M}}$ as
\begin{equation}
	P_W
	= P_W^{({\rm{R}})} + P_W^{({\rm{E}})},
	\label{eq:P_W_R_P_W_E}
\end{equation}
where
\begin{equation}
	P_W^{({\rm{R}})}
	= - {\cal{R}}^{ij}_R \frac{\partial B^j}{\partial x^i},
	\label{eq:P_W_R_def}
\end{equation}
\begin{equation}
	P_W^{({\rm{E}})}
	= - {\bf{E}}_{\rm{M}} \cdot \mbox{\boldmath$\Omega$}.
	\label{eq:P_W_E_def}
\end{equation}

\subsubsection{Cross-helicity generation by velocity and magnetic-field strains}
	If substitute the expression for the Reynolds and turbulent Maxwell stresses into $P_W^{({\rm{R}})}$ (\ref{eq:P_W_R_def}), we have
\begin{equation}
	P_W^{({\rm{R}})}
	= + \nu_{\rm{K}} {\cal{S}}^{ij}{\cal{M}}^{ij}
	- \nu_{\rm{M}} ({\cal{M}}^{ij})^2.
	\label{eq:P_W_R_exp_sec4}
\end{equation}
Since $- \mbox{\boldmath${\cal{M}}$}^2 < 0$, the second or $\nu_{\rm{M}}$-related term always contributes to reduce the magnitude of the turbulent cross helicity. The first or $\nu_{\rm{K}}$-related term may work for increasing the turbulent cross helicity. Since the turbulent or eddy viscosity is always positive ($\nu_{\rm{K}} > 0$), the production of turbulent cross helicity depends on the sign of $\mbox{\boldmath${\cal{S}}$}: \mbox{\boldmath${\cal{M}}$} = \{ {{\cal{S}}^{ij} {\cal{M}}^{ij}} \}$. If the sign of $\mbox{\boldmath${\cal{S}}$}: \mbox{\boldmath${\cal{M}}$} > 0$, a positive cross helicity is generated, while $\mbox{\boldmath${\cal{S}}$}: \mbox{\boldmath${\cal{M}}$} < 0$, a negative turbulent cross helicity is generated as
\begin{equation}
	\left\{ {
	\begin{array}{ll}
		P_W^{({\rm{R}})} > 0\;\; 
		&\mbox{for}\;\; 
		\mbox{\boldmath${\cal{S}}$}: \mbox{\boldmath${\cal{M}}$} > 0,\\
		P_W^{({\rm{R}})} < 0\;\;
		&\mbox{for}\;\;
		\mbox{\boldmath${\cal{S}}$}: \mbox{\boldmath${\cal{M}}$} < 0.
		\rule{0.ex}{5.ex}
	\end{array}
  } \right.
  \label{eq:P_W_R_pos_neg}
\end{equation}

\begin{figure}[htb]
  \centering
  \includegraphics[width= 0.8 \columnwidth]{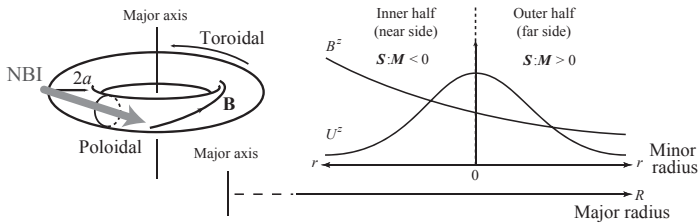}
  \caption{Cross-helicity generation due to the mean velocity and magnetic-field strains.}
    \label{fig:ch_gen_due_to_mag_shear}
\end{figure}

This cross-helicity generation mechanism works when the momentum is injected with a velocity shear to the configuration with a mean magnetic-field shear. The neutral beam injection (NBI) in the toroidal direction in a fusion device may be the case of this turbulence cross-helicity generation. If the neutral beam is externally injected in the central minor axis region, the turbulent cross helicity at the outer half  (far side) may be positive, while the turbulent cross helicity at the inner half region (near side) is negative (Fig.~\ref{fig:ch_gen_due_to_mag_shear}).

\subsubsection{Cross-helicity generation by vorticity and electric-current}
\begin{figure}[htb]
  \centering
  \includegraphics[width= 0.8 \columnwidth]{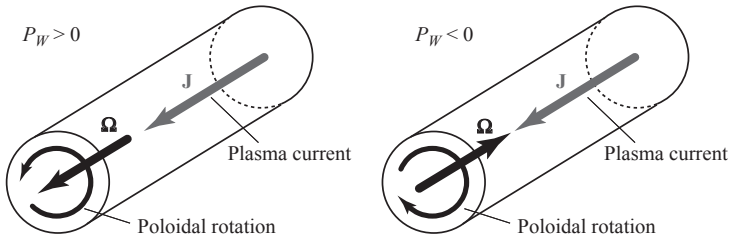}
  \caption{Cross-helicity generation due to the mean vorticity $\mbox{\boldmath$\Omega$}$ (poloidal rotation) and the mean electric-current density ${\bf{J}}$ (toroidal plasma current).}
    \label{fig:ch_gen_due_to_J_Omega}
\end{figure}

	If we substitute the expression for the turbulent EMF (\ref{eq:emf_exp_sec3}) into the production rate $P_W^{({\rm{E}})}$ (\ref{eq:P_W_E_def}), we have
\begin{equation}
	P_W^{({\rm{E}})}
	= + \beta {\bf{J}} \cdot \mbox{\boldmath$\Omega$}
	- \alpha {\bf{B}} \cdot \mbox{\boldmath$\Omega$} 
	- \gamma \mbox{\boldmath$\Omega$}^2.
	\label{eq:P_W_E_exp}
\end{equation}
The third or $\gamma$-related term always work for decreasing the magnitude of the cross helicity, $W$ or $\gamma (= \tau W)$ since $- \mbox{\boldmath$\Omega$}^2 < 0$. The second or $\alpha$-related term suggests the increase or decrease of the turbulent cross helicity due to the magnetic field generated by the $\alpha$ effect. Depending on the sign of the turbulent residual helicity $\alpha (= \tau H)$ coupled with the mean-field structure ${\bf{B}} \cdot \mbox{\boldmath$\Omega$}$, the turbulent cross helicity is generated or destroyed by the $\alpha$ effect. This interaction between the cross helicity and helicity plays an important role in the oscillations and reversal of the magnetic field and the cross helicity. This point will be argued in more detail in the cross-helicity dynamo in oscillatory magnetic field in stars in Section~5. The first or $\beta$-related term is expected to play the central role in the cross-helicity generation since the turbulent magnetic diffusivity $\beta$ exists even without any breakage of mirror-symmetry linked to $\alpha$ and $\gamma$, and represents the primary effect of turbulence. Since the turbulent magnetic diffusivity is always positive ($\beta > 0$), the sign of the cross-helicity generation is determined by the sign of ${\bf{J}} \cdot \mbox{\boldmath$\Omega$}$. 
\NOTE{
In cylindrical plasma configuration, which is often adopted as an approximation of toroidal plasma configuration, the poloidal plasma rotation is represented by the azimuthal or axial mean vorticity $\mbox{\boldmath$\Omega$}$, and the toroidal plasma current is represented by the azimuthal or axial mean electric current ${\bf{J}}$ (Fig.~\ref{fig:ch_gen_due_to_J_Omega}). Whether positive or negative turbulent cross helicity is generated is determined by whether the mean electric-current density and the mean vorticity is positively aligned or negatively aligned. If the mean electric-current density is parallel to the mean vorticity in the sense ${\bf{J}} \cdot \mbox{\boldmath$\Omega$} > 0$, a positive cross helicity is generated $P_W^{({\rm{E}})} > 0$, while they are antiparallel (${\bf{J}} \cdot \mbox{\boldmath$\Omega$} < 0$), a negative cross helicity is generated as
\begin{equation}
	\left\{ {
	\begin{array}{ll}
		P_W^{({\rm{E}})} > 0\;\; 
		&\mbox{for}\;\; 
		{\bf{J}} \cdot \mbox{\boldmath$\Omega$} > 0,\\
		P_W^{({\rm{E}})} < 0\;\;
		&\mbox{for}\;\;
		{\bf{J}} \cdot \mbox{\boldmath$\Omega$} < 0.
		\rule{0.ex}{5.ex}
	\end{array}
	} \right.
	\label{eq:P_W_E_pos_neg}
\end{equation}
This suggests that in the toroidal or cylindrical plasma configuration, a positive or negative turbulent cross helicity is systematically produced by the poloidal rotation coupled with the plasma current.
}

As we see later in Section~7, this production mechanism based on the coupling of the mean vorticity $\mbox{\boldmath$\Omega$}$ with the mean electric-current density ${\bf{J}}$ plays a dominant role in the cross-helicity generation in the turbulent magnetic reconnection.

\subsubsection{Cross-helicity generation by turbulence inhomogeneity along mean magnetic field}
In addition to the production rates related to the coupling of the mean velocity and magnetic-field strains, $P_W^{({\rm{R}})} \sim \mbox{\boldmath${\cal{S}}$} : \mbox{\boldmath${\cal{M}}$}$, and the coupling of the mean vorticity and electric-current density, $P_W^{({\rm{E}})} \sim \mbox{\boldmath$\Omega$} \cdot {\bf{J}}$, the inhomogeneity of turbulence along the mean magnetic field, $({\bf{B}} \cdot \nabla) K$, provides a possibility of positive or negative cross-helicity generation depending on the sign of $({\bf{B}} \cdot \nabla) K$. This generation mechanism of turbulent cross helicity is totally different from the production-rate-related mechanisms given above by $P_W^{({\rm{R}})}$ and $P_W^{({\rm{E}})}$. As we saw in (\ref{eq:mean_fld_P_W_def_sec4}), the production-rate-related mechanisms $P_W^{({\rm{R}})}$ and $P_W^{({\rm{E}})}$ have the corresponding counterparts in the equation of the mean-field cross helicity ${\bf{U}} \cdot {\bf{B}}$ as
\begin{equation}
	P_{\cal{W}}^{({\rm{R}})}
	= + {\cal{R}}^{ij} \frac{\partial B^j}{\partial x^i}
	= - P_{W}^{({\rm{R}})},
	\label{eq:mean_fld_P_W_R_P_W_R}
\end{equation}
\begin{equation}
	P_{\cal{W}}^{({\rm{E}})}
	= + {\bf{E}}_{\rm{M}} \cdot \mbox{\boldmath$\Omega$}
	= - P_W^{({\rm{E}})}.
	\label{eq:mean_fld_P_W_E_P_W_E}
\end{equation}
This means that the turbulent cross helicity generation expressed by $P_W^{({\rm{R}})}$ and $P_W^{({\rm{E}})}$ are originated from the cascade of the mean-field cross helicity ${\bf{U}} \cdot {\bf{B}}$ to the turbulent cross helicity $\langle {{\bf{u}}' \cdot {\bf{b}}'} \rangle$. This is not the case for the turbulent cross-helicity generation $({\bf{B}} \cdot \nabla) K$. In the mean-field cross-helicity equation, we have $({\bf{B}} \cdot \nabla) {\cal{K}} = ({\bf{B}} \cdot \nabla) \langle{{\bf{U}}^2 + {\bf{B}}^2} \rangle/2$ but not in the form of $- ({\bf{B}} \cdot \nabla) K$. This is because this mechanism is not due to the production rate of the cross helicity, but is originated from the transport or flux rate of the cross helicity across the boundaries written as $\nabla \cdot (K{\bf{B}})$ in the turbulent cross helicity equation and $\nabla \cdot ({\cal{K}}{\bf{B}})$ in the mean-field cross helicity equation.

\begin{figure}[htb]
  \centering
  \includegraphics[width= 0.6 \columnwidth]{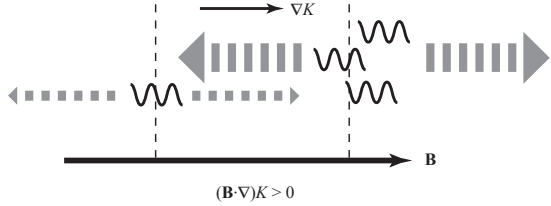}
  \caption{Cross-helicity generation due to inhomogeneity of turbulence and asymmetry of Alfv\'{e}n wave propagation.}
    \label{fig:ch_gen_inhomo_along_B}
\end{figure}

The cross-helicity generation due to the turbulence inhomogeneity along the mean magnetic field, $({\bf{B}} \cdot \nabla) K$, is a part of the transport rate, and is regarded as a flux across the boundaries. This mechanism is directly related to the asymmetry of the Alfv\'{e}n wave propagation along the mean magnetic field ${\bf{B}}$. For simplicity of discussion, we consider the turbulent energy is constituted by an assembly of Alfv\'{e}n waves, which propagate in parallel and antiparallel direction along the mean magnetic field ${\bf{B}}$. The inhomogeneity of turbulence along ${\bf{B}}$ corresponds to the inhomogeneous distribution of the Alfv\'{e}n-wave packets along ${\bf{B}}$. For instance, if the number of the Alfv\'{e}n-wave packets increases along the mean magnetic-field direction, the turbulent energy increases along ${\bf{B}}$, $({\bf{B}} \cdot \nabla) K > 0$ (Fig.~\ref{fig:ch_gen_inhomo_along_B}). In the domain, due to the difference between the parallel and anti-parallel propagations of the Alfv\'{e}n-wave packets, the waves propagating in the anti-parallel to ${\bf{B}}$ is more than the counterparts propagating in the parallel direction. This dominance of the Alfv\'{e}n waves propagating in the anti-parallel direction to the mean magnetic field leads to a positive cross helicity in the domain. In this sense, the part of the transport rate in the cross-helicity evolution equation, $({\bf{B}} \cdot \nabla) K$, represents the generation of the turbulent cross helicity associated with the asymmetric Alfv\'{e}n-wave propagation. This is also the case for the generation of the mean-field cross-helicity ${\bf{U}} \cdot {\bf{B}}$. If the anti-parallel propagating Alfv\'{e}n wave is dominant, the part of the transport rate of the mean-field cross helicity, $({\bf{B}} \cdot \nabla) {\cal{K}}$, becomes positive, and represents a positive cross-helicity generation at large scales. This makes marked difference between the cross-helicity generation mechanisms related to the production rates $P_W^{({\rm{R}})}$ and $P_W^{({\rm{E}})}$.

\begin{figure}[htb]
  \centering
  \includegraphics[width= 0.6 \columnwidth]{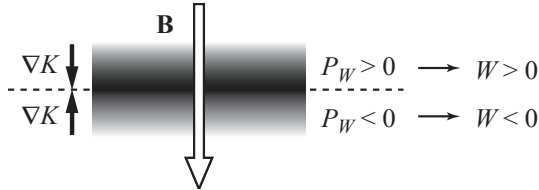}
  \caption{Segregation of turbulent cross helicity due to the cross-helicity generation mechanism due to turbulence inhomogeneity along the global magnetic field $({\bf{B}} \cdot \nabla) K$.}
    \label{fig:ch_segreg_disk}
\end{figure}

This cross-helicity generation mechanism based on the turbulence inhomogeneity along the mean magnetic field is expected to play an essential role in some astrophysical configuration. If the global magnetic field thrust through a plasma gas disk, where the magnitude of turbulence is non-uniform with respect to the depth, we have non-zero $({\bf{B}} \cdot \nabla) K$ (Fig.~\ref{fig:ch_segreg_disk}). If the level of turbulence is higher in the midplane region than those in the peripheral planes (north and south regions of the disk), a distribution of positive and negative turbulent cross helicity in the northern and southern halves of the disk is realized. This mechanism is expected to work for the segregation of turbulent cross helicity in an astrophysical disk.

\subsection{Cross-helicity generation in the compressible MHD}
In the previous subsection (Section~4.2), we listed the cross-helicity generation mechanisms in the incompressible MHD turbulence. They are $P_W^{({\rm{R}})}$ (\ref{eq:P_W_R_def}), $P_W^{({\rm{E}})}$ (\ref{eq:P_W_E_def}), and $({\bf{B}} \cdot \nabla) K$ (\ref{eq:T_K_B}). If we look at these generation mechanisms, we see they require a presence of the mean magnetic field ${\bf{B}}$. We should note that the third term in $P_W^{({\rm{E}})}$ (\ref{eq:P_W_E_exp}), $- \gamma \mbox{\boldmath$\Omega$}^2$, always works for reducing the magnitude of the turbulent cross helicity.

	It is worth noting that, as we saw in Section~3.2.1 and 3.2.2, both the $\alpha$ effect, $\alpha {\bf{B}}$, and the turbulent diffusivity effect, $\beta \nabla \times {\bf{B}}$, need the presence of the mean magnetic field ${\bf{B}}$ and its spatial variation represented by ${\bf{J}} = \nabla \times {\bf{B}}$ for the induction of magnetic fluctuations $\delta {\bf{b}}'$, which constitutes of the turbulent electromotive force through $\langle {{\bf{u}}' \times \delta{\bf{b}}'} \rangle$. However, generation of the transport coefficients $\alpha$ and $\beta$ does not require the presence of the mean magnetic field ${\bf{B}}$. 

	As the generation mechanisms of the turbulent cross helicity, $P_W (= P_W^{({\rm{R}})} + P_W^{({\rm{E}})})$ and $({\bf{B}} \cdot \nabla) K$, show, in the presence of the mean magnetic field ${\bf{B}}$, the turbulent cross helicity is generated. Then, coupled with the mean absolute vorticity (relative vorticity and rotation), the cross helicity effect certainly works for the turbulent electromotive force (EMF). In the sense that the cross-helicity effect can contribute to the counter balancing to the turbulent magnetic diffusivity, the cross-helicity dynamo can be regarded as a battery effect. In order for the cross-helicity effect to be a self-exited dynamo, we need a mechanism that produces the turbulent cross helicity without resorting to the presence of the mean magnetic field ${\bf{B}}$. Relaxing the incompressible constraint may be one of the ways to generate cross helicity even in the absence of the mean magnetic field.

	In order to see the effect of variable density, we leave from the magnetic field measured in the Alfv\'{e}n-speed unit, and return to the one measured in the original natural unit. We define the turbulent cross helicity as
\begin{equation}
	W_\ast
	= \langle {{\bf{u}}' \cdot {\bf{b}}'_\ast} \rangle
	\label{eq:comp_W_def_sec4}
\end{equation}
where ${\bf{b}}_\ast$ is the magnetic field measured with the original unit, while ${\bf{b}}' (= {\bf{b}}'_\ast/(\mu_0 \rho)^{1/2})$ is the magnetic field measured with the Alfv\'{e}n speed unit. The evolution equation of the turbulent cross helicity $W_\ast$ in the compressible MHD is written as
\begin{eqnarray}
	\frac{DW_\ast}{Dt}
	&\equiv& \left( {
		\frac{\partial}{\partial t} + {\bf{U}} \cdot \nabla
	} \right) W_\ast
	\nonumber\\
	&=& - \frac{1}{2} \left\langle {
		u'{}^i u'{}^j
		- \frac{1}{\mu_0 \overline{\rho}} b'_\ast{}^i b'_\ast{}^j
	} \right\rangle
	\left( {\frac{
		\partial B_\ast^j}{\partial x^i}
		+ \frac{\partial B_\ast^i}{\partial x^j}
		} \right)
	- \langle {{\bf{u}}' \times {\bf{b}}'_\ast} \rangle 
	\cdot \mbox{\boldmath$\Omega$}
	\nonumber\\
	&&- (\gamma - 1) \frac{1}{\overline{\rho}}
		\langle {\rho' {\bf{b}}'_\ast} \rangle \cdot \nabla E
	- (\gamma - 1) \frac{1}{\overline{\rho}}
		\langle {e' {\bf{b}}'_\ast} \rangle \cdot \nabla \overline{\rho}
	- \frac{1}{\overline{\rho}}
    	\langle {\rho' {\bf{b}}'_\ast} \rangle 
    	\cdot \frac{D{\bf{U}}}{Dt}
	\nonumber\\
	&& - W_\ast \nabla \cdot {\bf{U}}
	+ {\bf{B}}_\ast \cdot \nabla \left\langle {{\bf{u}}'{}^2/2} \right\rangle
	+ \langle {{\bf{f}}' \cdot {\bf{b}}'_\ast} \rangle
	\nonumber\\
	&& - \varepsilon_{W_\ast}
	+ T'_{W_\ast}
	+ {\rm{R.T.}},
	\label{eq:comp_W_eq}
\end{eqnarray}
where, $E$ is the mean internal energy, $e'$ the internal energy fluctuation, $\overline{\rho}$ the mean density, $\rho'$ the density fluctuation, and ${\bf{f}}'$ the fluctuation part of the external force if any. In (\ref{eq:comp_W_eq}), the first two terms and the seventh term on the right-most side correspond to the production rates, $P_W^{({\rm{R}})}$ (\ref{eq:P_W_R_def}) and $P_W^{({\rm{E}})}$ (\ref{eq:P_W_E_def}), and the transport-rate-related to the inhomogeneity along the mean magnetic field, $({\bf{B}} \cdot \nabla) K$ (\ref{eq:T_W_B}) in the incompressible case. The third to fifth terms are compressibility-originated production rates, which do not directly originated from the mean magnetic field ${\bf{B}}$. They are written in the form of the turbulent fluxes $\langle {\rho' {\bf{b}}'} \rangle$ and $\langle {\rho' {\bf{b}}'} \rangle$ coupled with the mean internal-energy gradient $\nabla E$, the mean density gradient $\nabla \overline{\rho}$, and the mean velocity variation along the fluid motion $D{\bf{U}}/Dt$. These production terms are expected to play an important role in the turbulent cross-helicity generation in strongly compressible cases. The sixth term $- W \nabla \cdot {\bf{U}}$ represents the mean dilatation effect in the turbulent cross-helicity generation, and does not depend on the mean magnetic field. This indicates that a positive turbulent cross helicity is generated in the mean contraction case ($\nabla \cdot {\bf{U}} < 0$) and a negative cross helicity is generated in the mean expansion case ($\nabla \cdot {\bf{U}} > 0$). The eighth term $\langle {{\bf{f}}' \cdot {\bf{b}}'_\ast} \rangle$ represents the turbulent cross-helicity generation by external force. The dissipation term and the transport term other than ${\bf{B}} \cdot \nabla \langle {{\bf{u}}'{}^2} \rangle/2$ is denoted by $\varepsilon_{W_\ast}$ and $T'_{W_\ast}$, respectively, and their detailed expressions are suppressed here. The final ${\rm{R.T.}}$ denotes the residual terms due to the higher-order correlations.

	As the production terms intrinsic to the compressibility, we have $- (\gamma - 1)/\overline{\rho}\langle {\rho' {\bf{b}}'_\ast} \rangle \nabla E$, $- (\gamma - 1)/\overline{\rho}\langle {e' {\bf{b}}'} \rangle \nabla \overline{\rho}$, and $- 1/\overline{\rho}\langle {\rho' {\bf{b}}'} \rangle D{\bf{U}}/Dt$. These production mechanisms indicate that, in the compressible MHD case, a positive or negative cross helicity can be produced by the turbulent fluxes coupled with the mean internal-energy gradient, mean density gradient, etc.\ without resorting to the mean magnetic field.

\subsection{Evaluation of cross-helicity dissipation rate}
The dissipation rate of the cross helicity is defined by (4.18). We can formally derive the equation of the cross helicity from the equations of the fluctuating velocity (\ref{eq:fluct_u_eq}) and magnetic field (\ref{eq:fluct_b_eq}).
\begin{eqnarray}
	&&\frac{D \varepsilon_W}{Dt}
	\equiv \left( {
	\frac{\partial}{\partial t} + {\bf{U}} \cdot \nabla
	} \right) \varepsilon_W 
	\nonumber\\
	&& \hspace{24pt}= (\nu + \lambda) \left\langle {
	\frac{\partial u'{}^a}{\partial x^c}
	\frac{\partial b'{}^b}{\partial x^c}
	- \frac{\partial b'{}^a}{\partial x^c}
	\frac{\partial u'{}^b}{\partial x^c}
	} \right\rangle 
	\frac{\partial U^a}{\partial x^b}
	\nonumber\\
	&& \hspace{24pt} + (\nu + \lambda) \left\langle {
	\frac{\partial u'{}^a}{\partial x^c} b'{}^b
	- u'{}^b \frac{\partial b'{}^a}{\partial x^c}
	} \right\rangle 
	\frac{\partial^2 U^a}{\partial x^b \partial x^c}
	\nonumber\\
	&& \hspace{24pt} + (\nu + \lambda) \left\langle {
	\frac{\partial u'{}^b}{\partial x^a}
	\frac{\partial^2 u'{}^b}{\partial x^a \partial x^c}
	+ \frac{\partial b'{}^b}{\partial x^a}
	\frac{\partial^2 b'{}^b}{\partial x^a \partial x^c}
	} \right\rangle B^c
	\nonumber\\
	&& \hspace{24pt} + (\nu + \lambda) \left\langle {
	\frac{\partial u'{}^c}{\partial x^a}
	\frac{\partial u'{}^c}{\partial x^b}
	- \frac{\partial b'{}^c}{\partial x^a}
	\frac{\partial b'{}^c}{\partial x^b}
	} \right\rangle \frac{\partial B^a}{\partial x^b}
	\nonumber\\
	&& \hspace{24pt} - (\nu + \lambda) \left\langle {
	\frac{\partial u'{}^a}{\partial x^c}
	\frac{\partial u'{}^a}{\partial x^c}
	- \frac{\partial b'{}^a}{\partial x^c}
	\frac{\partial b'{}^a}{\partial x^c}
	} \right\rangle \frac{\partial B^a}{\partial x^b}
	\nonumber\\
	&& \hspace{24pt} - (\nu + \lambda) \left\langle {
	\frac{\partial u'{}^a}{\partial x^c} u'{}^b
	- \frac{\partial b'{}^a}{\partial x^c} b'{}^b
	} \right\rangle 
	\frac{\partial^2 B^a}{\partial x^b \partial x^c}
	\nonumber\allowdisplaybreaks\\
	&& \hspace{24pt} - (\nu + \lambda) \left\langle {
	\frac{\partial u'{}^b}{\partial x^a}
	\frac{\partial u'{}^c}{\partial x^a}
	\frac{\partial b'{}^b}{\partial x^c}
	} \right\rangle
	- (\nu + \lambda) \left\langle {
	\frac{\partial b'{}^b}{\partial x^a}
	\frac{\partial u'{}^c}{\partial x^a}
	\frac{\partial u'{}^b}{\partial x^c}
	} \right\rangle
	\nonumber\\
	&& \hspace{24pt} + (\nu + \lambda) \left\langle {
	\frac{\partial u'{}^b}{\partial x^a}
	\frac{\partial b'{}^c}{\partial x^a}
	\frac{\partial b'{}^b}{\partial x^c}
	} \right\rangle
	+ (\nu + \lambda) \left\langle {
	\frac{\partial b'{}^b}{\partial x^a}
	\frac{\partial b'{}^c}{\partial x^a}
	\frac{\partial b'{}^b}{\partial x^c}
	} \right\rangle
	\nonumber\allowdisplaybreaks\\
	&& \hspace{24pt} - (\nu + \lambda) \left\langle {
		(u'{}^c \pm b'{}^c)
		\frac{\partial}{\partial x^c} 
		\left( {
			\frac{\partial u'{}^b}{\partial x^a}
        	\frac{\partial b'{}^b}{\partial x^a}
		} \right)
	} \right\rangle
	\nonumber\\
	&& \hspace{24pt} + (\nu + \lambda) \frac{\partial}{\partial x^c} 
	\left\langle {
		\frac{1}{2} b'{}^c \frac{\partial}{\partial x^a}
	\left( {u'{}^b \pm b'{}^b} \right)
	} \right\rangle
	\nonumber\allowdisplaybreaks\\
	&& \hspace{24pt} - (\nu + \lambda) \left\langle {
		\frac{\partial b'{}^b}{\partial x^a}
		\frac{\partial^2 p_{\rm{M}}'}{\partial x^a \partial x^b}
	}\right\rangle
	+ (\nu + \lambda) 
	\frac{\partial^2}{\partial x^c \partial x^c}
	\varepsilon_W
	\nonumber\\
	&& \hspace{24pt} - (\nu + \lambda) \frac{\partial}{\partial x^c} \left[{
	\nu \left\langle {
    	\frac{\partial u'{}^b}{\partial x^a}
		\frac{\partial^2 b'{}^b}{\partial x^a \partial x^c}
	}\right\rangle
	+ \lambda \left\langle {
		\frac{\partial b'{}^b}{\partial x^a}
		\frac{\partial^2 u'{}^b}{\partial x^a \partial x^c}
	}\right\rangle
	}\right]
	\nonumber\\
	&& \hspace{24pt} - (\nu + \lambda)^2 \left\langle{
		\frac{\partial^2 u'{}^b}{\partial x^a \partial x^c}
		\frac{\partial^2 b'{}^b}{\partial x^a \partial x^c}
	}\right\rangle.
	\label{eq:epsW_eq_exact}
\end{eqnarray}
This shows how mean-field and its inhomogeneities contribute to the dissipation of the turbulent cross helicity, but the structure of the equation is very complicated as compared with the equation of the turbulent cross helicity (\ref{eq:K_W_evol_eq}) with (\ref{eq:P_W_def_sec4})-(\ref{eq:T_W_def_sec4}). One reason of this complexity arises from the fact that, unlike the case of the total amount of the cross helicity, the total amount of the cross-helicity dissipation rate $\varepsilon_W$ is not at all the conserved quantity.

\subsubsection{Algebraic model of the cross-helicity dissipation rate}
Under the intractable complexity of the exact dissipation equation (\ref{eq:epsW_eq_exact}), one way to evaluate the cross-helicity dissipation rate $\varepsilon_W$ is to express it in an algebraic form as
\begin{equation}
	\varepsilon_W
	= C_W \frac{W}{\tau}
	= C_W \frac{\varepsilon}{K} W,
	\label{eq:epsW_algeb_model_sec4}
\end{equation}
where $\tau$ is the eddy turnover time and $C_W$ is the model constant. As for the eddy turnover time, we adopt the simplest possible model in terms of the turbulent energy $K$ and its dissipation rate $\varepsilon$ as
\begin{equation}
	\tau = \frac{K}{\varepsilon}.
	\label{eq:tau_model_sec4}
\end{equation}
In (\ref{eq:epsW_algeb_model_sec4}), we consider that the dissipation of the turbulent cross helicity $W$ is proportional to $W$ divided by the characteristic timescale of turbulence $\tau$.

	There is some constraint on the cross-helicity dissipation rate and related model constant. First of all, there is a mathematical constraint on the magnitude of $W$. Since $({\bf{u}}' \pm {\bf{b}}')^2 \ge 0$, the magnitude of $W$ is bounded by $K$ as
\begin{equation}
	\frac{\|{W}\|}{K}
	= \frac{\|\langle {{\bf{u}}' \cdot {\bf{b}}'} \rangle\|}
		{\langle {
			{\bf{u}}'{}^2 + {\bf{b}}'{}^2
		} \rangle/2}
	\le 1.
	\label{eq:W_over_K_ineq}
\end{equation}
Due to this relation, we also have a constraint on the magnitude of $\varepsilon_W$. From (\ref{eq:K_W_evol_eq}) for $K$ and $W$, we have
\begin{eqnarray}
	\frac{D}{Dt} \frac{W}{K}
    &=& \frac{W}{K} \left( {
      \frac{1}{W} \frac{DW}{Dt} - \frac{1}{K} \frac{DK}{Dt}
    } \right)
    \nonumber\\
    &=& \frac{W}{K} \left( {\frac{1}{W} P_W - \frac{1}{K} P_K} \right)
	- \frac{W}{K} \left( {
		\frac{1}{W} \varepsilon_W - \frac{1}{K}   \varepsilon
    } \right)
    \nonumber\\
	&&+ \frac{W}{K} \left( {\frac{1}{W} T_W - \frac{1}{K} T_K} \right).
	\label{eq:W_over_K_eq}
\end{eqnarray}
This is a generic equation of the normalized cross helicity $W/K$, and should be satisfied with in any cases. In the case of homogeneous turbulence with no production and transport rates, (\ref{eq:W_over_K_eq}) is reduced to
\begin{equation}
	\frac{\partial}{\partial t} \frac{W}{K}
	= - \left( {
		\frac{1}{W} \varepsilon_W 
    - \frac{1}{K} \varepsilon} \right)
	\frac{W}{K}.
	\label{eq:W_over_K_eq_homo}
\end{equation}
We assume that the parenthesized quantity on the r.h.s.\ does not depend on $W/K$. This is the case if we adopt the simplest algebraic models for the energy and cross-helicity dissipation rates as
\begin{equation}
	\varepsilon = \frac{K}{\tau}
	\label{eq:eps_K_over_tau_sec4}
\end{equation}
and (\ref{eq:epsW_algeb_model_sec4}). In order to avoid the exponential growth of $W/K$ in time, and to satisfy the condition $W/K \le 1$, we have an equality: 
\begin{equation}
	\frac{\|\varepsilon_W\|}{\varepsilon}
	> \frac{\|W\|}{K},
	\label{eq:epsW_over_eps_bound_sec4}
\end{equation}
which is equivalent to the constraint on the model constant $C_W$ in (\ref{eq:epsW_algeb_model_sec4}) as
\begin{equation}
	C_W > 1.
	\label{eq:CW_bound_sec4}
\end{equation}

\subsubsection{Modelling the evolution equation of the cross-helicity dissipation rate}
In homogeneous MHD turbulence, the evolution of the cross helicity had been argued in the context that whether a positive or negative cross helicity becomes dominant depends on which sign of  cross helicity prevails at the initial stage \citep{dob1980a,dob1980b}. However, further investigation showed that the MHD turbulence has much more diverse possibilities including the scaled cross-helicity behavior \citep{tin1986,str1991}.

	In the context of inhomogeneous turbulence with large-scale velocity and magnetic-field shears, the evolution of the cross helicity depends on several large-scale inhomogeneities, and the cross-helicity dissipation rate may be subject to such inhomogeneity effects. Here, we incorporate such inhomogeneity effects through constructing the evolution equation of the turbulent cross-helicity dissipation rate.

	With the aid of the multiple-scale renormalized perturbation expansion theory, we derive an equation of the cross-helicity dissipation rate on the theoretical basis. In the framework of the multiple-scale renormalized perturbation expansion theory, the turbulent cross helicity $W$ is expressed as
\begin{eqnarray}
	\langle {{\bf{u}}' \cdot {\bf{b}}'} \rangle
	&=& \int d{\bf{k}} \langle {
		u'{}^\ell({\bf{k}};\tau) b'{}^\ell({\bf{k}};\tau)
	} \rangle
	\nonumber\\
	&=& \int d{\bf{k}} \left[ {
		\langle {u'_{00}{}^\ell b'_{00}{}^\ell} \rangle
		+ \langle {u'_{00}{}^\ell b'_{01}{}^\ell} \rangle
		+ \langle {u'_{01}{}^\ell b'_{00}{}^\ell} \rangle
		+ \cdots
	} \right.
	\nonumber\\
	&& \hspace{40pt} \left. {
	+ \delta \left( {
		\langle {u'_{10}{}^\ell b'_{00}{}^\ell} \rangle
		+ \langle {u'_{00}{}^\ell b'_{10}{}^\ell} \rangle
		+ \cdots
	} \right)
	+ \cdots
	} \right].
	\label{eq:ub_pert_exp_wave}
\end{eqnarray}
From the analysis, the turbulent cross helicity is expressed as
\begin{equation}
	W
	= 2 I_0 \{ {Q_{ub}} \}
	- I_0 \left\{ {
	G_{\rm{S}}, \frac{D}{Dt} (Q_{ub} + Q_{bu})
	} \right\},
	\label{eq:tsdia_W_exp_sec4}
\end{equation}
where we have adopted the abbreviated forms of the spectral and time integrals defined by
\begin{equation}
	I_n\{ {A} \}
	= \int d{\bf{k}}\ k^{2n} A(k, {\bf{x}}; \tau, \tau, t),
	\label{eq:abbrev_int_InA_sec4}
\end{equation}
\begin{equation}
	I_n\{ {A,B} \}
	= \int d{\bf{k}}\ \int_{-\infty}^\tau\!\!\! d\tau_1\ k^{2n} 
		A(k, {\bf{x}}; \tau, \tau_1, t)
		B(k, {\bf{k}}; \tau, \tau_1, t).
	\label{eq:abbrev_int_InAB_sec4}
\end{equation}
In (\ref{eq:tsdia_W_exp_sec4}), $Q_{ub}$ and $Q_{bu}$ are spectral correlation functions representing the basic or lowest-order field cross helicity, and $G_{\rm{S}}$ is the Green's function.

	We assume that the correlation and Green's functions in the inertial range are expressed as
\begin{equation}
	Q_{ub}(k,{\bf{x}}; \tau, \tau',t)
	= \sigma_W({\bf{k}},{\bf{x}};\tau) 
		\exp [ {- \omega_W}(k,{\bf{x}};t) \|\tau - \tau'\| ],
	\label{eq:Qub_form_assume_sec4}
\end{equation}
\begin{equation}
	G_{{\rm{S}}}(k,{\bf{x}};\tau,\tau',t)
	= \theta(\tau - \tau') 
		\exp [ {- \omega_{\rm{S}}}(k,{\bf{x}};t) (\tau - \tau') ],
	\label{eq:GS_form_assume_sec4}
\end{equation}
respectively. Here $\theta(\tau)$ is the Heaviside's step function [$\theta(\tau) = 1$ for $\tau > 0$ and $\theta = 0$ for $\tau < 0$]. Here $\sigma_W(k,{\bf{x}};\tau)$ is the power spectrum of the cross helicity $W$, and $\omega_W$ and $\omega_{\rm{S}}$ represent the frequencies or timescales of fluctuations. As for the spectrum of $W$ as well as that of the turbulent energy $K$, we assume
\begin{equation}
	\sigma_K 
	= \sigma_{K0} \varepsilon^{2/3} k^{-11/3},
	\label{eq:K_spectrum_sec4}
\end{equation}
\begin{equation}
	\sigma_W 
	= \sigma_{W0} \varepsilon^{-1/3} \varepsilon_W({\bf{x}};t) k^{-11/3},
	\label{eq:W_spectrum_sec4}
\end{equation}
and for the timescales,
\begin{equation}
	\omega_{\rm{S}}(k,{\bf{x}};t)
	= \omega_{{\rm{S}}0} \varepsilon^{1/3} k^{2/3} 
	= \tau_{\rm{S}}^{-1},
	\label{eq:K_timescale_sec4}
\end{equation}
\begin{equation}
	\omega_{W}(k,{\bf{x}};t)
	= \omega_{W0} \varepsilon_W^{1/3} k^{2/3} 
	= \tau_{W}^{-1},
	\label{eq:W_timescale_sec4}
\end{equation}
where $\omega_{{\rm{S}}0}$, $\omega_{K0}$, and $\omega_{W0}$ are the numerical constants. Expression~(\ref{eq:W_spectrum_sec4}) is based on the assumption that the spectrum of the cross helicity depends on the wavenumber $k$, the cross-helicity transfer/dissipation rate $\varepsilon_W$ and the energy transfer/dissipation rate $\varepsilon$. Alternatively, (\ref{eq:K_spectrum_sec4}) and (\ref{eq:W_spectrum_sec4}) are based on the assumption that the ratio of the energy spectrum to the energy transfer/dissipation rate, $\sigma_K / \varepsilon$, and that of the cross-helicity spectrum to the cross-helicity transfer/dissipation rate, $\sigma_W / \varepsilon_W$, have the same dependence on the energy transfer/dissipation rate $\varepsilon$ and wavenumber $k$ as
\begin{equation}
	\frac{\sigma_K}{\varepsilon}
	= \frac{\sigma_W}{\varepsilon_W}
	= \varepsilon^{-1/3} k^{-11/3}.
	\label{eq:transfer_over_diss_sec4}
\end{equation}
Considering $\varepsilon^{-1/3} k^{-2/3}$ gives a timescale, (\ref{eq:transfer_over_diss_sec4}) corresponds to assuming the energy $K (= \int \sigma_K d{\bf{k}})$ divided by $\varepsilon$ and $W (= \int \sigma_W d{\bf{k}})$ divided by $\sigma_W$ give the same timescale $\tau \sim \varepsilon^{-1/3} k^{-2/3}$.

	Using the correlation and Green's functions (\ref{eq:Qub_form_assume_sec4}) and (\ref{eq:GS_form_assume_sec4}), $W$ is calculated as
\begin{eqnarray}
	W
	&=& 2 \int d{\bf{k}}\ 
	\varepsilon^{-1/3}({\bf{x}};t) 
	\varepsilon({\bf{x}};t) k^{-11/3}
	\nonumber\\
	&&- 2 \int d{\bf{k}}\ \left[ {
		\frac{1}{\omega_{\rm{S}} + \omega_W}
		\frac{D\sigma_W}{Dt}
		-  \frac{\sigma_W}{(\omega_{\rm{S}} + \omega_W)^2}
		\frac{D\omega_W}{Dt}
	} \right].
	\label{eq:W_in_terms_of_Q_G_sec4}
\end{eqnarray}
From the inertial-range form (\ref{eq:K_spectrum_sec4})-(\ref{eq:W_timescale_sec4}), this is rewritten as
\begin{eqnarray}
	W 
	&=& 4 \cdot 2\pi \varepsilon^{-1/3} \varepsilon_W 
		\int_{\|{\bf{k}}\| \ge k_{\rm{C}}} \!\!\!\!\!\!dk\ 
		k^{-5/3}
	\nonumber\\
	&-& 4 \cdot 2 \pi \frac{\sigma_{W0}}{\omega_{\rm{sw}}
		\varepsilon_{\rm{sw}}^{1/3}} 
		\int_{\|{\bf{k}}\|\ge k_{\rm{C}}}\!\!\!\!\!\! dk\ 
		k^{4/3} \frac{D}{Dt} \left[ {
			\varepsilon^{-1/3}({\bf{x}};t) 
			\varepsilon_W({\bf{x}};t) 
		k^{-11/3}
		} \right]
	\nonumber\\
	&+& 4 \cdot 2 \pi \frac{\sigma_{W0} \omega_{W0}}
		{(\omega_{\rm{sw}}\varepsilon_{\rm{sw}}^{1/3})^2}
		\varepsilon^{-1/3} \varepsilon_W
		\int_{\|{\bf{k}}\|\ge k_{\rm{C}}}\!\!\!\!\!\! dk\ 
		k^{-3} \frac{D}{Dt} \left[ {
			\varepsilon_W^{1/3}({\bf{x}};t) k^{2/3}
		} \right],
	\label{eq:W_exp_inertial_form_sec4}
\end{eqnarray}
where $k_{\rm{C}}$ is the cut-off wavenumber. Here, we have introduced a synthesized timescale $\tau_{\rm{SW}}$, defined by
\begin{equation}
	\frac{1}{\tau_{\rm{SW}}}
	= \frac{1}{\tau_{\rm{S}}} + \frac{1}{\tau_{W}}
	=  \left( {\omega_{\rm{S}0} \varepsilon^{1/3}
	+ \omega_{W0} \varepsilon_{W}^{1/3}} \right) k^{2/3}
	\equiv \omega_{\rm{sw}} 
        \varepsilon_{\rm{sw}}^{1/3} k^{2/3}.
    \label{eq:synthe_timescale_sec4}
\end{equation}
It is worth while to note that $\omega_{\rm{SW}}$ and $\varepsilon_{\rm{SW}}$ appear only in the combination of $\omega_{\rm{SW}} \varepsilon_{\rm{SW}}^{1/3}$, which gives a timescale. Each of $\omega_{\rm{SW}}$ and $\varepsilon_{\rm{SW}}$ does not have a definite meaning. With this point in mind, hereafter we denote
\begin{equation}
	A_W(\omega_{\rm{S}0}, \omega_{W0})
	\equiv \frac{\omega_{W0} \varepsilon_W^{1/3}} 
	{\omega_{\rm{sw}} \varepsilon_{\rm{sw}}^{1/3}}
	= \frac{\omega_{W0} \varepsilon_W^{1/3}} 
	{\omega_{\rm{S}0} \varepsilon^{1/3}
	+ \omega_{W0} \varepsilon_W^{1/3}}
	= \frac{\tau_{\rm{SW}}}{\tau_{W}}
	\label{eq:Aw_def_sec4}
\end{equation}
for simplicity of notation.

\NOTE{
The cut-off wavenumber $k_{\rm{C}}$ introduced in (\ref{eq:W_exp_inertial_form_sec4}) is the lower bounder of the spectral integral. This wavenumber is directly related to the largest eddy size of turbulent motions, $\ell_{\rm{C}}$, as
\begin{equation}
	\ell_{\rm{C}} = 2\pi / k_{\rm{C}}.
	\label{eq:ell_C_def_sec4}
\end{equation}
This length scale $\ell_{\rm{C}}$ is considered to dominantly contribute to and determines the turbulent MHD energy and cross helicity. Here, we should note the following point. The correlation length of the cross helicity may be different from that of the MHD energy. The characteristic lengths of turbulence represent the scales of the largest turbulent motion corresponding to the scale with the largest magnitudes of spectra of MHD energy and cross helicity. Length scales difference between the MHD energy and cross helicity arise from a specific mechanism that produces considerably different spectral forms for the energy and cross helicity. This is not the case, for instance, for the solar-wind turbulence. In this sense, adopting the same correlation lengths for both the energy and cross helicity, as $\ell_K \sim \ell_W (\equiv \ell_{\rm{C}})$, is not unrealistic assumption. Related to this point, extended discussions of correlation length scales and timescales by \citet{mat1994,hos1995} are important.
}

We calculate the Lagrange derivatives and the spectral integrals in (\ref{eq:W_exp_inertial_form_sec4}). If we denote the scaled wavenumber as
\begin{equation}
	s = k/k_{\rm{C}},
	\label{eq:scaled_wave_num_sec4}
\end{equation}
we obtain
\begin{eqnarray}
	W 
	&=& 4 \cdot (2\pi)^{1/3} \sigma_{W0} 
		\varepsilon^{-1/3} \varepsilon_W k_{\rm{C}}^{2/3} 
		\int_{s \ge 1} \!\!\!ds\ s^{-5/3}
	\nonumber\\
	&-& 4 \cdot 2 \pi \frac{\sigma_{W0}}
		{\omega_{\rm{sw}}\varepsilon_{\rm{sw}}^{1/3}} 
		k_{\rm{C}}^{7/3}
		\int_{s\ge 1}\!\!\! ds\ 
		s^{-7/3} \frac{D}{Dt} \left[ {
		\varepsilon^{-1/3}({\bf{x}};t) 
		\varepsilon_W({\bf{x}};t) 
		k_{\rm{C}}^{-11/3}
	} \right]
	\nonumber\\
	&+& 4 \cdot 2 \pi \frac{\sigma_{W0} \omega_{W0}}
		{(\omega_{\rm{sw}} \varepsilon_{\rm{sw}}^{1/3})^2}
		\varepsilon^{-1/3} \varepsilon_W k_{\rm{C}}^{-2}
		\int_{s \ge 1}\!\!\! ds\ s^{-7/3} 
		\frac{D}{Dt} \left[ {
		\varepsilon_W^{1/3}({\bf{x}};t) k_{\rm{C}}^{2/3}
	} \right].
	\label{eq:W_exp_in_terms_of_kC_sec4}
\end{eqnarray}
In terms of $\ell_{\rm{C}}$ (\ref{eq:ell_C_def_sec4}), this can be rewritten as
\begin{eqnarray}
	W
	&=& 6 \cdot (2\pi)^{1/3} \sigma_{W0}
		\varepsilon^{-1/3} \varepsilon_W 
		\ell_{\rm{C}}^{2/3}
	\nonumber\\
	&+& \frac{1}{(2 \pi)^{1/3}} 
		\frac{\sigma_{W0}}{\omega_{\rm{sw}} 
		\varepsilon_{\rm{sw}}^{1/3}}
		\varepsilon^{-1/3} \varepsilon_W 
		\ell_{\rm{C}}^{4/3}
	\left\{ {
		\frac{1}{\varepsilon} \frac{D\varepsilon}{Dt}
		- \left[ {
			3 - A_W(\omega_{\rm{S}0}, \omega_{W0})
		} \right] \frac{1}{\varepsilon_W} 
		\frac{D\varepsilon_W}{Dt}
	} \right.
	\nonumber\\
	&& \left. {
	\hspace{40pt} + \left[ {11 
			- 2 A_W(\omega_{\rm{S}0}, \omega_{W0})
		} \right] \frac{1}{\ell_{\rm{C}}} 
		\frac{D\ell_{\rm{C}}}{Dt}
	} \right\}.
	\label{eq:W_exp_in_terms_of_ellC_sec4}
\end{eqnarray}
As we see in (\ref{eq:K_spectrum_sec4}), the spectrum of turbulent energy $K$ is assumed to be expressed in terms of the dissipation rate $\varepsilon$ and the wavenumber $k$. In the hydrodynamic turbulence modeling, we can construct a closed system of model equations from a combination of any two variables among $K_u (= \langle {{\bf{u}}'{}^2} \rangle/2)$, its dissipation rate $\varepsilon_u$, and $\ell_{\rm{C}} (= 2\pi/k_{\rm{C}})$. In order to keep the transferability of the model among these variables, $K_u$, $\varepsilon_u$, and $\ell_{\rm{C}}$ should be linked with each other by an algebraic relation. Otherwise we cannot expect any transferability among the model variables. This is also the case for the MHD turbulence modeling. If we retain the first term in (\ref{eq:W_exp_in_terms_of_ellC_sec4}), we get an algebraic expression for the turbulent cross helicity as
\begin{equation}
	W
	= 6 \cdot (2\pi)^{1/3} \sigma_{W0}
		\varepsilon^{-1/3} \varepsilon_W 
		\ell_{\rm{C}}^{2/3},
	\label{eq:W_exp_algeb_sec4}
\end{equation}
or equivalently,
\begin{equation}
	\ell_{\rm{C}} 
	= 6^{-3/2} (2\pi)^{-1/2} \sigma_{W0}^{-3/2}
	\varepsilon^{1/2} \varepsilon_W^{-3/2} W^{3/2}.
	\label{eq:ellC_exp_algeb_sec4}
\end{equation}
Equation~(\ref{eq:W_exp_algeb_sec4}), corresponding to the spectral expression (\ref{eq:W_spectrum_sec4}), assures an algebraic relation among $W$, $\varepsilon_W$, $\varepsilon$, and $\ell_{\rm{C}} (= 2\pi/k_{\rm{C}})$.

	Using (\ref{eq:ellC_exp_algeb_sec4}), we transfer the expression (\ref{eq:W_exp_in_terms_of_ellC_sec4}) expressed in terms of $\varepsilon$, $\varepsilon_W$, and $\ell_{\rm{C}}$ into the expression expressed in terms of $W$, $\varepsilon_W$, $\varepsilon$. As a result, we have
\begin{equation}
	\frac{D\varepsilon_W}{Dt}
	= C_{1} (\omega_{\rm{S}0}, \omega_{W0})
		\frac{\varepsilon_W}{\varepsilon} \frac{D\varepsilon}{Dt}
	+ C_{2}(\omega_{\rm{S}0}, \omega_{W0})
		\frac{\varepsilon_W}{W} \frac{DW}{Dt}
	\label{eq:epsW_eq_form_sec4}
\end{equation}
with
\begin{equation}
	C_{1}(\omega_{\rm{S}0}, \omega_{W0})
	= \frac{13 - A_W(\omega_{\rm{S}0}, \omega_{W0})}
	{39 - 8 A_W(\omega_{\rm{S}0}, \omega_{W0})},\;\;
	C_{2}(\omega_{\rm{S}0}, \omega_{W0})
	= \frac{33 - 6 A_W(\omega_{\rm{S}0}, \omega_{W0})}
	{39 - 8 A_W(\omega_{\rm{S}0}, \omega_{W0})}.
	\label{eq:C1_C2_exps_sec4}
\end{equation}
If the timescale associated with the cross helicity is similar to the one with the Green's function as
\begin{equation}
	\omega_{\rm{S}0} \varepsilon^{1/3} 
	\simeq \omega_{W0} \varepsilon_W^{1/3},
	\label{eq:simlar_omegaS_omegaW}
\end{equation}
$A_W(\omega_{{\rm{S}}0},\omega_{W0})$ is estimated as
\begin{equation}
	A_W(\omega_{\rm{S}0}, \omega_{W0}) \simeq 1/2.
	\label{eq:Aw_approx_sec4}
\end{equation}
Then, we have model coefficients $C_1$ and $C_2$ (4.62) as
\begin{equation}
	C_{1}(\omega_{\rm{S}0}, \omega_{W0}) \simeq \frac{12}{35} \simeq 0.34,\;\;
	C_{2}(\omega_{\rm{S}0}, \omega_{W0}) \simeq \frac{6}{7} \simeq 0.86.
	\label{eq:C1_C2_eval_sec4}
\end{equation}
We finally obtain the evolution equation of the cross-helicity dissipation ($\varepsilon_W$ equation) as
\begin{equation}
	\frac{D\varepsilon_W}{Dt}
	= C_{\varepsilon_W 1} \frac{\varepsilon_W}{K} P_K
	- C_{\varepsilon_W 2} \frac{\varepsilon_W}{K} \varepsilon
	+ C_{\varepsilon_W 3} \frac{\varepsilon_W}{W} P_W
	- C_{\varepsilon_W 4} \frac{\varepsilon_W}{W} \varepsilon_W
	\label{eq:epsW_model_eq_sec4}
\end{equation}
with the model constants
\begin{eqnarray}
	&&C_{\varepsilon_W 1} = C_1 C_{\varepsilon 1}
	= 0.34 \times 1.4 = 0.48,
	\nonumber\\
	&&C_{\varepsilon_W 2} = C_1 C_{\varepsilon 2}
	= 0.34 \times 1.9 = 0.65,
	\nonumber\\
	&&C_{\varepsilon_W 3} = C_{\varepsilon_W 4} = C_2
	= 0.86.
	\label{eq:CepsW1_CepsW2_CepsW3_eval_sec4}
\end{eqnarray}
Here we have used
\begin{equation}
	C_{\varepsilon 1} = 1.4,\;\;
	C_{\varepsilon 2} = 1.9
	\label{eq:Ceps1_Ceps2_val}
\end{equation}
from the standard $\varepsilon$ equation.

	In the present formulation, the evolution equation of the cross-helicity dissipation ($\varepsilon_W$ equation) is derived from the theoretical analysis of the velocity and magnetic-field fluctuations. Reflecting the dependence of the cross-helicity spectrum both on the $\varepsilon$ and $\varepsilon_W$ (\ref{eq:GS_form_assume_sec4}), the $\varepsilon_W$ equation depends on the $W$ equation and $\varepsilon$ equation as (\ref{eq:W_exp_algeb_sec4}) and (\ref{eq:Aw_approx_sec4}).

	From the algebraic relation (\ref{eq:W_exp_algeb_sec4}), we write the cross-helicity dissipation rate $\varepsilon_W$ as
\begin{equation}
	\varepsilon_W
	= 6^{-1} \cdot (2\pi)^{-1/3} \sigma_{W0}^{-1}
	\frac{W}{\tau}
	\label{eq:epsW_algeb_exp_sec4}
\end{equation}
with
\begin{equation}
	\tau
	= \varepsilon^{-1/3} \ell_{\rm{C}}^{2/3}.
	\label{eq:tau_exp_sec4}
\end{equation}
Equation~(\ref{eq:epsW_algeb_exp_sec4}) corresponds to the simplest algebraic model of $\varepsilon_W$ (\ref{eq:epsW_algeb_model_sec4}). This means that the simple algebraic model can be regarded as the lowest-order evaluation of the turbulent cross helicity $W$ in the framework of the multiple-scale renormalized perturbation analysis (\ref{eq:tsdia_W_exp_sec4}).

\section{Cross-helicity effect in stellar dynamos}

\subsection{Oscillatory and migratory dynamo}
\begin{figure}[htb]
  \centering
  \includegraphics[width= 0.7 \columnwidth]{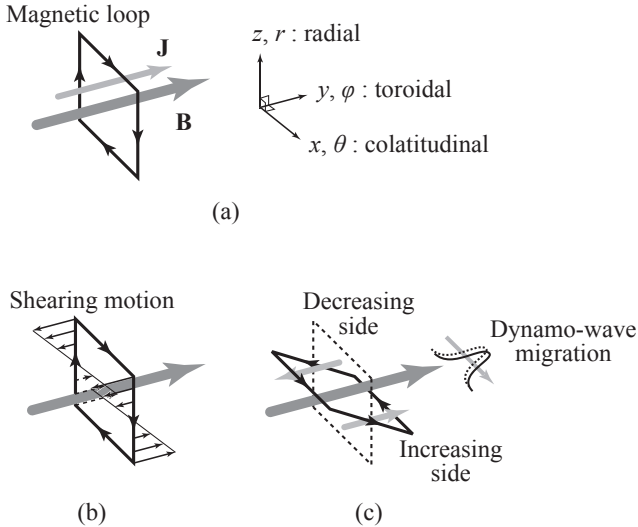}
  \caption{Schematic view of dynamo-wave migration. (a) A magnetic-field loop is generated by the action of the $\alpha$ effect in a local Cartesian coordinate system $(x,y,z)$. The case with the positive $\alpha$ is depicted in this figure. (b) Tilting of the magnetic loop due to a shearing motion. (c) Migration of the magnetic field.}
    \label{fig:dynamo_wave_migration}
\end{figure}

Combination of the helicity or $\alpha$ effect and the differential rotation or $\Omega$ effect has been investigated in the context of the polarity reversal of the solar magnetic fields. One of the representative oscillatory dynamo models is the Parker equations with the $\alpha - \Omega$ dynamo \citep{par1955}. \citet{par1955} considered a dynamo mechanism constituted of non-uniform rotation and small-scale cyclonic or tornado-like fluid motions. The non-uniform rotation generates the toroidal magnetic field from the poloidal one (the so-called $\Omega$ effect). With the aid of the cyclonic fluid motions, the poloidal magnetic field is regenerated from the toroidal one in the form of a loop magnetic field ($\alpha$ effect). This mechanism is called the $\alpha$ or helicity dynamo, of which elaborated works have been done \citep{mof1978,kra1980,mof2019}. The equation of the mean magnetic field is written as
\begin{equation}
	\frac{\partial {\bf{B}}}{\partial t}
	= \nabla \times \left( {{\bf{U}} \times {\bf{B}}} \right)
	+ \nabla \times \left( {
		- \beta \nabla \times {\bf{B}}
		+ \alpha {\bf{B}}
	} \right).
	\label{eq:mean_B_eq_sec5}
\end{equation}
In the Parker model, the $\alpha$ effect produces a poloidal (latitudinal and radial) magnetic-field component from a toroidal (azimuthal) magnetic-field component through twisting turbulent motions (helicity), and a toroidal magnetic-field component is produced by the differential rotation as shown in Fig.~\ref{fig:dynamo_wave_migration}. We consider a local Cartesian coordinate $(x,y,z)$, where $x$, $y$, and $z$ correspond to the colatitudinal $\theta$, the azimuthal $\phi$, and the radial $r$ directions, respectively. We assume that the system is homogeneous in the azimuthal direction $\partial / \partial \phi \simeq \partial / \partial y = 0$, and the mean velocity is only in the azimuthal or toroidal direction as
\begin{equation}
	{\bf{U}}
	= (U^x, U^y, U^z)
	= (0, U, 0)
	\label{uni_azimuthal_vel_assump_sec5}
\end{equation}
with the velocity shear in the co-latitudinal and radial directions as $\partial U/\partial x \simeq (1/r) \partial U /\partial \theta$ and $\partial U / \partial z \simeq \partial U / \partial r$. The magnetic field is decomposed into the toroidal (azimuthal) and poloidal (radial and colatitudinal) components as
\begin{eqnarray}
	{\bf{B}}
	&=& {\bf{B}}_{\rm{tor}} + {\bf{B}}_{\rm{pol}}
	\nonumber\\
	&=& (0, B^y, 0) 
	+ \nabla \times \left( {0, A^y, 0} \right)
	\nonumber\\
	&=& (0, B^y, 0) 
	+ \left( {
	- \frac{\partial A^y}{\partial z}, 
	0, 
	\frac{\partial A^y}{\partial x}} \right).
	\label{eq:torB_polB_decomp}
\end{eqnarray}
Hereafter we drop suffix $y$ for the brevity of notation. The evolution equation of the azimuthal magnetic vector potential $A^\phi (\equiv A)$, which represents the poloidal magnetic field, is solved simultaneously with the azimuthal component of the magnetic field $B^\theta (\equiv B)$, which represents the toroidal or azimuthal magnetic field.

	In this Cartesian coordinate system, the simplest Parker equations can be written as
\begin{equation}
	\frac{\partial A}{\partial t}
	=  \beta \left( {
		\frac{\partial^2 A}{\partial z^2}
		+ \beta \frac{\partial^2 A}{\partial x^2}
	} \right)
	+ \alpha B,
	\label{eq:parker_pol_fld_eq_sec5}
\end{equation}
\begin{equation}
	\frac{\partial B}{\partial t}
	= \beta \left( {
		\frac{\partial^2 B}{\partial z^2}
		+ \frac{\partial^2 B}{\partial x^2}
	} \right)
	- \frac{\partial U}{\partial z} 
	\frac{\partial A}{\partial x}
	- \frac{\partial U}{\partial x} 
	\frac{\partial A}{\partial z}.
	\label{eq:parker_tor_fld_eq_sec5}
\end{equation}
Here, for the sake of simplicity, the helicity or $\alpha$ effect, $\alpha {\bf{B}}$, is included only in the poloidal-field equation (\ref{eq:parker_pol_fld_eq_sec5}). This is because we assume that the $\Omega$ or differential-rotation effect, $\nabla \times ({\bf{U}} \times {\bf{B}}) = ({\bf{B}} \cdot \nabla) {\bf{U}}$, is the sole generation mechanism of the toroidal-field equation (\ref{eq:parker_tor_fld_eq_sec5}). 

This simplest system of equations may give an oscillatory poloidal and toroidal magnetic field. In the reality of the solar or stellar dynamo, the transport coefficients $\alpha$ and $\beta$ depend on the statistical properties of turbulence, and how much differential rotation $G \equiv \partial U/\partial z$ is present in the stellar system should be determined by the nonlinear interaction between the mean fields and turbulence. However, in order to understand the basic behavior of the system of equations (\ref{eq:parker_pol_fld_eq_sec5}) and (\ref{eq:parker_tor_fld_eq_sec5}), it would be meaningful to examine the behavior of $A$ and $B$ under regarding $\alpha$, $\beta$ and $G$ as parameters. Suitably chosen parameters $\alpha$, $\beta$, and $G$ provide an oscillatory magnetic field mimicking the solar polar reversal activity. The behavior depends on the values of these parameters. For instance, it was shown through the examination of a cyclonic dynamo-wave behavior that the dynamo wave migrates from the high-latitude toward the low-latitude regions if the radial gradient of the angular velocity multiplied by the $\alpha$ coefficient is negative. In light of the migration of the sunspot in the equator-ward direction with time, this indicates that the angular velocity in the solar convection zone should satisfy
\begin{equation}
	\alpha \frac{d\Omega}{dr} < 0\;\;
	\left( {\alpha \frac{d\Omega}{dz} < 0} \right)
	\label{parker_criterion_migration_sec5}
\end{equation}
(Parker's criterion for the migratory dynamo).

	In the framework of the $\alpha - \Omega$ model, the physics of this criterion can be understood as follows. If the $\alpha$ coefficient is positive (which corresponds to a negative kinetic helicity) in the northern hemisphere, a magnetic loop is generated by the $\alpha$ effect so that the electric current density ${\bf{J}}$ associated with the magnetic loop may be aligned parallel (not antiparallel) to the toroidal magnetic field ${\bf{B}}$ [Fig.~\ref{fig:dynamo_wave_migration}(a)]. The rotation rate decreases in the radially outward direction ($d\Omega/dr < 0$. Because of this velocity shear in the $r$ direction, the shallower (larger $r$) part of a magnetic loop moves faster in the toroidal or negative $\phi$ direction as compared with the deeper (smaller $r$) part, leading to a tilting of the magnetic loop [Fig.~\ref{fig:dynamo_wave_migration}(b)]. As this consequence, the toroidal or $\phi$ component of the magnetic field decreases in the smaller $\theta$ region while the one in the larger $\theta$ region increases [Fig.~\ref{fig:dynamo_wave_migration}(c)]. Consequently, the pattern of the magnetic field migrates from the high-latitude (small $\theta$) toward low-latitude (large $\theta$) regions.

\subsection{Oscillatory magnetic field with the cross-helicity effect}
As we see in (\ref{eq:emf_expression}) of Section~2.4 and (\ref{eq:u_times_b_gamma_Omega}) of Section~3.2.3, in the presence of the mean vortical or rotational motion, a finite turbulent cross helicity ($W \ne 0$) can induce the electromotive force (EMF) in the direction of the mean absolute vorticity. If we include this cross-helicity effect into the EMF expression, the mean magnetic field equation is written as
\begin{equation}
	\frac{\partial {\bf{B}}}{\partial t}
	= \nabla \times 
	\left( {{\bf{U}} \times {\bf{B}}} \right)
	+ \nabla \times \left( {
		- \beta \nabla \times {\bf{B}}
		+ \alpha {\bf{B}}
		+ \gamma \mbox{\boldmath$\Omega$}
	} \right).
	\label{eq:mean_B_eq_w_gamma_sec5}
\end{equation}
where the $\gamma$-related term represents the cross-helicity effect in the EMF. The turbulent cross helicity $W$ itself is expected to oscillate and change its sign with the mean magnetic field. As this consequence, we have to consider the evolution equation of the cross helicity. Following the evolution equation of the turbulent cross helicity, we model the evolution equation for the transport coefficient $\gamma$ as
\begin{eqnarray}
	\frac{\partial \gamma}{\partial t}
	&=&  \beta \nabla^2 \gamma
	- \tau {\bf{E}}_{\rm{M}} \cdot \mbox{\boldmath$\Omega$}
	\nonumber\\
	&=& \beta \nabla^2 \gamma
	- \alpha \tau {\bf{B}} \cdot \mbox{\boldmath$\Omega$}
	+ \beta \tau (\nabla \times {\bf{B}}) \cdot \mbox{\boldmath$\Omega$}
	- \gamma \tau \mbox{\boldmath$\Omega$}^2.
	\label{eq:simple_dynamo_model_w_gamma_sec5}
\end{eqnarray}
Here the first term or $\beta \nabla^2 \gamma$ arises from the diffusion term $\nabla \cdot (\nu_{\rm{T}}/\sigma_W) \nabla W$ in the cross-helicity transport rate. The fourth or $\gamma$-related term always works for reducing the magnitude of $\gamma$ since $- \tau \mbox{\boldmath$\Omega$}^2 < 0$. The third or $\beta$-related term contributes to generation of $\gamma$ through the coupling of the mean electric current with the mean vorticity. 

The second or $\alpha$-related term $- \alpha \tau {\bf{B}} \cdot \mbox{\boldmath$\Omega$}$ plays a very important role in the oscillation of the cross helicity. This term gives us the possibility of the cross helicity generation once the poloidal magnetic field  is generated from the toroidal magnetic field through the $\alpha$ effect. Since this term plays a key role in the oscillation of the cross helicity, we argue the role of this term further in the context of the solar dynamo.

	We assume that the dominant dynamo is cross-helicity effect, and the $\alpha$ effect is perturbation to the reference state. We write the mean magnetic field and the mean electric-current density as
\begin{subequations}\label{eq:pert_alpha_flds}
\begin{equation}
	{\bf{B}} = {\bf{B}}_0 + {\bf{B}}_1,
	\label{eq:pert_B}
\end{equation}
\begin{equation}
	{\bf{J}} = {\bf{J}}_0 + {\bf{J}}_1,
	\label{eq:pert_J}
\end{equation}
\end{subequations}
where ${\bf{B}}_0$ and ${\bf{J}}_0$ are the reference state due to the cross-helicity effect, and ${\bf{B}}_1$ and ${\bf{J}}_1$ are the perturbation or modulation fields due to the $\alpha$ effect. Substituting (\ref{eq:pert_alpha_flds}) into the mean induction equation, we have
\begin{equation}
	\frac{\partial {\bf{B}}_0}{\partial t}
	= \nabla \times ({\bf{U}} \times {\bf{B}}_0 
		- \beta {\bf{J}}_0 
		+ \gamma \mbox{\boldmath$\Omega$}),
	\label{eq:pert_alpha_B0_eq}
\end{equation}
\begin{equation}
	\frac{\partial {\bf{B}}_1}{\partial t}
	= \nabla \times ({\bf{U} \times {\bf{B}}_1}
		+ \alpha {\bf{B}}_0
		- \beta {\bf{J}}_1).
	\label{eq:pert_alpha_B1_eq}
\end{equation}
The reference-field equation (\ref{eq:pert_alpha_B0_eq}) has a special solution for the stationary state as
\begin{equation}
	{\bf{B}}_0 = \frac{\gamma}{\beta} {\bf{U}}.
	\label{eq:pert_alpha_B0_sol}
\end{equation}
Substituting (\ref{eq:pert_alpha_B0_sol}) into (\ref{eq:pert_alpha_B1_eq}), we have the modulation-field equation as
\begin{equation}
	\frac{\partial {\bf{B}}_1}{\partial t}
	= \nabla \times \left({
			{\bf{U} \times {\bf{B}}_1}
			+ \frac{\alpha\gamma}{\beta}{\bf{U}}
			- \beta {\bf{J}}_1
		} \right).
	\label{eq:pert_alpha_B1_eq}
\end{equation}
Here, we approximate the mean velocity ${\bf{U}}$ in the polar coordinate system$(r, \theta, \phi)$ by the toroidal velocity as ${\bf{U}} = (U^r, U^\theta, U^\phi) \simeq (0, 0, U^\phi)$. We further assume the axisymmetry of ${\bf{U}}$ and ${\bf{B}}$. In the low latitude region, the radial magnetic field is assumed to be small ($B^r \simeq 0$) and the latitudinal gradient of the toroidal velocity is also small ($\partial U^\phi / \partial \theta \simeq 0$). Under these approximations, we have
\begin{equation}
	\nabla \times ({\bf{U}} \times {\bf{B}}_1)
	\simeq \left( {
		0, 
		0, 
		B_1^r \frac{\partial U^\phi}{\partial r}
		+ B_1^\theta \frac{1}{r} \frac{\partial U^\phi}{\partial \theta}
	} \right)
	\simeq \left( {0, 0, 0} \right).
	\label{eq:vanishing_U_times_B1}
\end{equation}
In this case, the modulation-field equation~(\ref{eq:pert_alpha_B1_eq}) has an approximate special solution for the stationary state as
\begin{equation}
	{\bf{J}}_1 
	= \frac{\alpha}{\beta} {\bf{B}}_0
	= \frac{\alpha\gamma}{\beta^2} {\bf{U}}.
	\label{eq:J1_special_sol}
\end{equation}
This field corresponds to the modulation poloidal field ${\bf{B}}_1$ generated from the toroidal field ${\bf{B}}_0$ by the $\alpha$ effect.

\begin{figure}[t]
  \centering
  \includegraphics[width= 1.0 \columnwidth]{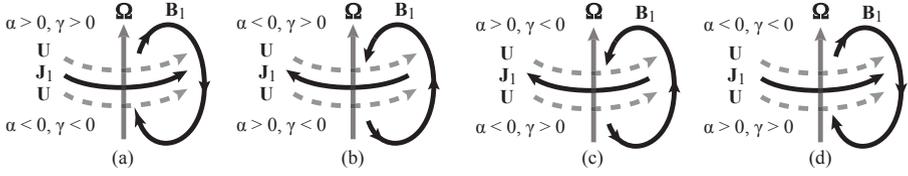}
  \caption{Signs of the cross-helicity production rate $P_W^{(\alpha)}$ due to the $\alpha$ effect. (a) ${\bf{B}}_1 \cdot \mbox{\boldmath$\Omega$} > 0$ for $\alpha > 0$ and $\gamma > 0$ for the northern hemisphere (same for the other cases), (b) ${\bf{B}}_1 \cdot \mbox{\boldmath$\Omega$} < 0$ for $\alpha < 0$ and $\gamma > 0$, (c)  ${\bf{B}}_1 \cdot \mbox{\boldmath$\Omega$} < 0$ for $\alpha > 0$ and $\gamma < 0$, (d) ${\bf{B}}_1 \cdot \mbox{\boldmath$\Omega$} > 0$ for $\alpha < 0$ and $\gamma < 0$.}
    \label{fig:alpha_effect_on_gamma}
\end{figure}

	We first assume that the turbulent cross helicity is positive ($\gamma > 0$) in the northern hemisphere. In this case, from (\ref{eq:J1_special_sol}), the direction of the modulation electric-current density ${\bf{J}}_1$ is determined by the sign of $\alpha$. They are parallel to the mean velocity ${\bf{U}}$ for $\alpha > 0$, and antiparallel to ${\bf{U}}$ for $\alpha < 0$. As this consequence, the modulation magnetic field ${\bf{B}}_1$ is parallel to $\mbox{\boldmath$\Omega$}$  (${\bf{B}}_1 \cdot \mbox{\boldmath$\Omega$} > 0$) for $\alpha > 0$ [Fig.~\ref{fig:alpha_effect_on_gamma} (a)] and antiparallel to $\mbox{\boldmath$\Omega$}$ (${\bf{B}}_1 \cdot \mbox{\boldmath$\Omega$} < 0$) for $\alpha < 0$ [Fig.~\ref{fig:alpha_effect_on_gamma} (b)]. This leads to invariantly negative $P_W^{(\alpha)} = - \alpha {\bf{B}}_1 \cdot \mbox{\boldmath$\Omega$}$ irrespective of the sign of $\alpha$ as 
\begin{equation}
	P_W^{(\alpha)}
	= - \alpha {\bf{B}}_1 \cdot \mbox{\boldmath$\Omega$} < 0\;\;\;
	\mbox{for}\;\;\; \alpha \substack{>\\ <} 0, \gamma > 0.
	\label{eq:PW_alpha_neg_for_gamma_pos}
\end{equation}
	On the other hand, if the cross helicity is negative ($\gamma < 0$) in the northern hemisphere,  we have $ P_W^{\alpha} = - \alpha {\bf{B}}_1 \cdot \mbox{\boldmath$\Omega$} > 0$ irrespective of the sign of $\alpha$ as
\begin{equation}
	P_W^{(\alpha)}
	= - \alpha {\bf{B}}_1 \cdot \mbox{\boldmath$\Omega$} > 0\;\;\;
	\mbox{for}\;\;\; \alpha \substack{>\\ <} 0, \gamma < 0
	\label{eq:PW_alpha_pos_for_gamma_neg}
\end{equation}
[Fig.~\ref{fig:alpha_effect_on_gamma} (c) and (d)].

From (\ref{eq:PW_alpha_neg_for_gamma_pos}) and (\ref{eq:PW_alpha_pos_for_gamma_neg}). the production term $P_W^{(\alpha)} = - \alpha {\bf{B}}_1 \cdot \mbox{\boldmath$\Omega$}$ always generates the turbulent cross helicity whose sign is opposite to the original sign of the turbulent cross helicity. This may leads to the oscillation of cross helicity.
	
	With this consideration, we construct a set of evolution equations constituted by the mean magnetic-field equation and the cross helicity equation.  We see from (\ref{eq:mean_B_eq_w_gamma_sec5}) and (\ref{eq:simple_dynamo_model_w_gamma_sec5}) that the simplified dynamo equations are written in the form
\begin{equation}
	\frac{\partial A}{\partial t}
	= \beta \left( {
		\frac{\partial^2 A}{\partial x^2}
		+ \frac{\partial^2 A}{\partial z^2}
	} \right)
	+ \alpha B,
	\label{eq:pol_fld_eq_w_gamma}
\end{equation}
\begin{eqnarray}
	\frac{\partial B}{\partial t}
	&=& \beta \left( {
		\frac{\partial^2 B}{\partial x^2}
		+ \frac{\partial^2 B}{\partial z^2}
	} \right)
	- \frac{\partial U}{\partial z} \gamma
	- \frac{\partial U}{\partial x} \gamma
	- \frac{\partial^2 U}{\partial z^2} 
	\frac{\partial \gamma}{\partial z}
	\nonumber\\
	&&- \frac{\partial U}{\partial z}
		\frac{\partial \gamma}{\partial z}
	- \frac{\partial U}{\partial x} 
		\frac{\partial \gamma}{\partial x}
	- \frac{\partial U}{\partial z} 
		\frac{\partial A}{\partial x}
	- \frac{\partial U}{\partial x} 
		\frac{\partial A}{\partial z},
	\label{eq:tor_fld_eq_w_gamma}
\end{eqnarray}
\begin{eqnarray}
	\frac{\partial \gamma}{\partial t}
	&=& \beta \left( {
		\frac{\partial^2 \gamma}{\partial x^2}
		+ \frac{\partial^2 \gamma}{\partial z^2}
	} \right)
	- \alpha \tau \left( {
		\frac{\partial U}{\partial x} 
		\frac{\partial A}{\partial x}
	- \frac{\partial U}{\partial z} 
		\frac{\partial A}{\partial z}
	} \right)
	\nonumber\\
	&&+ \beta \tau \left( {
		\frac{\partial U}{\partial x}
		\frac{\partial B}{\partial x}
		- \frac{\partial U}{\partial z} 
		\frac{\partial B}{\partial z}
	} \right)
	- \gamma \tau \left[ {
		\left( {\frac{\partial U}{\partial x}} \right)^2 
		+ \left( {\frac{\partial U}{\partial z} } \right)^2
	}\right].
	\label{eq:gamma_eq_w_gamma}
\end{eqnarray}	
As in (\ref{eq:parker_tor_fld_eq_sec5}), we dropped the $\alpha$ effect in the toroidal field equation (\ref{eq:tor_fld_eq_w_gamma}). Note that this system of equations is reduced to the standard Parker equations (\ref{eq:parker_pol_fld_eq_sec5}) and (\ref{eq:parker_tor_fld_eq_sec5}) if the cross-helicity effect vanishes ($\gamma = 0$).

	We further assume some symmetries with respect to the equator ($\theta = \pi/2, x=0$) for the mean flow speed $U$, $\alpha$, and the turbulent magnetic diffusivity $\beta$: the mean velocity speed $U$ is symmetric with respect to the equator, the helicity $\alpha$ is antisymmetric and vanishes at $\theta = \pi/2, x=0$, and the turbulent magnetic diffusivity $\beta$ is spatially uniform. As for the radial or $z$ dependence of $A$, $B$, and $\gamma$, we assume the form of $\exp(ikz)$, and for the mean velocity, $\partial U/\partial z = k_u U$. Then the system of equations (\ref{eq:pol_fld_eq_w_gamma})-(\ref{eq:tor_fld_eq_w_gamma}) is reduced to a much simplified form, and we perform an eigenvalue analysis for the normal mode of the simplified system of equations.

\begin{figure}[htb]
  \centering
  \includegraphics[width= 0.6 \columnwidth]{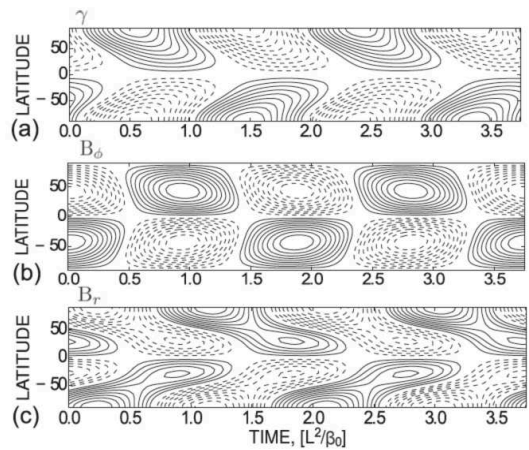}
  \caption{Butterfly diagram (spatiotemporal evolution of the cross helicity and magnetic field) of the simplified dynamo model with the cross-helicity effect incorporated. The contours of (a) the turbulent cross helicity represented by $\gamma$ [contour range of -0.5, 0.5], (b) toroidal magnetic field [-0.8, 0.8], and (c) poloidal magnetic field [-0.2, 0.2]. The solid and dashed contours denote the positive and negative values, respectively. Redrawn from \citet{yok2016b}.}
    \label{fig:ch_solar_dynamo_butterfly}
\end{figure}

	The spatial distribution of (a) the cross helicity $\gamma$, (b) the toroidal magnetic field $B^\phi$, and (c) the poloidal magnetic field $B^r$ are plotted against time in Fig.~\ref{fig:ch_solar_dynamo_butterfly}. The plots of (b) the toroidal magnetic field $B^\phi$ and (c) the poloidal magnetic field $B^r$ show that magnetic fields are generated first at the higher latitude region, then migrate to the lower latitude region as time goes by. Comparing three contours, we clearly see the causal relation among $\gamma$, $B^\phi$, and $B^r$. First a positive cross helicity is generated, then a positive toroidal magnetic field starts to be generated due to the cross-helicity effect. Then, due to the $\alpha$ effect, a positive poloidal field starts to be generated. Once a positive poloidal magnetic field is generated, the magnitude of the positive cross helicity starts decreasing, and finally a negative cross helicity shows up. This negative cross helicity starts generating a negative toroidal magnetic field, which is the reversal of the toroidal magnetic field. A polarity reversal cycle proceeds as this.
	
\subsection{Cross-helicity effect in cool stars}
	In the framework of the standard $\alpha - \Omega$ dynamo, whether we can obtain a toroidal magnetic field and its polarity reversal observed in butterfly diagram depend on the particular configuration of the differential rotation is required. In the absence of differential rotation, no toroidal field can be effectively generated in this framework. However, it is known that the differential rotation of some fully convective stars is very small. Observations show that the surface differential rotation of the fast rotating M-dwarf, with a strong magnetic field of 10 kG strength, is very small \citep{don2008,mor2008}. The direct numerical simulations (DNSs) show that the differential rotation is strongly quenched in the fully convective stars with the generated magnetic field \citep{bro2008}. In the absence of the differential rotation, a very strong toroidal magnetic field observed in the fully convective stars cannot be explained by the $\alpha$--$\Omega$ dynamo. As we saw in Section~2 for the cross-helicity effect, the essential requisites for the dynamo effect are the rotation (absolute vorticity) and the cross helicity in turbulence. So, this effect is expected to work even in stars with a solid-body rotation. With this expectation, the role of cross-helicity dynamo was investigated in the context of fully convective stars \citep{pip2018}.

\begin{figure}[htb]
  \centering
  \includegraphics[width= 0.7 \columnwidth]{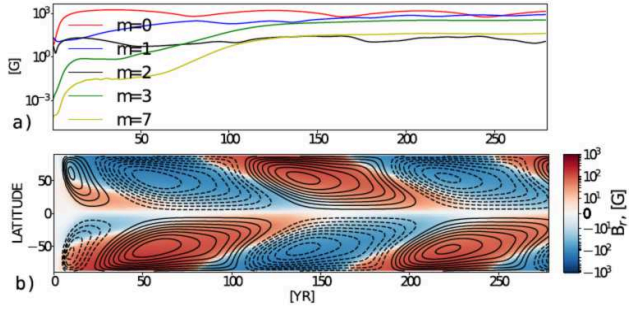}
  \caption{Cross-helicity dynamo model applied to a cool star. Magnetic field growth of several toroidal modes (upper) and the butterfly diagram (lower) of the toroidal magnetic field (contour) and poloidal magnetic field $B^r$ (color). Redrawn from \citet{pip2018}.}
    \label{fig:cool_star_butterfly}
\end{figure}

\begin{figure}[htb]
  \centering
  \includegraphics[width= 0.7 \columnwidth]{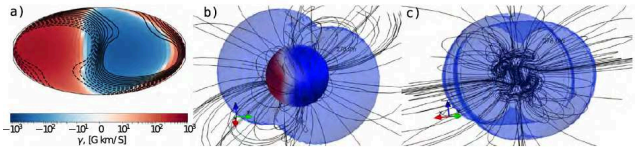}
  \caption{Spatial distribution of the turbulent cross helicity $\gamma$ in color and radial magnetic field in contour (left), strength of the toroidal magnetic field for axisymmetric and non-axisymmetric toroidal magnetic field (middle), and the axisymmetric toroidal magnetic field (right). Redrawn from \citet{pip2018}}
    \label{fig:cool_star_ch_mag_config}
\end{figure}

	The eigenvalue solutions are analyzed for the system of simplified equations on the basis of the equation of the mean magnetic field (\ref{eq:mean_B_eq_w_gamma_sec5}) and that of the turbulent cross helicity (\ref{eq:simple_dynamo_model_w_gamma_sec5}) with the turbulent EMF implemented with the cross-helicity effect. The spherical harmonics decomposition is employed for the toroidal and poloidal magnetic fields and for the turbulent cross helicity. The generated magnetic-field configuration depends on the magnetic-field generation mechanisms. In the absence of the $\Omega$ or differential rotation effect, the helicity and cross-helicity effects, and combination of them are the magnetic-field generation mechanisms. The results of the eigenvalue analysis for the $\alpha^2 \gamma^2$ model are shown in Figs.~\ref{fig:cool_star_butterfly} and \ref{fig:cool_star_ch_mag_config}. The time evolutions of a few modes ($m=0$: axisymmetric, otherwise: non-axisymmetric) are shown in Fig.~\ref{fig:cool_star_butterfly} (a). In the $\alpha^2 \gamma^2$ model, the axisymmetric mode ($m=0$) can be dominant or comparable with the non-axisymmetric modes. This makes strong contrast with the $\alpha^2$ and $\gamma^2$ dynamos, where the non-axisymmetric toroidal magnetic field is more preferable. The time--latitude (butterfly) diagram is shown in Fig.~\ref{fig:cool_star_butterfly} (b), where poloidal (radial) field at the surface is shown in color and the toroidal field at the radius of $r = 0.75 R$ ($R$: the surface radius) is shown in contour. The time--latitude evolution of the axisymmetric toroidal field is similar to the solar one. Figure~\ref{fig:cool_star_ch_mag_config} shows snapshots of the radial magnetic-field and cross-helicity distributions in the stationary phase of dynamo. Figure~\ref{fig:cool_star_ch_mag_config} (a) shows the cross-helicity distribution at the surface in color, and the radial magnetic-field distribution at the surface in contour (range of $\pm$ 1 kG). This figure shows the dominance of the non-axisymmetric magnetic field and the dominance of the non-axisymmetric cross-helicity distribution. The patterns of the axisymmetric and non-axisymmetric magnetic field are shown in Fig.~\ref{fig:cool_star_ch_mag_config} (b) and (c).
	
	These results indicate that the axisymmetric magnetic field, which is not preferred either in the pure $\alpha^2$ or pure $\gamma^2$ dynamo model in the rapid solid-body rotation case, can be generated in the $\alpha^2-\gamma^2$ dynamo model. The cross helicity produced by the non-axisymmetric magnetic field contributes to the generation of the axisymmetric toroidal magnetic field. This finding provides us with the possibility that the axisymmetric dipole-like magnetic field at strengths of several kG can be generated by the cross-helicity effect. This is a new dynamo scenario that can be applicable to the generation of an axisymmetric magnetic field in solid-body rotation cool stars.

\section{Flow generation by cross helicity}
\NOTE{
So far, we have discussed the cross-helicity effect in the context of the magnetic-field induction. The cross helicity is the correlation between the velocity and magnetic field. As the cross-helicity dynamo discussed in Sections~3 and 5 represents the induction of the magnetic fields by flows, flows may be induced by magnetic fields in the presence of the turbulent cross helicity. In this section, the cross-helicity effect in the linear and angular momentum transport is disccussed.
}

\subsection{Reynolds and turbulent Maxwell stresses}
The Reynolds and turbulent Maxwell stresses represent the turbulent momentum flux in the mean velocity equation. As the expression of the Reynolds and turbulent Maxwell stresses (\ref{eq:rey_turb_max_str_exp}) shows, in addition to the eddy-viscosity effect represented by $\nu_{\rm{K}} \mbox{\boldmath${\cal{S}}$} (= \nu_{\rm{K}} \{ {{\cal{S}}^{ij}} \})$, we have the contribution from $\nu_{\rm{M}} \mbox{\boldmath${\cal{M}}$} (= \nu_{\rm{M}} \{ {{\cal{M}}^{ij}} \})$ and $\mbox{\boldmath$\Gamma$} \mbox{\boldmath$\Omega$}_\ast (= \{ {\Gamma^i \Omega^j} \})$:
\begin{equation}
	\langle {u'{}^i u'{}^j - b'{}^i b'{}^j} \rangle
	=  + \frac{2}{3} K_{\rm{R}} \delta^{ij}
	- \nu_{\rm{K}} {\cal{S}}^{ij}
	+ \nu_{\rm{M}} {\cal{M}}^{ij}
	+ [\Gamma^i \Omega_\ast^j + \Gamma^j \Omega_\ast^i]_{\rm{D}},
	\label{eq:rey_turb_max_strs_exp_sec6}
\end{equation}
\NOTE{where the first term on the r.h.s.\ is the contribution from the diagonal part of ${\mbox{\boldmath${\cal{R}}$}}$ and is expressed by the turbulent MHD residual energy $K_{\rm{R}} [\equiv (\langle {{\bf{u}}'{}^2} \rangle - \langle {{\bf{b}}'{}^2} \rangle)/2]$. The importance of the turbulent cross helicity in the context of the turbulent MHD residual energy $K_{\rm{R}}$ is discussed in \citet{hei2023}.} 

In (\ref{eq:rey_turb_max_strs_exp_sec6}), the eddy viscosity $\nu_{\rm{K}}$ is linked to the turbulent energy $K$. On the other hand, the transport coefficients $\nu_{\rm{M}}$ and $\mbox{\boldmath$\Gamma$}$ are linked to the turbulent cross helicity $W$ and the gradient of the turbulent helicity $\nabla H$, respectively. We see from (\ref{eq:rey_turb_max_strs_exp_sec6}) or (\ref{eq:rey_turb_max_str_exp}) that the turbulent cross helicity $W (= \langle {{\bf{u}}' \cdot {\bf{b}}'} \rangle)$ coupled with the mean magnetic-field strain $\mbox{\boldmath${\cal{M}}$}$ and the inhomogeneous turbulent kinetic helicity,  $\nabla H (= \nabla \langle {{\bf{u}}' \cdot \mbox{\boldmath$\omega$}'} \rangle)$ coupled with the mean absolute vorticity $\mbox{\boldmath$\Omega$}_\ast$ contribute to the linear and angular momentum transport by counter-balancing the eddy viscosity.

	The inhomogeneous helicity $\nabla H$ effect in the momentum-transport suppression and large-scale flow generation have been theoretically and numerically investigated in hydrodynamic turbulence \citep{yok1993,yok2016a,yok2023}. Firstly, the inhomogeneous helicity effect was applied to a turbulent swirling pipe flow. Turbulent swirling flow is an axial pipe flow accompanied by a circumferential flow around the pipe axis. Without the circumferential flow, because of the strong momentum transport by turbulence, the mean axial flow in pipe flow shows a very flat profile in the most part of the flow except for the near wall boundary layer. This is marked contrast with the parabolic axial flow profile in a laminar pipe flow. In the presence of circumferential flow, the mean axial velocity profile shows a dent near the central axis region. In this sense, in the turbulent swirling flow, an inhomogeneous mean flow structure is sustained even in very strong turbulence. The configuration of turbulent swirling flow is simple, but such a simple flow configuration cannot be reproduced by a standard turbulence model with the eddy-viscosity representation. Main reason of this deficiency is attributed to the overestimate of the turbulent viscosity effect. This dent profile of the mean axial velocity in turbulent swirling flow has been successfully reproduced by the eddy-viscosity model supplemented by the inhomogeneous helicity ($\nabla H$) effect. The suppression of turbulent momentum transport by the inhomogeneous helicity effect was validated with the aid of turbulence model simulation.

	More recent validation of the inhomogeneous helicity effect in hydrodynamic turbulence has been performed with the aid of direct numerical simulation (DNS). There, global flow induction in helical turbulence in a uniformly rotating triple periodic box was investigated. In addition to the rotation, an inhomogeneous turbulent helicity is externally injected during the whole period of simulation by forcing. Starting with no mean-flow initial condition, the DNSs show that a global mean flow in the rotation direction is induced by the coupling of the inhomogeneous turbulent helicity and rotation. It is also shown that at the early development stage, where the mean velocity strain is absent, the spatial distribution of the Reynolds stress is in good agreement with the that of the inhomogeneous turbulent helicity coupled with the rotation. At the developed stage, where the mean velocity reaches its stationary state, the balancing between the eddy-viscosity term and the inhomogeneous-helicity term is observed.

These turbulent model simulation and DNSs clearly show that the eddy-viscosity representation of the Reynolds stress is not at all enough, and the inhomogeneous turbulent helicity effect should be included in the modeling of the turbulent flows lacking mirror-symmetry.

\subsection{Cross-helicity effect in momentum transport}
In the MHD turbulence, the cross-helicity effect coupled with the mean magnetic-field strain, $\nu_{\rm{M}} \mbox{\boldmath${\cal{M}}$} = \nu_{\rm{M}} \{ {{\cal{M}}^{ij}} \}$, enters the Reynolds and turbulent Maxwell stresses expression (\ref{eq:rey_turb_max_strs_exp_sec6}). The eddy-viscosity term $\nu_{\rm{K}} \mbox{\boldmath${\cal{S}}$} = \nu_{\rm{K}} \{ {\cal{S}}^{ij} \}$ always contributes to destroying large-scale inhomogeneous structures or enhancing turbulent mixing by increasing the effective viscosity. In contrast to the eddy viscosity, the cross-helicity-related effect $\nu_{\rm{M}} \mbox{\boldmath${\cal{M}}$}$ may contribute to forming large-scale flow structures or suppressing turbulent mixing against the eddy viscosity.

Since the Reynolds and turbulent Maxwell stresses are second-order tensors, it is not necessarily straightforward to consider their physical origin and consequence. Here, these properties are considered in terms of a vector called the vortexmotive force or pondero-motive force $\langle {{\bf{u}}' \times \mbox{\boldmath$\omega$}'} \rangle$, which appears in the mean vortex equation as
\begin{equation}
	\frac{\partial \mbox{\boldmath$\Omega$}}{\partial t}
	= \cdots
	+ \nabla \times \langle {{\bf{u}}' \times \mbox{\boldmath$\omega$}'} \rangle
	+ \cdots.
	\label{eq:mean_vorticity_eq_sec6}
\end{equation}
By uncurling this, we may guess that the mean velocity ${\bf{U}}$ is subject to the vortexmotive force as
\begin{equation}
	\frac{\partial {\bf{U}}}{\partial t}
	= \cdots + \langle {{\bf{u}}' \times \mbox{\boldmath$\omega$}'} \rangle 
	+ \cdots.
	\label{eq:mean_vel_eq_sec6}
\end{equation}
although the actual ${\bf{U}}$ should be determined by the boundary conditions with an undetermined potential function associated with the uncurling. Note that the vortexmotive force $\langle {{\bf{u}}' \times \mbox{\boldmath$\omega$}'} \rangle$ and the Reynolds stress $\mbox{\boldmath${\cal{R}}$}_{\rm{K}} = \{ {{\cal{R}}_{\rm{K}}^{ij}} \} = \{ {\langle {u'{}^i u'{}^j} \rangle} \}$ are linked with each other by the exact relation
\begin{equation}
	\langle {
		{\bf{u}}' \times \mbox{\boldmath$\omega$}'
	} \rangle^i
	= - \frac{\partial {\cal{R}}^{ij}}{\partial x^j}
	+ \frac{\partial K}{\partial x^i}.
	\label{eq:vmf_rey_rel_sec6}
\end{equation}

\begin{figure}[htb]
  \centering
  \includegraphics[width= 0.7 \columnwidth]{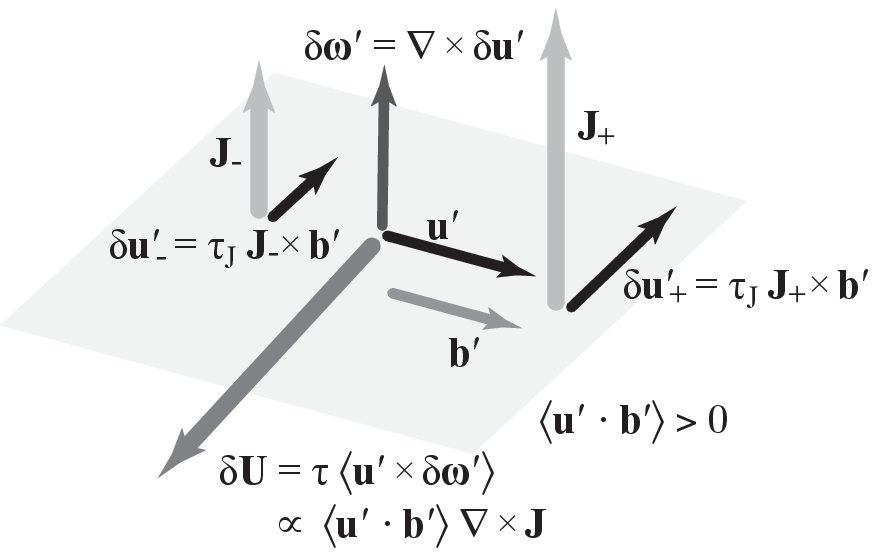}
  \caption{Global flow generation by turbulent cross helicity through the fluctuating Lorentz force ${\bf{J}} \times {\bf{b}}'$.}
    \label{fig:ch_momentum_trans}
\end{figure}

Let us consider a fluid element fluctuating in the plane perpendicular to the mean electric-current density ${\bf{J}}$ (Fig.~\ref{fig:ch_momentum_trans}). The magnitude of the mean electric-current density itself is non-uniformly distributed. We assume a case where the velocity and magnetic-field fluctuations are positively correlated with each other (positive cross helicity: $\langle {{\bf{u}}' \cdot {\bf{b}}'} \rangle > 0$).

Due to a part of the fluctuating Lorentz force, ${\bf{J}} \times {\bf{b}}'$, a velocity fluctuation $\delta {\bf{u}}'$ is induced as
\begin{equation}
	\delta {\textbf{u}}' 
	= \tau_J {\textbf{J}} \times {\textbf{b}}',
	\label{delu_tau_J_times_b_sec6}
\end{equation}
where $\tau_J$ is the timescale associated with the motion induced by the fluctuating Lorentz force.
Associated with this velocity fluctuation variation $\delta {\bf{u}}'$, the fluctuating vorticity $\delta \mbox{\boldmath$\omega$}'$ is given as
\begin{eqnarray}
	\delta \mbox{\boldmath$\omega$}'
	&=& \nabla \times \delta {\textbf{u}}'
	\nonumber\\
	&=& \tau_J \nabla \times ({\textbf{J}} \times {\textbf{b}}')
	\simeq  \tau_J ({\textbf{b}}' \cdot \nabla) {\textbf{J}}.
	\label{eq:delomega_tau_b_grad_J_sec6}
\end{eqnarray}
This indicates that the fluctuating vorticity is induced if the magnetic fluctuation feels an inhomogeneous mean electric-current density. The direction of the fluctuating vorticity is in the direction of the mean electric current ${\bf{J}}$, either parallel or antiparallel, depending on the mean electric-current density distribution (inhomogeneity of ${\bf{J}}$).

	In the combination of the induced velocity and vorticity fluctuations, a large-scale flow is induced by the vortexmotive force $\langle {{\bf{u}}' \times \delta \mbox{\boldmath$\omega$}'} \rangle$. Since the turbulent vortexmotive force $\langle {{\bf{u}}' \times \mbox{\boldmath$\omega$}'} \rangle$ may contribute to the generation of the mean velocity in the sense
\begin{equation}
	\frac{\partial {\bf{U}}}{\partial t}
	\sim \langle {{\bf{u}}' \times \mbox{\boldmath$\omega$}'} \rangle,
	\label{eq:delU_vmf_sec6}
\end{equation}
we expect $\langle {{\bf{u}}' \times \delta \mbox{\boldmath$\omega$}'} \rangle$ induces a large-scale flow. In the presence of the inhomogeneous mean electric-current density $\nabla {\bf{J}}$ a global flow can be generated by the turbulent cross helicity effect as
\begin{equation}
	\delta{\textbf{U}}
	= \tau \langle {
		{\textbf{u}}' \times \delta \mbox{\boldmath$\omega$}'
	} \rangle
	\propto \langle {{\textbf{u}}' \cdot {\textbf{b}}'} \rangle 
		\nabla \times {\textbf{J}}
	= - \langle {{\textbf{u}}' \cdot {\textbf{b}}'} \rangle \nabla^2 {\bf{B}}
	\label{eq:delU_-W_laplace_B_sec6}
\end{equation}
in the direction of $\nabla \times {\bf{J}}$. This result suggests that if the magnetic-field strength is spatially concentrated in some region (in the sense stronger than the surroundings as $\nabla^2 {\bf{B}} < 0$), a global flow $\delta {\bf{U}}$ in the direction parallel to the mean magnetic field ${\bf{B}}$ can be induced in the field-concentrated region if the turbulent cross helicity is positive there ($\langle {{\bf{u}}' \cdot {\bf{b}}'} \rangle > 0$). In case that the turbulent cross helicity is negative there ($\langle {{\bf{u}}' \cdot {\bf{b}}'} \rangle < 0$), the induced global flow $\delta {\bf{U}}$ direction is antiparallel to the mean magnetic field.

These arguments indicate that the turbulent cross helicity coupled with the mean magnetic-field shear may induce a large-scale flow. The direction of generated flow depends on the sign of the local turbulent cross helicity and the spatial distribution of the mean magnetic field represented by the mean magnetic-field shear. As we see in (\ref{eq:delU_-W_laplace_B_sec6}), in the location where the mean magnetic field is more prominent than the surroundings ($\nabla^2 {\bf{B}} < 0$), global flow is induced in the direction parallel to the mean magnetic field ${\bf{B}}$ for positive turbulent cross helicity ($\langle {{\bf{u}}' \cdot {\bf{b}}'} \rangle > 0$), and in the direction antiparallel to ${\bf{B}}$ for negative turbulent cross helicity ($\langle {{\bf{u}}' \cdot {\bf{b}}'} \rangle < 0$). Considering that the cross helicity is correlation between the velocity and magnetic field, this result is natural. It is important to note that this global flow generation does not occur for the uniform mean magnetic field.

\subsection{Poloidal flow generation in reversed shear (RS) mode confinement}
\begin{figure}[htb]
  \centering
  \includegraphics[width= 0.5 \columnwidth]{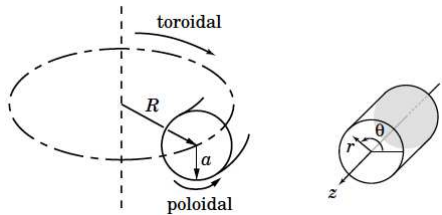}
  \caption{Tokamak's geometry ($R$: major radius of tokamak's torus, $a$: minor radius, $r$: radial coordinate) and the local cylindrical coordinate system $(r, \theta, z)$. The toroidal and poloidal directions in the torus geometry
correspond to the $z$ and $\theta$-$r$ directions, respectively, in the cylindrical approximation.}
    \label{fig:rs_torus_plasma}
\end{figure}

In Section~6.2, we saw that a non-zero cross helicity in turbulence coupled with a non-trivial spatial distribution of mean electric-current density $\bf{J}$, can induce a global flow. Here we see an example of this effect in the context of the torus fusion plasma (Fig.~\ref{fig:rs_torus_plasma}) \citep{yos1999}. 

\begin{figure}[htb]
  \centering
  \includegraphics[width= 0.9 \columnwidth]{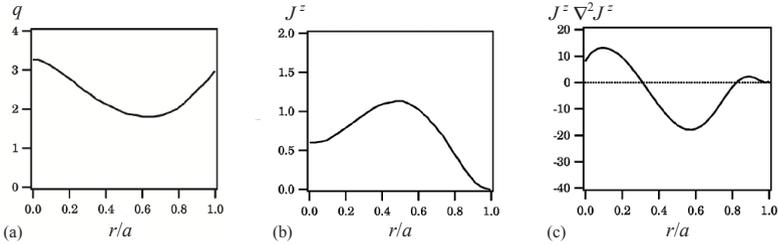}
  \caption{Radial profiles of (a) the safety factor $q$, (b) the plasma or toroidal electric-current density $J^z$, and (c) the non-uniformity of the mean electric-current density represented by $J^z \nabla^2 J^z$ in the reversed shear (RS) mode confinement.}
    \label{fig:rs_mode_radial_profiles}
\end{figure}

It is known that the confinement of plasma is greatly improved in devices with a reversed or negative magnetic shear in the core region \citep{fuj1997,fuj2001}. In this reversed shear (RS) mode, it is also observed an associated poloidal rotation in the minimum $q$ region [Fig.~\ref{fig:rs_mode_radial_profiles} (a)]. Here $q$ is the safety factor defined by
\begin{equation}
	q = \frac{rB^z}{RB^\theta},
	\label{eq:RS_safety_factor}
\end{equation}
where $r$ and $R$ are the minor and major radii and $B^z$ and $B^\theta$ correspond to the toroidal and poloidal magnetic field, respectively. 

We approximate the torus geometry by cylindrical coordinate ($r, \theta, z$), where the toroidal and poloidal directions in the torus geometry correspond to $z$ and $r$-$\theta$ directions. We assume the axisymmetry along the central minor axis ($\partial/\partial \theta =0$) of the statistical quantities and neglect $z$ dependence ($\partial / \partial z = 0$). Under this assumption, the mean fields are written as
\begin{subequations}
\begin{equation}
	{\bf{B}} = \left( {0, B^\theta(r,t), B^z(r,t)} \right),
	\label{eq:RS_B_compo}
\end{equation}
\begin{equation}
	{\bf{J}} = \left( {0, J^\theta(r,t), J^z(r,t)} \right),
	\label{eq:RS_J_compo}
\end{equation}
\begin{equation}
	{\bf{U}} = \left( {0, U^\theta(r,t), U^z(r,t)} \right),
	\label{eq:RS_U_compo}
\end{equation}
\end{subequations}
which lead to the Lorentz force
\begin{equation}
	{\bf{J}} \times {\bf{B}}
	= \left( {
    B^z J^\theta - B^\theta J^z,\;
    0,\;
    0
  } \right).
  \label{eq:RS_mean_lorentz_compo}
\end{equation}

We substitute the Reynolds and turbulent Maxwell stresses $\mbox{\boldmath${\cal{R}}$}$ expression (\ref{eq:rey_turb_max_strs_exp_sec6}), with the inhomogeneous helicity effect $\mbox{\boldmath$\Gamma$} \mbox{\boldmath$\Omega$}$ dropped, into the mean velocity equation. If we approximate that the transport coefficients $\nu_{\rm{K}}$ and $\nu_{\rm{M}}$ are locally uniform, and neglect the spatial derivatives of them, the momentum equation is written as
\begin{equation}
	\frac{\partial {\bf{U}}}{\partial t}
	= - \nabla \left( {P_{\rm{M}} + \frac{2}{3} K_{\rm{R}}} \right)
	+ \frac{(U^\theta)^2}{r} {\bf{e}}_r
	+ {\bf{J}} \times {\bf{B}}
	+ \nu_{\rm{K}} \nabla^2 {\bf{U}}
	- \nu_{\rm{M}} \nabla^2 {\bf{B}},
	\label{eq:RS_U_eq_w_R}
\end{equation}
where the molecular viscosity $\nu$ was dropped as compared with the turbulent viscosity $\nu_{\rm{K}}$.

The $z$ component of the mean vorticity $\Omega^z$, which represents the poloidal rotation, obeys
\begin{equation}
	\frac{\partial \Omega^z}{\partial t}
	= \nu_{\rm{K}} \nabla^2 \Omega^z
	- \nu_{\rm{M}} \nabla^2 J^z,
	\label{eq:RS_Omegz_eq}
\end{equation}
where $J^z$ is the $z$ component of the mean electric-current density. In the absence of the cross-helicity effect $\nu_{\rm{M}}$, the mean axial vorticity $\Omega^z$ decays due to the eddy viscosity $\nu_{\rm{K}}$ effect. The cross-helicity effect $\nu_{\rm{M}}$ gives a possibility to generate the large-scale poloidal rotation. We focus our attention on the $\nu_{\rm{M}}$ effect. Substituting $\nu_{\rm{M}}$ expression (\ref{eq:nuM_exp}) into (\ref{eq:RS_Omegz_eq}) we have
\begin{equation}
	\frac{\partial \Omega^z}{\partial t}
	= - \frac{5 C_\gamma}{7} \frac{K}{\varepsilon} W \nabla^2 J^z
	+ R_{\Omega 1},
	\label{eq:RS_Omegz_W_eq}
\end{equation}
where $R_{\Omega 1}$ denotes all the other terms. As for the cross helicity $W$ evolution, we assume that the production rate is dominantly due to $P_W^{({\rm{E}})}$ (\ref{eq:P_W_E_def}), which is approximated as
\begin{equation}
	P_W^{({\rm{E}})} \simeq \beta {\bf{J}} \cdot \mbox{\boldmath$\Omega$}.
	\label{eq:RS_Pw_approx}
\end{equation}
Then, the evolution equation of $W$ is written as
\begin{equation}
	\frac{\partial W}{\partial t}
	= \beta J^z \Omega^z + R_W
	= C_\beta \frac{K^2}{\varepsilon} J^z \Omega^z + R_W,
	\label{eq:RS_W_eq}
\end{equation}
where $R_W$ denotes all the other terms. From (\ref{eq:RS_Omegz_W_eq}) and (\ref{eq:RS_W_eq}), the mean vorticity equation is written as
\begin{equation}
	\frac{\partial^2 \Omega^z}{\partial t^2}
	- \left( {
		- \frac{5 C_\beta C_\gamma}{7} 
			\frac{K^3}{\varepsilon^2} J^z \nabla^2 J^z
	} \right) \Omega^z
	= R_{\Omega 2},
	\label{eq:RS_Omegz_Omegz_eq}
\end{equation}
where $R_{\Omega 2}$ represents all the remaining contributions and is not discussed here.

	This indicates that the large-scale vorticity may grow if the quantity in the parenthesis of (\ref{eq:RS_Omegz_Omegz_eq}) is positive as
\begin{equation}
	\chi_\Omega^2
	\equiv - \frac{5 C_\beta C_\gamma}{7} 
    	\frac{K^3}{\varepsilon^2} J^z \nabla^2 J^z
	> 0,
	\label{eq:RS_growth_cond}
\end{equation}
where $\chi_\Omega$ denotes the growth rate of $\Omega^z$.

	In the RS mode configuration, corresponding to the minimum $q$ profile in the core region, the radial distribution of the mean electric-current $J^z$ is given as Fig.~\ref{fig:rs_mode_radial_profiles} (b). The corresponding spatial distribution of $J^z \nabla^2 J^z$ is given as Fig.~\ref{fig:rs_mode_radial_profiles} (c). It follows from (\ref{eq:RS_growth_cond}) that the global poloidal rotation is induced in the region where $J^z \nabla^2 J^z < 0$.

	We performed a numerical simulation of the $K - \varepsilon - W$ turbulence model, where in addition to the mean velocity equation with the Reynolds and turbulent Maxwell stresses, the equations of the turbulent MHD energy $K$, its dissipation rate $\varepsilon$, and the turbulent cross helicity $W$ are simultaneously solved. At the initial stage we have no cross helicity $W=0$ in the entire domain of the simulation, then at some time (the cross-helicity onset time $t = 80$) we externally inject a cross helicity proportional to the turbulent MHD energy $K$ ($W / K = 0.5$) in the whole region of $r$.

\begin{figure}[htb]
  \centering
  \includegraphics[width= 0.5 \columnwidth]{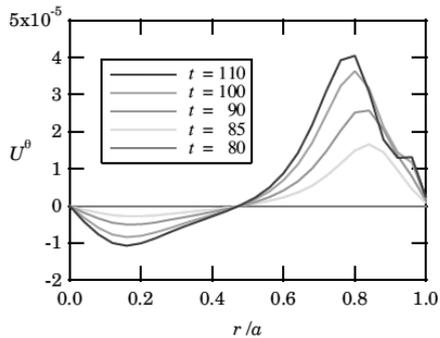}
  \caption{Spatiotemporal evolution of the poloidal velocity $U^\theta$. The turbulent cross helicity $W$ proportional to the turbulent MHD energy $K$ is set in at the cross-helicity onset time ($t = 80$).}
    \label{fig:rs_mode_gen_vel}
\end{figure}

	Our numerical simulation shows that before the cross-helicity onset time, no mean velocity is generated at all. However, once the cross helicity is set in, the mean poloidal velocity is generated in the region where $J^z \nabla^2 J^z < 0$ (Fig.~\ref{fig:rs_mode_gen_vel}). 

	There is much room for improving the present simplest turbulence model \cite{yok2008}, but this example clearly shows that the presence of the turbulent cross helicity $W$ coupled with the non-uniform spatial distribution of the mean electric-current density ${\bf{J}}$ can induce a global poloidal flow.

\section{Cross-helicity effects in magnetic reconnection}
\NOTE{
Magnetic reconnection provides an important situation where the cross-helicity effects both in the momentum transport and dynamo play a key role. From the viewpoint of the turbulent transport, the field configuration of the magnetic reconnection is regarded as a situation where turbulence is self-excited by the inhomogeneities of the magnetic field and velocity, and the flow and magnetic field are affected and determined by the turbulent transport. The evolution equation of the turbulent cross helicity (\ref{eq:K_W_evol_eq}) with (\ref{eq:P_W_def_sec4}) and (\ref{eq:T_W_def_sec4}) shows that in the presence of the mean magnetic field, the turbulent cross helicity is ubiquitously generated by the production mechanism related to the cross-helicity cascade or the transport mechanism related to the asymmetry of Alfv\'{e}n wave propagation. This suggests that turbulent cross helicity is ubiquitously present in the magnetic reconnection configurations. The role of the turbulent cross helicity in magnetic reconnection has not been fully investigated in the previous studies. This point will be discussed in this section.
}

\subsection{Necessity of turbulent reconnection}
	Magnetic reconnection is the most efficient mechanism to convert the magnetic energy into the kinetic energy and eventually into heat by Ohmic dissipation. It is considered to be one of the main mechanisms to the jet formation, mass ejection, and particle acceleration in astrophysical objects. Magnetic reconnection is a fundamental process also in the Sun, both in the interior and atmosphere. In the former, reconnection of magnetic field plays an essential role in forming a magnetic field configuration in dynamo in the convective zone or the tachocline just below the convective zone, while in the latter, reconnection is the core process for the solar flare and a possible mechanism for the coronal heating.

	The simplest reconnection model was proposed by \citet{swe1958} and \citet{par1957}. In this Sweet--Parker model, the reconnection is considered to take place in a thin current sheet with the thickness $\Delta$ and the width $L$ between the oppositely directed reconnection magnetic fields. In the case of uniform density $\rho = \rho_0$, the mass conservation gives
\begin{equation}
	u_{\rm{in}} L = u_{\rm{out}} \Delta,
	\label{eq:mass_conserv_sec7}
\end{equation}
where $u_{\rm{in}}$ is the inflow speed and $u_{\rm{out}}$ is the out flow speed. If we assume that the all the magnetic energy is converted to the kinetic energy, the conservation of the energy flux is given as
\begin{equation}
	\frac{1}{2 \mu_0} b_{\rm{in}}^2 u_{\rm{in}} L 
	= \frac{1}{2} \rho_0 u_{\rm{out}}^2 u_{\rm{out}} \Delta.
	\label{eq:conserv_en_flux_sec7}
\end{equation}
From (\ref{eq:mass_conserv_sec7}) and (\ref{eq:conserv_en_flux_sec7}), the outflow speed is evaluated as
\begin{equation}
	u_{\rm{out}}
	= \frac{1}{(\mu_0 \rho_0)^{1/2}} b_{\rm{in}}
	\equiv V_{\rm{A}},
	\label{eq:outflow_eval_sec7}
\end{equation}
where $V_{\rm{A}}$ is the Alfv\'{e}n speed. Namely, the out-flow speed is the Alfv\'{e}n speed of the reconnection magnetic field. 

	One way to evaluate the reconnection rate is to calculate the inflow Mach number $M_{\rm{in}}$: how much inflow $u_{\rm{in}}$ is induced for a given reconnection magnetic field $b_{\rm{in}}$. Using (\ref{eq:mass_conserv_sec7}) and (\ref{eq:outflow_eval_sec7}), the inflow Mach number is expressed as
\begin{equation}
	M_{\rm{in}}
	= \frac{u_{\rm{in}}}{V_{\rm{A}}}
	= \frac{u_{\rm{in}}}{u_{\rm{out}}}
	= \frac{\Delta}{L}.
	\label{eq:inflow_mach_sec7}
\end{equation}
This suggests that in order to obtain the more enhanced reconnection rate, the larger $\Delta$ and the smaller $L$ are preferable.

	In the Sweet--Parker model, the magnetic field entering the diffusion region $({\bf{u}}_{\rm{in}} \cdot \nabla) {\bf{b}}_{\rm{in}}$ is diffused solely by the magnetic diffusivity $\eta \nabla^2 {\bf{b}}_{\rm{in}}$. In this case, we have $b_{\rm{in}} u_{\rm{in}} / \Delta \sim \eta b_{\rm{in}} / \Delta^2$, resulting in
\begin{equation}
	\Delta = \eta / u_{\rm{in}}.
	\label{eq:mole_diff_depth_sec7}
\end{equation}
Substituting this into (7.4), we obtain
\begin{equation}
	M_{\rm{in}} = S_{\rm{L}}^{-1/2},
	\label{eq:inflow_mach_lundquist_sec7}
\end{equation}
where $S_{\rm{L}}$ is the Lundquist number defined by
\begin{equation}
	S_{\rm{L}} = \frac{V_{\rm{A}}L}{\eta}
	\label{eq:lundquist_def_sec7}
\end{equation}
(the magnetic Reynolds number with the velocity replaced by the Alfv\'{e}n speed of the reconnection magnetic field). In the astrophysical and space physics phenomena, the Lundquist number is ubiquitously huge (e.g.\ $S_{\rm{L}} > 10^{12}$). Consequently, the reconnection rate of the Sweet--Parker model (\ref{eq:inflow_mach_lundquist_sec7}) is too slow ($M_{\rm{in}} < 10^{-6}$) for elucidating a fast reconnection observed in the astrophysics and space physics. For instance, the reconnection rate in the solar flare is observed to be $M_{\rm{in}} = 10^{-3}-10^{-1}$.

	In order to alleviate this drawback of the Sweet--Parker model, several reconnection models, which are potentially able to elucidate fast reconnection, have been proposed. For example, the Petschek model of magnetic reconnection, where the slow shock plays an essential role, realizes a tiny diffusion region
with effectively reducing $L$. 

	Turbulence reconnection is one of such approaches. Firstly, turbulent motions of the medium affect the statistical and dynamic properties of magnetic reconnection through the enhanced and inhomogeneous turbulent diffusivity. In the presence of turbulence, magnetic-field lines are subject to meandering motions. As this result, the effective thickness of diffusion becomes thicker as well as the enhancement of the effective diffusivity. As we see in (\ref{eq:inflow_mach_sec7}), this contributes to the enhanced reconnection rate. Two dimensional direct numerical simulations (DNSs) of the magnetic reconnection in externally injected turbulence suggest that the reconnection rate increases as the level of injected turbulence increases. The reconnection-rate dependence on the Lundquist number $S_{\rm{L}}$, which is $M_{\rm{in}} \propto S_{\rm{L}}^{-1/2}$ for the simplest Sweet--Parker scaling, disappears as the injected turbulence level increases \citep{lou2009}. This DNS observation implies that, in the presence of strong turbulence, the reconnection rate depends not on the molecular magnetic diffusivity $\eta$ but on the non-linear dynamics of turbulence.

	Investigating magnetic reconnection with imposing turbulence by external forcing is one way to  see the turbulence effects on the reconnection rate. At the same time, it is true that the magnetic-field and velocity configurations associated with magnetic reconnection are highly nonuniform. In the presence of strong shear of the magnetic and velocity at large scales, represented by the electric-current density and vortical motions, turbulence is expected to be generated by these inhomogeneous large-scale fields. Actually, the mean-field shears contribute to the generation of turbulence as free energy sources through their coupling with the turbulent fluxes such as $- \langle {- u'{}^i u'{}^j + b'{}^i b'{}^j} \rangle (\partial U^j / \partial x^i)$, $- \langle {{\bf{u}}' \times {\bf{b}}'} \rangle \cdot {\bf{J}}$, etc. In this sense, turbulence is self-generated by the field configuration associated with the magnetic reconnection, without resorting to the external forcing.

\subsection{System of model equations of mean- and turbulent-fields}
	In this section, we present a non-linear turbulence modelling approach for treating such a self-generated turbulence and its effects on the magnetic reconnection. In this approach, the reconnection of the large-scale magnetic field subject to the turbulent transport is investigated. At the same time, the evolution of the turbulence is simultaneously considered and solved under the influence of the configurations of the mean magnetic and velocity fields.
Mean-field equations with turbulent fluxes

	The mean-field equations describing the magnetic reconnection are given by the mass, momentum, energy, and magnetic induction ones as
\begin{equation}
	\frac{\partial \langle {\rho} \rangle}{\partial t}
	+ \nabla \cdot (\langle {\rho} \rangle {\bf{U}})
	= 0,
	\label{eq:mean_mass_eq_sec7}
\end{equation}
\begin{equation}
	\frac{\partial}{\partial t}
	\langle {\rho} \rangle U^i
	+ \frac{\partial}{\partial x_j} \langle {\rho} \rangle U^j U^i
	= - \frac{\partial}{\partial x^i} \langle {\rho} \rangle P
	+ \frac{\partial}{\partial x^j} \mu {\cal{S}}^{ji}
	+ \langle {\rho} \rangle \left( {
		{\bf{J}} \times {\bf{B}}
	} \right)^i
	- \frac{\partial}{\partial x^j} \left( {
		\langle {\rho} \rangle {\cal{R}}^{ij}
	} \right),
	\label{eq:mean_vel_eq_sec7}
\end{equation}
\begin{eqnarray}
	&&\frac{\partial}{\partial t}
	\left\{ {
		\langle {\rho} \rangle \left[ {
		\frac{P}{\gamma - 1}
		+ \frac{1}{2} \left( {{\bf{U}}^2 + {\bf{B}}^2} \right)
		+ \frac{1}{2} \left\langle {{\bf{u}}'{}^2 + {\bf{b}}'^2} \right\rangle
		} \right]
	} \right\}
	\nonumber\\
	&&= - \nabla \cdot \left\{ {
		\left[ {
		\frac{\gamma}{\gamma - 1} P
		+ \frac{1}{2} \left( {
			{\bf{U}}^2 + \langle {{\bf{u}}'{}^2} \rangle
		} \right)
	} \right] \langle {\rho} \rangle {\bf{U}}
	} \right.
	\nonumber\\
	&& \left. {
	+ \langle {\rho} \rangle 
	\langle {
		({\bf{U}} \cdot {\bf{u}}') {\bf{u}}'
	} \rangle
	+ \langle {\rho} \rangle {\bf{E}} \times {\bf{B}}
	} \right\},
	\label{wean_en_eq_sec7}
\end{eqnarray}
\begin{equation}
	\frac{\partial {\bf{B}}}{\partial t}
	= - \nabla \times {\bf{E}}.
	\label{eq:mean_mag_ind_eq_sec7}
\end{equation}
From the Ohm's law, the mean electric field ${\bf{E}}$ is given as
\begin{equation}
	{\bf{E}}
	= - {\bf{U}} \times {\bf{B}}
	+ \eta {\bf{J}}
	- {\bf{E}}_{\rm{M}}.
	\label{eq:mean_ohm_eq_sec7}
\end{equation}
Here the turbulent fluxes $\mbox{\boldmath${\cal{R}}$}$ and ${\bf{E}}_{\rm{M}}$ are given as (\ref{eq:emf_expression}) and (\ref{eq:rey_turb_max_str_exp}), respectively.

	On the other hand, the equations of the turbulent kinetic energy $K$ (\ref{K_def_sec4}), the turbulent cross helicity $W$ (\ref{eq:W_def_sec4}), and the energy dissipation rate $\varepsilon$ (\ref{eq:eps_K_def_sec4}) are given as
\begin{equation}
	\left( {
		\frac{\partial}{\partial t}
		+ {\bf{U}} \cdot \nabla
	} \right) K
	= - {\bf{E}}_{\rm{M}} \cdot {\bf{J}}
	- {\cal{R}}^{ij} \frac{\partial U^i}{\partial x^j}
	- \varepsilon_K
	+ {\bf{B}} \cdot \nabla W
	+ \nabla \cdot \left( {
		\frac{\nu_{\rm{K}}}{\sigma_{\rm{K}}} \nabla K
	} \right),
	\label{eq:K_eq_sec7}
\end{equation}
\begin{equation}
	\left( {
		\frac{\partial}{\partial t}
		+ {\bf{U}} \cdot \nabla
	} \right) W
	= - {\bf{E}}_{\rm{M}} \cdot \mbox{\boldmath$\Omega$}
	- {\cal{R}}^{ij} \frac{\partial B^i}{\partial x^j}
	- \varepsilon_W
	+ {\bf{B}} \cdot \nabla K
	+ \nabla \cdot \left( {
		\frac{\nu_{\rm{K}}}{\sigma_{\rm{W}}} \nabla W
	} \right),
	\label{eq:W_eq_sec7}
\end{equation}
\begin{equation}
	\left( {
		\frac{\partial}{\partial t}
		+ {\bf{U}} \cdot \nabla
	} \right) \varepsilon
	= C_{\varepsilon 1} \frac{\varepsilon}{K} P_K
	- C_{\varepsilon 2} \frac{\varepsilon}{K} \varepsilon
	+ \nabla \cdot \left( {
		\frac{\nu_{\rm{K}}}{\sigma_{\rm{\varepsilon}}} \nabla \varepsilon
	} \right).
	\label{eq:eps_eq_sec7}
\end{equation}

For the sake of brevity, we consider the simplest possible case with no variable density. Further, we assume there is no helicity effects in $\mbox{\boldmath${\cal{R}}$}$ and ${\bf{E}}_{\rm{M}}$. Then they are reduced to
\begin{equation}
	\mbox{\boldmath${\cal{R}}$}_{\rm{D}}
	= - \nu_{\rm{K}} \mbox{\boldmath${\cal{S}}$}
	+ \nu_{\rm{M}} \mbox{\boldmath${\cal{M}}$},
	\label{eq:rey_turb_max_strs_dev_wo_hel_sec7}
\end{equation}
\begin{equation}
	{\bf{E}}_{\rm{M}}
	= - \beta \nabla \times {\bf{B}}
	+ \gamma \mbox{\boldmath$\Omega$}.
	\label{eq:emf_wo_alpha_sec7}
\end{equation}
In this case, the mean-field equations are reduced to a much simpler form for the mean velocity ${\bf{U}}$ and magnetic field ${\bf{B}}$ as
\begin{equation}
	\frac{\partial {\bf{U}}}{\partial t}
	= {\bf{U}} \times \mbox{\boldmath$\Omega$}
	+ {\bf{J}} \times {\bf{B}}
	+ \nu_{\rm{K}} \nabla^2 \left( {
		{\bf{U}} - \frac{\gamma}{\beta} {\bf{B}}
	} \right)
	- \nabla \left( {
		P 
		+ \frac{1}{2} {\bf{U}}^2
		+ \frac{1}{2} \langle {{\bf{b}}'{}^2} \rangle
		+ \frac{2}{3} K_{\rm{R}}
	} \right),
	\label{eq:mean_vel_eq_wo_hel_sec7}
\end{equation}
\begin{equation}
	\frac{\partial {\bf{B}}}{\partial t}
	= \nabla \times ({\bf{U}} \times {\bf{B}})
	- \nabla \times [ {
		\beta \nabla \times {\bf{B}}
	}]
	+ \nabla \times (\gamma \mbox{\boldmath$\Omega$}),
	\label{eq:mean_mag_eq_wo_alpha_sec7}
\end{equation}
where the molecular viscosity and magnetic diffusivity, $\nu$ and $\eta$, have been dropped since their magnitude are much smaller than the turbulent counterparts $\nu_{\rm{K}}$ and $\beta$.

	Taking curl of (\ref{eq:mean_vel_eq_wo_hel_sec7}), we have the mean vorticity equation as
\begin{eqnarray}
	\frac{\partial \mbox{\boldmath$\Omega$}}{\partial t}
	&=& \nabla \times \left[ {
		\left( {
			{\bf{U}} - \frac{\gamma}{\beta} {\bf{B}}
		}\right) \times \mbox{\boldmath$\Omega$}
		+ \nu_{\rm{K}} \nabla^2 \left( {
		{\bf{U}} - \frac{\gamma}{\beta} {\bf{B}}
		} \right)
	} \right]
	\nonumber\\
	&&+ \nabla \times \left[ {
		{\bf{F}}
		+ \frac{1}{\beta} ({\bf{U}} \times {\bf{B}}) 
		\times {\bf{B}}
		- \frac{1}{\beta} \frac{\partial {\bf{A}}}{\partial t}
		\times {\bf{B}}
	} \right].
	\label{eq:mean_vort_eq_sec7}
\end{eqnarray}

\NOTE{
For the purpose of understanding the role of the turbulent cross helicity in momentum transport, we divide a mean-field quantities into two parts. The velocity ${\bf{U}}$ and the vorticity $\mbox{\boldmath$\Omega$}$ are divided as
\begin{equation}
	{\bf{U}} = {\bf{U}}_0 + \delta{\bf{U}},\;\;
	\mbox{\boldmath$\Omega$}
	= \mbox{\boldmath$\Omega$}_0
	+ \delta \mbox{\boldmath$\Omega$}.
	\label{eq:U_pert_Omega_pert_sec7}
\end{equation}
Here, ${\bf{U}}_0$ and $\mbox{\boldmath$\Omega$}_0$ are the mean vfields without the effect of the mean magnetic field ${\bf{B}}$, while $\delta {\bf{U}}$ and $\delta \mbox{\boldmath$\Omega$}$ are the mean fields representing the first-order effects of ${\bf{B}}$ through the turbulent cross helicity. Substituting (\ref{eq:U_pert_Omega_pert_sec7}) into (\ref{eq:mean_vort_eq_sec7}), we have the zeroth-order-field equation as
\begin{equation}
	\frac{\partial \mbox{\boldmath$\Omega$}_0}{\partial t}
	= \nabla \times \left( {
		{\bf{U}}_0 \times \mbox{\boldmath$\Omega$}_0
		+ \nu_{\rm{K}} \nabla^2 {\bf{U}}_0
		+ {\bf{F}}
	} \right),
	\label{eq:Omega0_eq_sec7}
\end{equation}
and the first-order-field equation as
\begin{equation}
	\frac{\partial \delta \mbox{\boldmath$\Omega$}}{\partial t}
	= \nabla \times \left[ {
		\left( {
			\delta {\bf{U}} 
			- \frac{\gamma}{\beta} {\bf{B}}
		}\right) \times \mbox{\boldmath$\Omega$}_0
	+ \nu_{\rm{K}} \nabla^2 \left( {
			\delta {\bf{U}} 
			- \frac{\gamma}{\beta} {\bf{B}}
		} \right)
	} \right].
	\label{eq:Omega1_eq_sec7}
\end{equation}
}

\NOTE{
	Equation~(\ref{eq:Omega0_eq_sec7}) represents the evolution of $\mbox{\boldmath$\Omega$}_0$, which is subject to the turbulent viscosity $\nu_{\rm{K}}$ and the external force ${\bf{F}}$. On the other hand, (\ref{eq:Omega1_eq_sec7}) represents the evolution of $\delta \mbox{\boldmath$\Omega$}$ arising from the turbulent cross-helicity effect. We see from (\ref{eq:Omega1_eq_sec7}) that, for a given mean vorticity structure $\mbox{\boldmath$\Omega$}_0$, a particular solution of the stationary state is given as
\begin{equation}
	\delta {\bf{U}}
	= \frac{\gamma}{\beta} {\bf{B}}
	= C_{W/K} \frac{W}{K} {\bf{B}}.
	\label{eq:delU_stat_sol_sec7}
\end{equation}
This solution suggests that a large-scale flow $\delta {\bf{U}}$ is induced in the direction of the mean magnetic field ${\bf{B}}$ as an effect of the turbulent cross helicity. The induced-flow direction is parallel to ${\bf{B}}$ for the positive turbulent cross helicity ($W>0$), and antiparallel to ${\bf{B}}$ for the negative turbulent cross helicity ($W<0$). Note that this mean-flow induction will not be observed in the absence of the turbulent cross helicity ($W=0$).
}

	A similar argument can be done for the mean induction equation (\ref{eq:mean_mag_eq_wo_alpha_sec7}). If we divide the mean magnetic field ${\bf{B}}$ and the mean electric-current density ${\bf{J}}$ as
\begin{equation}
	{\bf{B}} = {\bf{B}}_0 + \delta {\bf{B}}_0,\;\;
	{\bf{J}} = {\bf{J}}_0 + \delta {\bf{J}}_0,
	\label{eq:B_pert_J_pert_sec7}
\end{equation}
and substitute (\ref{eq:B_pert_J_pert_sec7}) into (\ref{eq:mean_mag_eq_wo_alpha_sec7}), we have
\begin{equation}
	\frac{\partial {\bf{B}}_0}{\partial t}
	= \nabla \times ({\bf{U}} \times {\bf{B}}_0)
	- \nabla \times \left( { \beta \nabla \times  {\bf{B}}_0} \right)
	\label{eq:B0_eq_sec7}
\end{equation}
\begin{equation}
	\frac{\partial \delta{\bf{B}}}{\partial t}
	= \nabla \times ({\bf{U}} \times \delta {\bf{B}})
	- \nabla \times \left[ { \beta
		\nabla \times  \left( {
			\delta {\bf{B}}
    		- \frac{\gamma}{\beta} {\bf{U}}
		} \right)
	} \right]
	\label{eq:B1_eq_sec7}
\end{equation}
for the first-order field.

	Equation~(\ref{eq:B0_eq_sec7}) represents the evolution of the mean magnetic field ${\bf{B}}_0$, which is subject to the turbulent magnetic diffusivity. On the other hand, (\ref{eq:B1_eq_sec7}) represents the mean magnetic-field induction $\delta {\bf{B}}$ due to the turbulent cross-helicity effect. Equation~(\ref{eq:B1_eq_sec7}) implies that 
\begin{equation}
	\delta {\bf{B}} = \frac{\gamma}{\beta} {\bf{U}}
	= C_{W/K} \frac{W}{K} {\bf{U}}
	\label{eq:delB_stat_sol_sec7}
\end{equation}
is a particular solution of the first-order magnetic field induction $\delta {\bf{B}}$ in the stationary state. This solution (\ref{eq:delB_stat_sol_sec7}) suggests that, for a given magnetic field ${\bf{B}}_0$, the mean magnetic field $\delta {\bf{B}}$ can be induced by the turbulent cross-helicity effect in the direction of the mean velocity ${\bf{U}}$. The direction of $\delta {\bf{B}}$ is parallel to ${\bf{U}}$ for positive turbulent cross helicity ($W>0$), and antiparallel to ${\bf{U}}$ for negative cross helicity ($W<0$). This induction will not be observed in the absence of the turbulent cross helicity ($W=0$).

	The velocity induction (\ref{eq:delU_stat_sol_sec7}) and magnetic-field induction (\ref{eq:delB_stat_sol_sec7}) are natural in the sense that the cross helicity represents the correlation between the velocity and magnetic field. It is important to remark that these mean-field inductions are mediated by the cross helicity in turbulent fields ($W = \langle {{\bf{u}}' \cdot {\bf{b}}'} \rangle$).

\begin{figure}[htb]
  \centering
  \includegraphics[width= 1.0 \columnwidth]{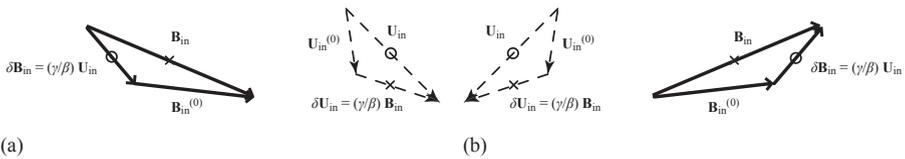}
  \caption{Modulations of the reconnection magnetic field and the inflow velocity due to the cross-helicity effect. Both in the regions with positive turbulent cross helicity ($\gamma >0$) (a) and with negative turbulent cross helicity ($\gamma < 0$) (b), the direction of the inflow velocity becomes less perpendicular to the current sheet, and the reconnection magnetic field ${\bf{B}}_{\rm{in}}$ becomes more oblique to the current sheet. Redrawn from \citet{yok2011b}.}
    \label{fig:modulation_flds_due_to_ch}
\end{figure}

	These mean-field inductions may make it possible to enhance the magnetic reconnection through the modulation of the inflow velocity and reconnection magnetic field (Fig.~\ref{fig:modulation_flds_due_to_ch}). In the presence of the turbulent cross-helicity effect (\ref{eq:delB_stat_sol_sec7}), the mean magnetic field $\delta {\bf{B}}_{{\rm{in}}}$ is induced. Then, the reference reconnection magnetic field without the cross-helicity effect, ${\bf{B}}_{\rm{in}}^{(0)}$, is modulated to ${\bf{B}}_{\rm{in}}$ as
\begin{equation}
	{\bf{B}}_{{\rm{in}}}
	= {\bf{B}}_{{\rm{in}}}^{(0)} + \delta {\bf{B}}_{\rm{in}}
	= {\bf{B}}_{{\rm{in}}}^{(0)}
	+ \frac{\gamma}{\beta} {\bf{U}}_{\rm{in}}^{(0)}.
	\label{eq:rec_mag_fld_modul}
\end{equation}
At the same time, due to the turbulent cross-helicity effect (\ref{eq:delU_stat_sol_sec7}), the mean velocity $\delta {\bf{U}}_{\rm{in}}$ is induced. Then, the reference inflow without the cross-helicity effect, ${\bf{U}}_{\rm{in}}^{(0)}$, is modulated to ${\bf{U}}_{\rm{in}}$ as
\begin{equation}
	{\bf{U}}_{\rm{in}}
	= {\bf{U}}_{\rm{in}}^{(0)} + \delta {\bf{U}}_{\rm{in}}
	= {\bf{U}}_{\rm{in}}^{(0)} 
	+ \frac{\gamma}{\beta} {\bf{B}}_{\rm{in}}^{(0)}.
	\label{eq:inflow_vel_modul}
\end{equation}
	
As Fig.~\ref{fig:modulation_flds_due_to_ch} shows, the inflow velocity and the reconnection magnetic field are modulated by the cross helicity effect. The modulation directions depend on the sign of the turbulent cross helicities. In both cases of (a) positive cross helicity ($\gamma > 0$) and (b) negative cross helicity ($\gamma < 0$), the modulated inflow velocities ${\bf{U}}_{\rm{in}}$ become less perpendicular to the current sheet than the reference inflow velocity ${\bf{U}}_{\rm{in}}^{(0)}$, while the modulated reconnection magnetic field ${\bf{B}}_{\rm{in}}$ becomes more oblique to the current sheet than the reference reconnection magnetic field ${\bf{B}}_{\rm{in}}^{(0)}$. This modulation leads to the larger value of $\Delta / L$ in the reconnection rate formula defined by the inflow Mach number $M_{\rm{in}}$ (\ref{eq:inflow_mach_sec7}).  As this consequence, the inflow velocity ${\bf{U}}_{\rm{in}}$ and reconnection magnetic field ${\bf{B}}_{\rm{in}}$, which are modulated by the turbulent cross-helicity effect, turn out to be suitable for the fast reconnection. This enhancement of reconnection rate is due to the modulation of the field configuration, and the enhancement rate relative to the reference state depends on the value of the normalized cross helicity $\gamma/\beta$ or equivalently $W/K$. Evaluation of this effect is seen in \citet{yok2011b}.

\subsection{Numerical simulations of reconnection model equations}
\NOTE{
In the previous subsection, we evaluated the mean velocity and magnetic-field modulations due to the turbulent cross-helicity effect, $\delta {\bf{U}}$ and $\delta {\bf{B}}$, in an perturbative expansion manner on the basis of the expressions of the turbulent fluxes. The results (\ref{eq:delU_stat_sol_sec7}) and (\ref{eq:delB_stat_sol_sec7}) imply that the turbulent cross helicity contributes to the fast reconnection through the mean-field modulations. However, in order to obtain the  mean and turbulence fields configurations in the general situation, we have to solve the system of model equations numerically.
}

\subsubsection{Turbulent magnetic reconnection model with the evolution of turbulence implemented}
\NOTE{
	The system of model equations consist of the mean-field equations (\ref{eq:mean_mass_eq_sec7})-(\ref{eq:mean_mag_ind_eq_sec7}) with the turbulent fluxes (Reynolds and turbulent Maxwell stresses and turbulent electromotive force),(\ref{eq:rey_turb_max_strs_dev_wo_hel_sec7}) and (\ref{eq:emf_wo_alpha_sec7}), implemented, and the equations of the turbulent statistical quantities (turbulent energy, turbulent cross helicity, and turbulent energy dissipation rate), (\ref{eq:K_eq_sec7})-(\ref{eq:eps_eq_sec7}). By solving both the mean fields and turbulence fields, self-consistency of the model simulation including the realizability and non-linear interaction between the mean and turbulence fields are assured.
}

	First series of model simulations were performed with the model equations. In these first simulations, instead of solving the $\varepsilon$ equation (\ref{eq:eps_eq_sec7}), the dissipation rates of the turbulent energy and cross helicity $\varepsilon_K$ and $\varepsilon_W$, are approximated as
\begin{equation}
	\varepsilon = \frac{K}{\tau},
	\label{eq:eps_algebra_model_sec7}
\end{equation}
\begin{equation}
	\varepsilon_W = C_W \frac{W}{\tau},
	\label{eq:epsW_algebra_model_sec7}
\end{equation}
where $C_W$ is a model constant \citep{hig2013,yok2013b}. In (\ref{eq:eps_algebra_model_sec7}) and (\ref{eq:epsW_algebra_model_sec7}), $\tau$ is the characteristic timescale of turbulence representing the turbulence relaxation, and is set as a constant throughout each simulation. To investigate the relationship between turbulence and reconnection, the value of $\tau$ is changed from simulation to simulation by using a parameter $C_\tau$ defined by
\begin{equation}
	\tau = C_\tau \tau_0.
	\label{eq:parameter_tau_sec7}
\end{equation}
Since $C_\tau$ determines the characteristic timescale of turbulence, it controls the dissipation rate of turbulent energy $\varepsilon_K$. If $C_\tau$ is much smaller than unity ($C_\tau \ll 1$), the dissipation rate of the turbulent energy is very large, so the turbulence level becomes very low as time goes by. On the other hand, if $C_\tau$ is much larger than unity ($C_\tau \gg 1$), the dissipation rate of the turbulent energy is very small, then the turbulence level becomes very high. We performed simulations with different $C_\tau$ values. As for the initial turbulent field, we set the turbulent energy $K = 1.0 \times 10^{-2}$ and $W=0$ everywhere. 

\begin{figure}[htb]
  \centering
  \includegraphics[width= 0.6 \columnwidth]{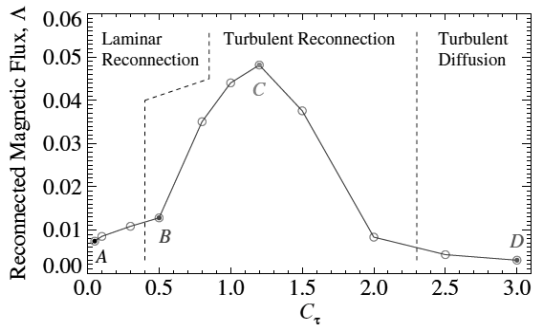}
  \caption{Reconnected magnetic-field flux $\Lambda$ (\ref{eq:rec_flux_def_sec7}) at $t/\tau_{\rm{A}} = 254$ against $C_\tau$. Redrawn from \citet{hig2013}}
    \label{fig:ch_momentum_transreconnected_mag_flux}
\end{figure}

	The efficiency of the reconnection process can be measured by the reconnected fluxes defined by
\begin{equation}
	\Lambda 
	= \int_{-L_x/2}^{L_x/2}\!\!\! dx \| B^z \|_{z=0}/(B_0^x L_x).
	\label{eq:rec_flux_def_sec7}
\end{equation}
The reconnected magnetic fluxes $\Lambda$ against $C_\tau$ are plotted in Fig.~\ref{fig:ch_momentum_transreconnected_mag_flux}. This plot shows that both in the domains $C_\tau \lesssim 0.5$ and $2.0 \lesssim C_\tau$, $\Lambda$ is small, while in-between these regions there is a region $0.5 \lesssim C_\tau \lesssim 2.0$ the reconnected magnetic-field flux $\Lambda$ where $\Lambda$ becomes much larger than the other regions. The small $C_\tau$ region may be called as the {\it laminar reconnection} region since the level of turbulence there is very low because of very short relaxation timescale of the turbulent energy. On the other hand, the large $C_\tau$ region may be called the {\it turbulent diffusion} region since the level of turbulence is very high there because of very long relaxation timescale of the turbulent energy. In-between these regions, there is a region with much larger $\Lambda$, which means higher reconnection rate. This region can be called the {\it turbulent reconnection} region.

\begin{figure}[htb]
  \centering
  \includegraphics[width= 0.8 \columnwidth]{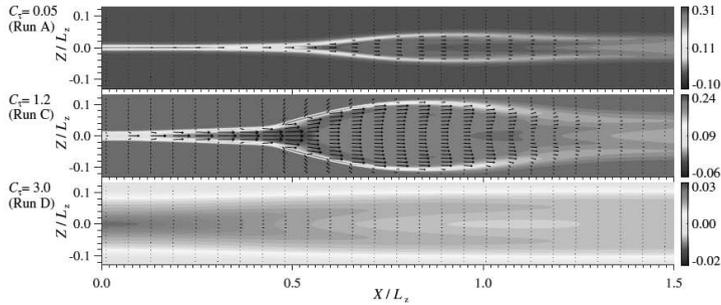}
  \caption{Contour plots of the $y$ component of the mean electric-current density at $t/\tau_{\rm{A}} = 254$ in the laminar reconnection ($C_\tau = 0.05$), the turbulent reconnection ($C_\tau = 1.2$), and the turbulent diffusion ($C_\tau = 3.0$). The black arrows denotes the flow velocity vector. Redrawn from \citet{hig2013}.}
    \label{fig:J_contour}
\end{figure}

In these three regimes, i.e., the laminar reconnection, the turbulent reconnection, and the turbulent diffusion regimes, the field configurations are fairly different with each other. The contour of $y$ component of the mean electric-current density at a time ($t/\tau_{\rm{A}} = 254$) in the laminar reconnection ($C_\tau = 0.05$), the turbulent reconnection ($C_\tau = 1.2$), and the turbulent diffusion ($C_\tau = 3.0$) regimes are shown in Fig.~\ref{fig:J_contour}.

The laminar reconnection regime (upper) shows Sweet--Parker like field configuration with some magnitude of the out flow. On the other hand, in the turbulent reconnection regime (middle), the field configuration is more Petschek-like. Namely, the thin reconnection region with small depth $\Delta$ is much more concentrated in the vicinity of $X=0$. In this sense, the width $L$ becomes much smaller than the counterpart of the laminar reconnection regime. The out-flow speed is much larger and has some $y$ component in the turbulent reconnection regime. This modulation of the flow velocity matches the tendency due to the cross-helicity effect. In the turbulent diffusion regime, the field is distributed in much broader manner due to the strong diffusivity due to turbulence. In this regime, we do not have any particular reconnection such as observed in the laminar and turbulent reconnection regimes. As this consequence, no strong outflow is observed in this regime. This can be interpreted that the level of turbulence is too high everywhere and the reconnection magnetic field is too much diffused to be reconnected.

We examined the role of the turbulent cross helicity $W = \langle {{\bf{u}}' \cdot {\bf{b}}'} \rangle$ in the turbulent magnetic reconnection, by artificially putting the turbulent cross helicity zero ($W = 0$) throughout the simulation, and compare the result with the counterpart with non-vanishing turbulent cross helicity case ($W \ne 0$). 

The turbulent cross helicity works for localizing the spatial distribution of the turbulent diffusivity $\beta$ effect in two ways; (i) one is localizing the spatial distribution of $\beta$ through the localization of the energy through the mechanism mentioned in Section~4.2, and (ii) the other is suppressing the turbulent diffusivity effect $\beta {\bf{J}}$ by the cross-helicity effect $\gamma \mbox{\boldmath$\Omega$}$. We argue these points here.

\begin{figure}[htb]
  \centering
  \includegraphics[width= 0.6 \columnwidth]{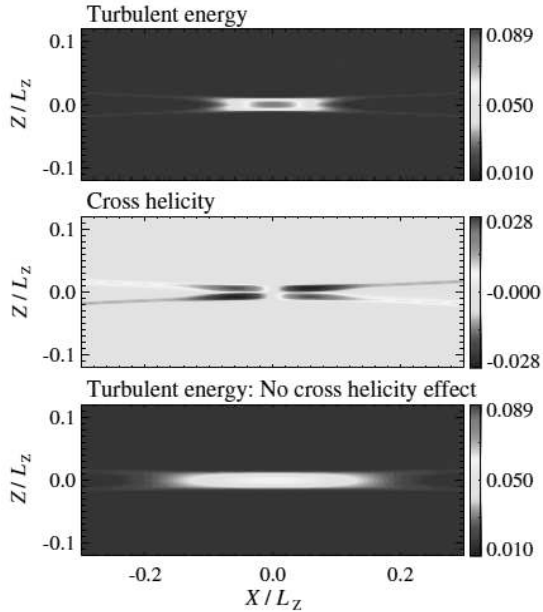}
  \caption{Spatial distribution of the turbulent MHD energy $K$ and cross helicity $W$. The turbulent MHD energy $K$ distribution with a finite turbulent cross helicity $W \ne 0$ (Upper). The turbulent cross helicity distribution (Middle), and the turbulent MHD energy distribution without the turbulent cross helicity $W = 0$. Redrawn from \citet{hig2013}.}
    \label{fig:turb_en_localization_by_ch}
\end{figure}

Figure~\ref{fig:turb_en_localization_by_ch} compares the spatial distributions of the turbulent MHD energy calculated with the $K$ equation (\ref{eq:K_eq_sec7}). The spatial distribution of $K$ in the case of simultaneously solving the turbulent cross helicity $W$ equation (\ref{eq:W_eq_sec7}) is shown in the upper plot. The corresponding turbulent cross helicity distribution is shown in the middle plot. As we expected from the production mechanism of $W$ associated with the coupling of the mean electric current density $\bf{J}$ and the mean vorticity $\mbox{\boldmath$\Omega$}$, $- \beta {\bf{J}} \cdot \mbox{\boldmath$\Omega$}$, the turbulent cross helicity shows the quadropole-like spatial distribution. The turbulent MHD energy distribution in the case of no turbulent cross helicity ($W = 0$) is shown in the lower plot. In this case, the turbulent cross helicity is artificially put null ($W = 0$) in the whole calculation domain throughout the whole calculation period. We clearly see marked difference in the $K$ distributions between the cases with $W \ne 0$ and with $W = 0$. In the presence of turbulent cross helicity $W \ne 0$, the turbulent MHD energy $K$ is much more concentrated in the vicinity of the X point. The magnitude of $K$ near the X point is much higher than the case with no cross helicity. Because the turbulent diffusivity $\beta$ is basically proportional to the turbulent MHD energy $K$, this results suggests that the turbulent magnetic diffusivity $\beta$ in the presence of the turbulent cross helicity is much more concentrated in the vicinity of the X point with much higher magnitude than the counterpart in the absence of the turbulent cross helicity.

This concentration or localization of the turbulent magnetic diffusivity in the vicinity of X point is caused by the $K$ generation mechanism due to the $W$ inhomogeneity along the mean magnetic field, $T_K^{({\rm{B}})} = ({\bf{B}} \cdot \nabla) W$ (\ref{eq:T_K_B}). The inhomogeneity of the turbulent cross helicity, corresponding to asymmetry of the Alfv\'{e}n wave propagations between the parallel and antiparallel directions, causes the imbalance of the energy flux across the boundaries (see the description in Section~4.1 and Figure~\ref{fig:en_gen_due_to_cross_helicity}). Then, a strong localization of the turbulent MHD energy can be induced in the inhomogeneous cross-helicity region near the X point.

\begin{figure}[htb]
  \centering
  \includegraphics[width= 0.5 \columnwidth]{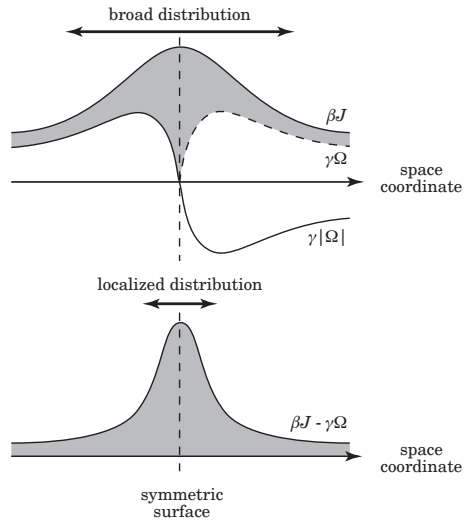}
  \caption{Schematic picture of the effective localization of the turbulent magnetic diffusivity $\beta {\bf{J}}$ due to the spatial distribution of the turbulent cross-helicity effect $\gamma \mbox{\boldmath$\Omega$}$. The spatial distributions of the turbulent magnetic-diffusivity effect $\beta {\bf{J}}$ and the turbulent cross-helicity effect $\gamma \mbox{\boldmath$\Omega$}$ (upper). The spatial distribution of the difference between the turbulent magnetic-diffusivity effect and the turbulent cross-helicity effect, $\beta {\bf{J}} - \gamma \mbox{\boldmath$\Omega$}$. Redrawn from \citet{yok2013b}.}
    \label{fig:localization_of turb_diff_by_ch}
\end{figure}

The other cross-helicity effect for the localization of the effective turbulent diffusivity effect may be caused by the suppression of the turbulent diffusivity effect $\beta {\bf{J}}$ due to the cross-helicity effect $\gamma \mbox{\boldmath$\Omega$}$. We see from the EMF expression given by (\ref{eq:emf_wo_alpha_sec7}) that the effective turbulent magnetic diffusivity effect $\beta {\bf{J}}$ can be altered by the turbulent cross-helicity effect $\gamma \mbox{\boldmath$\Omega$}$ as
\begin{equation}
	\beta {\bf{J}}
	\rightarrow
	\beta {\bf{J}} - \gamma \mbox{\boldmath$\Omega$}.
	\label{eq:reduct_eff_beta_effect_sec7}
\end{equation}
This suggests that in the presence of the turbulent cross helicity $W = \langle {{\bf{u}}' \cdot {\bf{b}}'} \rangle$ coupled with the mean vortical motion $\mbox{\boldmath$\Omega$} (= \nabla \times {\bf{U}})$, the turbulent magnetic diffusivity may be altered by the cross-helicity effect. Since the spatial distributions of the turbulent energy and the turbulent cross helicity are in general different, we have a possibility of the spatial localization of the effective turbulent magnetic diffusivity might occur. In a region where the magnitude of the turbulent cross helicity is relatively high, the suppression of $\|\beta {\bf{J}}\|$ by $\|\gamma \mbox{\boldmath$\Omega$}\|$ must be more effective. In this sense, the turbulent magnetic diffusivity effect is effectively suppressed there. On the other hand, in the region where the magnitude of turbulent cross helicity is relatively low, such a suppression effect must be reduced, resulting in relatively high turbulent magnetic diffusivity effect is sustained there. In Fig.~\ref{fig:localization_of turb_diff_by_ch}, the notion of the effective localization of the turbulent magnetic-diffusivity effect is schematically depicted.

Because of the anti-symmetric distribution of the turbulent cross helicity $W$ with respect to the coordinate, the turbulent cross-helicity effect should vanish in the vicinity of the X point. The sign reversal of the turbulent cross helicity is represented by the spatial distribution of the $\gamma \|\mbox{\boldmath$\Omega$}\|$. This shows marked contrast with the spatial distribution of the turbulent MHD energy $K$, which shows a broad positive-definite distribution throughout the space coordinate with a maximum magnitude near the X point. Since the sign of the turbulent cross helicity $W$ and consequently that of $\gamma$ is determined by the e mean vorticity $\mbox{\boldmath$\Omega$}$, the spatial distribution of $\gamma \mbox{\boldmath$\Omega$}$ is always positive [or equal to null at the X point] (Fig.~\ref{fig:localization_of turb_diff_by_ch} upper). This difference of the spatial distribution of $\beta J$ and $\gamma \Omega$ results in the effective localization of $\beta J - \gamma \Omega$. The strong turbulent magnetic diffusivity is confined to the vicinity of the X point, where suppression effect due to the cross helicity vanishes due to the sign reversal of the turbulent cross helicity.	

\begin{figure}[htb]
  \centering
  \includegraphics[width= 0.6 \columnwidth]{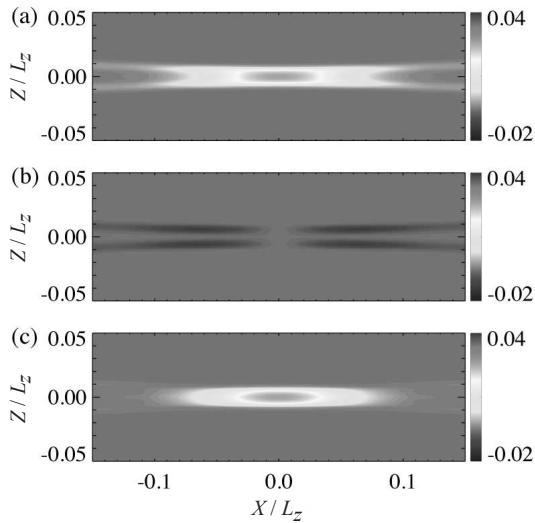}
  \caption{The balancing effect in the turbulent electromotive force (EMF). The spatial distributions of (a) the turbulent magnetic diffusivity effect, $\beta {\bf{J}}^z$, (b) the turbulent cross helicity effect, $\gamma \mbox{\boldmath$\Omega$}^z$, and (c) the difference between the turbulent magnetic diffusivity effect and the turbulent cross helicity effect, $\beta {\bf{J}}^z - \gamma \mbox{\boldmath$\Omega$}^z$. Redrawn from \citet{yok2013b}.}
    \label{fig:balancing_in_emf}
\end{figure}

Figure~\ref{fig:balancing_in_emf} shows the spatial distribution of the $z$ components of (a) the turbulent magnetic diffusivity effect, $- \beta {\bf{J}}$, (b) the turbulent cross-helicity effect, $\gamma \mbox{\boldmath$\Omega$}$, and (c) their summation, $- \beta {\bf{J}} + \gamma \mbox{\boldmath$\Omega$}$.

Reflecting the spatial distribution of the turbulent MHD energy $K$ represented by Fig.~\ref{fig:turb_en_localization_by_ch} (upper), the turbulent magnetic diffusivity effect $- \beta {\bf{J}}$ is broadly distributed in space including the long tails associated with the tails of the mean-electric current density in the four directions [see the contour of the mean electric-current density shown in Fig.~\ref{fig:J_contour} (middle)]. The cross helicity effect $\gamma \mbox{\boldmath$\Omega$}$ is relatively weak and vanishes in the vicinity of the $X$ point. This is expected from the anti-symmetric distribution of the turbulent cross helicity. Because of the relatively large distribution of the cross helicity in the tail regions, the $\gamma \mbox{\boldmath$\Omega$}$ certainly contributes to reducing the effects of $- \beta {\bf{J}}$ in such tail regions. As this reduction effect of the cross helicity, the turbulent magnetic diffusivity effect $- \beta {\bf{J}}$ is effectively localized near the X point region as conceptionally expected in Fig.~\ref{fig:localization_of turb_diff_by_ch}.

All these calculations of the turbulent MHD energy $K$ (and the turbulent magnetic diffusivity $\beta$) and the turbulent cross helicity $W$ (and the cross-helicity effect $\gamma$) were performed by simultaneously solving the $K$ and $W$ equations. But the timescale of turbulence $\tau$ was treated as a parameter as (\ref{eq:eps_algebra_model_sec7}) and (\ref{eq:epsW_algebra_model_sec7}) with (\ref{eq:parameter_tau_sec7}).

In order to solve the system of turbulence model equations in a self-consistent manner, the equation of the turbulence dissipation rate $\varepsilon$ has to be simultaneously solved \citep{wid2019}. In this work, the timescale of turbulence is determined self-consistently by the non-linear dynamics of turbulence through solving the $\varepsilon$ equations as well as the turbulent energy $K$ equation. The timescale is expressed in terms of the turbulent energy $K$ and its dissipation rate $\varepsilon$ by
\begin{equation}
	\tau = \frac{K}{\varepsilon}.
	\label{eq:timescale_K_eps_sec7}
\end{equation}	
In this sense, the level of turbulence is subject to the non-linear dynamics of the interaction between the turbulence and mean fields.

\begin{figure}[htb]
  \centering
  \includegraphics[width= 0.5 \columnwidth]{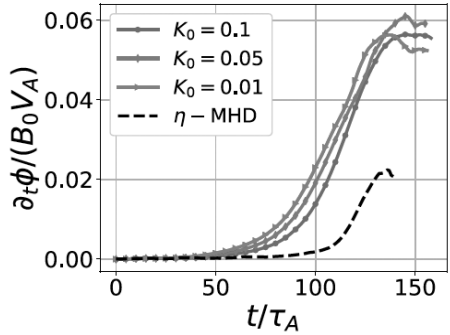}
  \caption{Temporal evolution of magnetic reconnection rate with variable initial turbulent energy. Temporal evolutions of the magnetic reconnection rate $\partial_t \phi / (B_0 V_{\rm{A}})$ with different initial turbulent energy levels ($K_0 = 0.01$, $0.05$, and $0.1$). The case without turbulence is also plotted as $\eta$-MHD case. Redrawn from \citet{wid2019}.}
    \label{fig:evol_rec_mag_flux_w_solv_eps}
\end{figure}

The temporal evolutions of the reconnection rate is examined with several initial level of turbulence (Fig.~\ref{fig:evol_rec_mag_flux_w_solv_eps}). The reconnection rate defined by the reconnected magnetic flux per time, $\partial_t \phi / (B_0 V_{\rm{A}})$, is plotted against time $t/\tau_{\rm{A}}$ for various initial turbulence levels ($K_0 = 0.01$, $0.05$, and $0.1$). Irrespective of the initial turbulence level, all the cases reaches the same level of reconnection rate, which is much higher than the resistive MHD case with no turbulence ($\eta$-MHD case). This is marked contrast with the previous simulations with the turbulence timescale $\tau$ being as a parameter (Fig.~\ref{fig:evol_rec_mag_flux_w_solv_eps}).	

In a self-consistent turbulence model, where the turbulence energy dissipation rate $\varepsilon$ is solved as well as the turbulent MHD energy $K$, the evolution of turbulence is subject to the non-linear dynamics of the turbulence itself. The timescale of turbulence is self-adjusted and the reconnection rate reaches at the same higher level (fast reconnection) irrespective of the initial turbulence level. This suggests that, if the turbulence is generated by inhomogeneity of the reconnection magnetic field and its energy is dissipated by the non-linear dynamics of itself, the turbulence level attains to a level which is appropriate to get a fast magnetic reconnection. The reconnection rate does not depend on the initial level of turbulence. This situation is entirely different from the case where the turbulence energy is injected by external forcing, or the case with the energy dissipation rate $\varepsilon$ is given as a parameter.

\begin{figure}[htb]
  \centering
  \includegraphics[width= 0.9 \columnwidth]{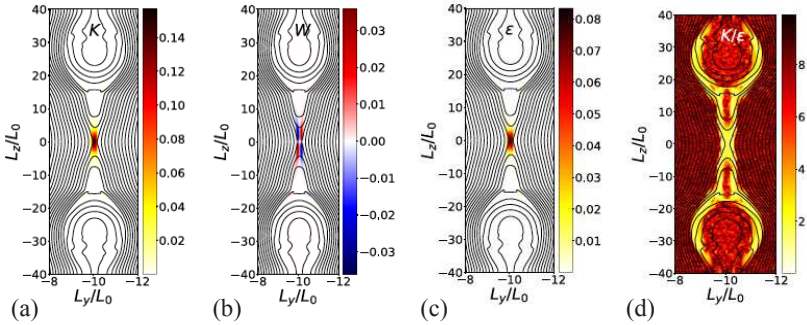}
  \caption{Spatial distributions of (a) the turbulent MHD energy $K$, (b) the turbulent cross helicity $W$, (c) the turbulent MHD energy dissipation rate $\varepsilon$ (second right), and (d) the turbulent timescale $\tau = K/\varepsilon$. Spatial profiles at the saturated magnetic reconnection at time $t/\tau_{\rm{A}} = 150$. The initial turbulence level is $K_0 = 0.01$, and the magnetic diffusivity is $\eta = 10^{-5}$. Redrawn from \citet{wid2019}.}
    \label{fig:spat_distr_K_W_eps_tau}
\end{figure}

However, the role of turbulent cross helicity in the magnetic reconnection is considered to be similar to the cases with the timescale given as a parameter. The turbulent cross helicity self-generated by the mean-field inhomogeneities contributes to localize the turbulent diffusivity effect, leading to a fast reconnection. This point can be seen in the spatial distributions of the turbulent MHD energy $K$, the turbulent cross helicity $W$, the turbulent MHD energy dissipation rate, and the turbulence timescale $\tau = K/\varepsilon$ in Fig.~\ref{fig:spat_distr_K_W_eps_tau}.

The spatial distributions of $K$ and $W$ are basically similar to the results in the simulations with the turbulence timescale $\tau$ given as a parameter (Fig.~\ref{fig:turb_en_localization_by_ch}). The spatial distribution of $\varepsilon$ is similar to the counterpart for $K$. This reflects the similarity of the $\varepsilon$ equation (\ref{eq:eps_eq_sec7}) in form with the $K$ equation (\ref{eq:K_eq_sec7}). However, the spatial distribution of the turbulence timescale defined by $\tau = K / \varepsilon$ is not uniform at all. This makes a certain difference from the case with constant parameter $\tau$. 

\begin{figure}[htb]
  \centering
  \includegraphics[width= 0.7 \columnwidth]{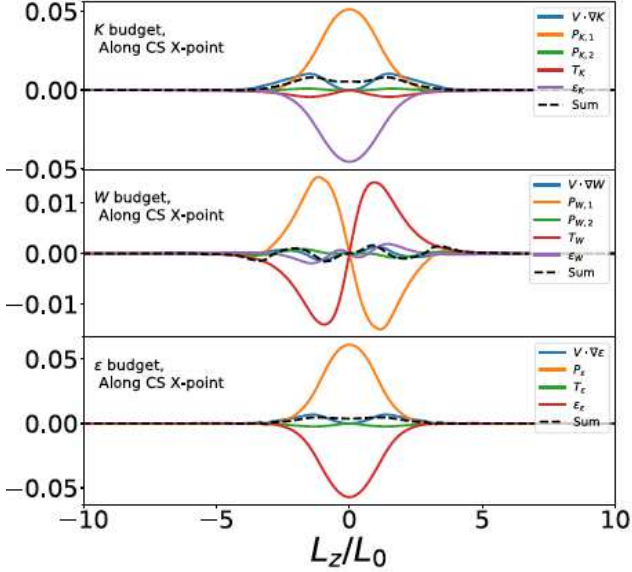}
  \caption{Budgets of the turbulent MHD energy $K$ (upper), cross helicity $W$ (middle), and the energy dissipation rate $\varepsilon$ (lower) are plotted along the current sheet including the X point. The budgets are plotted at time $\tau / \tau_{\rm{A}} = 120$, with the magnetic diffusivity $\eta = 10^{-4}$. Redrawn from \citet{wid2019}.}
    \label{fig:budgets_K_W_eps}
\end{figure}

The production, dissipation, and transport rates of the turbulent statistical quantities, $K$, $W$, and $\varepsilon$, obey the evolution equations (\ref{eq:K_eq_sec7}), (\ref{eq:W_eq_sec7}) and (\ref{eq:eps_eq_sec7}), respectively. The budgets of $K$, $W$ and $\varepsilon$ are shown in Fig.~\ref{fig:budgets_K_W_eps}.

In a similar way as we saw in the case with a parameter $\tau$, the generation of the turbulent MHD energy $K$ is mainly attributed to the production rate $P_{K,1} = - \beta {\bf{J}}^2$ in (\ref{eq:K_eq_sec7}) (Fig.~\ref{fig:budgets_K_W_eps}, upper). In addition, the transport term associated with the turbulent cross-helicity inhomogeneity along the mean magnetic field, $T_K = ({\bf{B}} \cdot \nabla) W$ in (\ref{eq:K_eq_sec7}), contributes to the localization of the $K$ distribution by the negative production of $K$ at the tail region of the $P_{K,1}$. This is one of the prominent effects of the turbulent cross helicity in enhancing the magnetic reconnection.

The generation of the turbulent cross helicity $W$ is mainly attributed to the production rate associated with the coupling of the mean vorticity and mean electric-current density $P_{W,1} = - \beta {\bf{J}} \cdot \mbox{\boldmath$\Omega$}$ in (\ref{eq:W_eq_sec7}) (Fig.~\ref{fig:budgets_K_W_eps}, middle). This effect is to some extent canceled by the transport term associated with the inhomogeneity of the turbulent MHD energy along the mean magnetic field, $T_W = ({\bf{B}} \cdot \nabla) K$ in (\ref{eq:W_eq_sec7}).

The turbulent cross helicity is ubiquitously present around the reconnection point because of the production mechanisms of the turbulent cross helicity associated with the reconnection magnetic field ${\bf{B}}$ and its spatial variations. This ubiquitous presence of the turbulent cross helicity around the reconnection point provides an environment for the cross helicity to work for the fast magnetic reconnection, through the cross-helicity effects in magnetic-field induction and the momentum and angular momentum transport. The turbulent cross helicity coupled with the mean velocity shear induces a mean magnetic field, and the turbulent cross helicity coupled with the mean magnetic-field shear induce a mean velocity. These cross-helicity effects are considered to contribute to the enhancement of the magnetic reconnection rate by changing the configurations of the reconnection magnetic-field and velocity configurations.

\NOTE{
	However, we should note that, in contrary to the numerical simulations presented above, the non-trivial cross-helicity configuration such as the quadrupole distribution is not reported in some numerical simulation \citep{now2022}. This difference should be related to the turbulence generation mechanism. In our numerical setup discussed above, the turbulence is self-generated by the inhomogeneous magnetic field and velocity configuration. In such a case, the non-trivial turbulent cross helicity distribution is naturally realized. This is fairly different from the case where turbulence is homogeneously generated by external forcing. In order to clearly understand the role of cross helicity in the fast reconnection in turbulence, we have to further investigate the conditions for the turbulent cross-helicity production in the turbulent reconnection.
}

\section{Concluding remarks}

From the viewpoint of transport, the primary effect of turbulence is enhancing the process of mixing. The eddy viscosity representation of the Reynolds stress in the mean momentum equation and the turbulent magnetic diffusivity in the turbulent EMF in the mean magnetic-field equation are representative turbulence effects, which enhance the process of mixing. On the other hand, we observe several amazing persistent large-scale structures and their generation in extremely strong turbulence. In order to elucidate such large-scale structures in turbulence, we need some mechanisms other than the eddy viscosity and diffusivity. These other mechanisms should contribute to {\it the counter diffusion} effect that counterbalances the turbulent viscosity and diffusion effects and suppresses the enhanced transport. The effects of pseudo-scalars such as the kinetic, current, and cross helicities, as well as the non-equilibrium effect associated with the coherent fluctuation motions (plumes, thermals, and jets), are the representative candidates for the counter diffusion mechanisms.

	In this paper, with the aid of a multiple-scale renormalized perturbation expansion theory, combination of the multiple-scale analysis and the direct-interaction approximation (DIA), the analytical expressions of the turbulent EMF and the Reynolds and turbulent Maxwell stresses are systematically obtained from the fundamental equations. As the direct consequence of the introduction of the non-mirror symmetric components of the lowest-order (or background) fluctuation fields, the pseudo-scalar statistical quantities as well as the pure-scalar ones, enter the expression of the turbulent fluxes. 
	
	The importance of the cross-helicity effects in the magnetic-field induction and momentum transport are stressed. In addition to the kinetic- and current-helicity effects in dynamo ($\alpha$ effect) and the inhomogeneous kinetic helicity effect in vortex dynamos, the cross-helicity effects enter the expressions of the turbulent EMF and the Reynolds and Maxwell stresses. These pseudo-scalar effects are expected to represent {\it the counter- or anti-diffusion} effect that balances with the eddy diffusivity and viscosity effects, which represent the enhancement of the effective transport due to turbulence.
\begin{equation}
	{\bf{E}}_{\rm{M}}
	= \overbrace{- \eta_{\rm{T}} {\bf{J}}}
		^{\substack{
			\mbox{turb.\ mag.}\\ 
			\mbox{diffusivity}}}
	\underbrace{+ \gamma \mbox{\boldmath$\Omega$}}
		_{\substack{
			\mbox{cross-hel.}\\ 
			\mbox{effect}}}
	\underbrace{+ \alpha {\bf{B}}}_{\substack{
			\mbox{hel.}\\ 
			\mbox{effect}}},
\end{equation}
\begin{equation}
	\mbox{\boldmath${\cal{R}}$}_{\rm{D}}
= \overbrace{
  - \nu_{\rm{K}} \mbox{\boldmath${\cal{S}}$}_{\rm{D}}
  }^{\substack{
      \mbox{turb.}\\
      \mbox{viscosity}
    }
  }
\underbrace{
  + \nu_{\rm{M}} \mbox{\boldmath${\cal{M}}$}_{\rm{D}}
  }_{
    \substack{
      \mbox{cross-hel.}\\
      \mbox{effect}
    }
  }
\underbrace{
  + (\mbox{\boldmath$\Gamma$} 
      \mbox{\boldmath$\Omega$}_\ast)_{\rm{D}}
  }_{
    \substack{
      \mbox{hel.}\\
      \mbox{effect}
    }
  },
\end{equation}
where the suffix ${\rm{D}}$ denotes the deviatoric or traceless part of a tensor.
	
\begin{figure}[htb]
  \centering
  \includegraphics[width= 0.7 \columnwidth]{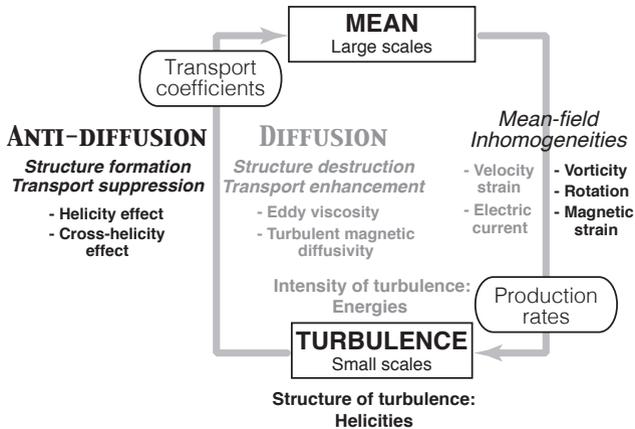}
  \caption{Self-consistent turbulence modeling of the mean and turbulence fields. The balance between the diffusion and anti-diffusion effects should be also treated in a self-consistent closure framework.}
    \label{fig:self-consistent_model}
\end{figure}

	Another point stressed in this article is the importance of a self-consistent and systematic treatment of the turbulence effect in the mean-field equations. In a self-consistent description of turbulent phenomena, the mutual interaction between the mean and turbulence fields should be simultaneously and fully considered (Fig.~\ref{fig:self-consistent_model}). In other words, this point is called the closure problem, and this has not been fully explored in the previous studies of dynamos and momentum transport. For instance, in some dynamo studies, quenching or saturation of dynamo effect has been discussed by considering each effect of the mean magnetic field, rotation, diffusivity, etc. However, such arguments are often based on picking up some specific terms (mean magnetic field, rotation, molecular diffusivity, viscosity, etc.). So, the validity of such quenching arguments highly depends on specific configurations considered.

	The transport coefficients in the expressions of the turbulent fluxes are determined by the statistical properties and dynamical behavior of the turbulence. We performed a multiple-scale renormalized perturbation expansion analysis on the MHD turbulence. We systematically obtained the expressions for the turbulent fluxes in the mean-field equations, including the analytical expressions of the transport coefficients. On the basis of these analytical expressions, the coefficients are represented by appropriate one-point turbulent statistical quantities such as the turbulent energy, kinetic, current, and cross helicities. By simultaneously considering and solving the transport equations of the statistical quantities, as well as the mean-field equations, a self-consistent system of model equations was constructed. In this framework, the turbulent transport coefficients in the mean-field equations are determined through the transport equations of the turbulent statistical quantities. These transport equations include the production, dissipation, and transport rates of the turbulent statistical quantities, which are subject to the mean-field inhomogeneities. As such, this turbulence model can self-consistently determine the evolutions of the mean- and fluctuation-fields. In this sense, the present approach provides a new framework that can treat global structure formation in extremely strong turbulence far beyond the previous heuristic turbulence modeling approach \citep{yok2018c}.

\backmatter

\bmhead{Acknowledgments}

The author would like to thank Pat Diamond for inviting him to write this article on the topic of the cross-helicity effects. This work was supported by the Japan Society of the Promotion of Science (JSPS) Grants-in-Aid for Scientific Research JP18H01212. Part of this work was done during his stay at Nordic Institute for Theoretical Physics (NORDITA) in May and June 2022. The author also would like to thank the Isaac Newton Institute for Mathematical Sciences, Cambridge, for support and hospitality during the programme DYT2 where work on this paper was undertaken. This work was supported by EPSRC grant no EP/R014604/1.

\section*{Declarations}
\NOTE{
The author has no conflict of interest.
}

\bibliographystyle{mnras}


\bibliography{sn-bibliography}


\end{document}